\documentclass[11pt,letterpaper,english]{article}
\usepackage{microtype}
\usepackage{float}
\usepackage{amsmath} 
\usepackage{amsthm}
\usepackage{mathrsfs}
\usepackage{graphicx} 
\usepackage{caption}
\usepackage{subcaption}
\usepackage{setspace}
\usepackage{amssymb}
\usepackage{multirow}
\usepackage{rotating}
\usepackage{array}
\usepackage[toc,page]{appendix}

\usepackage{booktabs}
\usepackage{multirow}
\usepackage{listings}
\usepackage{comment}
\usepackage[small,compact]{titlesec} 
\usepackage{siunitx} 
\usepackage{indentfirst}
\usepackage{amsfonts} 
\newcommand{\sym}[1]{\rlap{#1}} 

\usepackage{xr}
\usepackage{algorithm}
\usepackage{algpseudocode}
\usepackage{xcolor}
\usepackage{times}
\usepackage{subcaption}
\definecolor{darkblue}{rgb}{0.0, 0.0, 0.55}
\usepackage[hyphens]{url} 
\usepackage[colorlinks=true, pdfborder={0 0 0}, linkcolor=black,citecolor=darkblue,urlcolor=darkblue]{hyperref}
\usepackage{url}
\usepackage{ragged2e}


\usepackage[margin=2.5cm]{geometry}
\usepackage[sectionbib,round]{natbib}

\usepackage{listings}
\bibpunct[, ]{(}{)}{;}{a}{}{,}

\definecolor{brightlavender}{rgb}{0.75, 0.58, 0.89}
\newcommand{\hy}[1]{\textcolor{blue}{#1}}

\usepackage{graphicx} 
\theoremstyle{plain}

\newtheorem{definition}{Definition}

\newcommand{\squishlist}{
   \begin{list}{$\bullet$}
    { \setlength{\itemsep}{0pt} \setlength{\parsep}{1pt}
      \setlength{\topsep}{1pt} \setlength{\partopsep}{1pt}
      \setlength{\leftmargin}{1.5em} \setlength{\labelwidth}{1em}
      \setlength{\labelsep}{0.5em} } }

\newcommand{\squishlisttwo}{
   \begin{list}{$\bullet$}
    { \setlength{\itemsep}{0pt} \setlength{\parsep}{0pt}
      \setlength{\topsep}{0pt} \setlength{\partopsep}{0pt}
      \setlength{\leftmargin}{1em} \setlength{\labelwidth}{1.5em}
      \setlength{\labelsep}{0.5em} } }

\newcommand{\squishend}{
    \end{list}  }
\apptocmd{\thebibliography}{\RaggedRight}{}{}

\begin{document}

\title{Balancing Engagement and Polarization: Multi-Objective Alignment of News Content Using LLMs}

\author{\small Mengjie (Magie) Cheng$^+$\thanks{
We thank Simha Mummalaneni, Omid Rafieian, Jesse Shapiro, Zikun Ye, and Shunyuan Zhang for detailed feedback that has significantly improved the paper. Please address all correspondence to: macheng@hbs.edu, eofek@hbs.edu, and hemay@uw.edu.} \and \small Elie Ofek$^+$ \and \small Hema Yoganarasimhan$^\#$} 
\date{
    \small $^+$Harvard Business School\\%
    \small $^\#$University of Washington\\ [2ex]%
}

\maketitle
\begin{abstract}

We study how media firms can use LLMs to generate news content that aligns with multiple objectives -- making content more engaging while maintaining a preferred level of polarization/slant consistent with the firm's editorial policy. Using news articles from {\it The New York Times}, we first show that more engaging human-written content tends to be more polarizing. Further, naively employing LLMs (with prompts or standard Direct Preference Optimization approaches) to generate more engaging content can also increase polarization. This has an important managerial and policy implication: using LLMs without building in controls for limiting slant can exacerbate news media polarization. We present a constructive solution to this problem based on the Multi-Objective Direct Preference Optimization (MODPO) algorithm, a novel approach that integrates Direct Preference Optimization with multi-objective optimization techniques. We build on open-source LLMs and develop a new language model that simultaneously makes content more engaging while maintaining a preferred editorial stance. Our model achieves this by modifying content characteristics strongly associated with polarization but that have a relatively smaller impact on engagement. Our approach and findings apply to other settings where firms seek to use LLMs for content creation to achieve multiple objectives, e.g., advertising and social media.

\end{abstract}

\thispagestyle{empty}
\newpage
\setcounter{page}{1}
\section{Introduction}
\label{sec:intro}
\subsection{LLMs and Generative AI in News Media}


Recent advances in Large Language Models (LLMs) and generative AI are transforming content creation in the news and journalism industry. Prominent media companies such as the \textit{Associated Press}, \textit{Forbes}, \textit{The New York Times}, \textit{The Washington Post}, and \textit{ProPublica}, have all adopted AI to help refine or generate articles and automate reporting on crimes, financial markets, politics, sporting events, and foreign affairs \citep{marconi2020newsmakers}. For example, in September 2020, \textit{The Guardian} published an article entirely written by GPT-3 \citep{gudardian2020}. More recently, since January 2023, \textit{CNET} has been using automation technology to generate news articles and financial advice, accompanied by a disclaimer: ``This article was generated using automation technology and thoroughly edited and fact-checked by an editor on our editorial staff" \citep{cnet2023}. Other publishers, such as \textit{BuzzFeed} and \textit{Men’s Journal}, have adopted similar practices.\footnote{\textit{BuzzFeed} utilizes an AI assistant, Buzzy, and notes that articles are “collaboratively written" by its human staff and Buzzy \citep{buzzfeed2023}. \textit{Men’s Journal} explains its article generation approach as ``a curation of expert advice from Men’s Fitness, using deep-learning tools for retrieval combined with OpenAI’s LLMs for various stages of the workflow” \citep{wsj2023}.}

The use of generative AI in the media sector is appealing for several reasons. First, LLMs can rapidly create large volumes of content at a lower cost than humans. Indeed, the GPT-3 technical report \citep{brown2020language} shows that, according to human evaluations, LLMs can generate news articles that are often indistinguishable from those written by journalists. This efficiency presumably frees journalists to focus on more complex tasks, such as investigative reporting and in-depth analysis \citep{dorr2016mapping}.  Second, generative AI can be used to produce content that boosts reader interest and engagement. Increasingly, editors are using LLMs to refine rough drafts into more polished and compelling articles, which resonate with readers and improve the articles' appeal in the competitive media landscape \citep{brighamdeveloping}. As a result, the integration of generative AI into news production can not only streamline editorial workflows but also potentially bring tangible economic benefits -- attracting greater traffic, driving advertising revenue, and increasing subscription conversions \citep{nishal2024envisioning}.

\subsection{Challenges and Research Agenda}
While making content more appealing and engaging for readers is a goal most, if not all, media firms strive to achieve,
editors and journalists may seek to accomplish additional objectives. 
For example, many media outlets strive to portray a particular political position (or slant) in order to align with the preferences of their readership or to safeguard their reputation \citep{mullainathan_shleifer_2005, gentzkow2006media}. Notably, \emph{The New York Times} and \emph{CNN} are generally considered to be more left-leaning, while \emph{Fox News} is commonly regarded as being more right-leaning in its editorial stance \citep{flaxman2016filter}. 
Consequently, as media firms embrace LLMs in an effort to generate engaging content, they may need to consider whether other objectives, such as desired political stance are inadvertently impacted.\footnote{Throughout the paper, we use the terms editorial/political stance, preferred level of polarization, and extent of slant interchangeably. Intuitively, all these terms refer to the direction and degree of slant (right/left-leaning) preferred by the editorial team.} Indeed, recent research in computer science has uncovered instances whereby leveraging LLMs to achieve one objective may unintentionally interfere with other objectives. For example, work in the realm of chatbot safety reveals that training LLMs to provide answers that are helpful can sometimes result in output that is harmful to individuals or the public \citep{safe-rlhf, beavertails, liu2024enhancing}.\footnote{For instance, although an LLM may be trained to provide users with effective answers to any query, caution may be desired in response to requests that can put the user or others at risk of injury. Thus, prompts such as ``how to make a bomb'' or ``how to commit suicide'' should trigger a careful response that does not simply provide the user with instructions on how to accomplish these tasks.} These observations underscore the importance of carefully considering potential interactions between different generative objectives when deploying LLMs.

In the context of using LLMs for news article creation and refinement, ex-ante, it is not entirely clear what the nature of the interaction between the objectives of engagement and editorial stance will likely be. That is, will making content more engaging have any effect on how politically polarizing it becomes, and if so, will the level of polarization tend to increase or decrease?

In this paper, we study the problem of how media firms can use LLMs to align news content on multiple objectives. Specifically, we focus on how firms can make news content more engaging while concomitantly controlling the level of polarization/slant to keep it consistent with the firm's editorial stance. In the first part of our paper, we document a common unintended consequence of using LLMs to boost news content engagement: such efforts can often increase its polarization level (e.g., a moderately left-leaning article will become extremely left-leaning). This is because the two language objectives -- engagement and polarization -- tend to move together. Hence, simple prompt-based approaches designed to make content more engaging can also make it more polarizing. To address this challenge, in the second part of the study, we present a framework for multi-objective alignment of LLMs in media settings. We build on open-source LLMs and develop a new language model policy that simultaneously makes content more engaging while maintaining a preferred editorial stance (i.e., controlling the extent of polarization); we further establish the effectiveness of the proposed model on several metrics. Finally, in the third part of the paper, we examine how our language model, designed to accommodate multiple objectives, differs in its content creation strategies compared to simpler language models fine-tuned for a single objective (i.e., only maximizing engagement).

\subsection{Our Approach and Key Findings}
We use a large dataset consisting of 13,508 news articles from \textit{The New York Times}, published between January 1, 2019, and May 30, 2021, for our main study. These articles span a wide range of topics, including health, business, politics, and the Covid-19 pandemic. For each article, we first use Latent Dirichlet Allocation (LDA) analysis to obtain its topic distribution and then employ the ``GPT-4o-mini" model to assess both engagement and polarization levels, which we further validate using human ratings on a random subset of approximately 1,000 articles.

We find that more engaging articles in our corpus also tend to be more polarizing, even after controlling for the topics covered by the article. This suggests that certain language aspects, such as structure, flow, and tone, simultaneously affect how engaging the content is as well as how polarizing it is. To further explore the interplay between engagement and polarization, we conduct a simple exercise that mimics a common use case of LLMs by media firms: We provide a baseline version of each article to an off-the-shelf LLM (``GPT-4o-mini") and prompt it to rewrite these articles in a more (less) engaging fashion. We then reassess both engagement and polarization levels for the recreated articles. We find that the LLM-generated versions have higher (lower) engagement scores compared to the original versions. However, this increase (decrease) in engagement comes with an unintended consequence -- greater (lower) polarization, on average, relative to the original corpus. 

Furthermore, the magnitude of this unintended effect on polarization is positively correlated with the degree of change in engagement scores: articles that experience greater increases (or decreases) in engagement also exhibit proportional increases (or decreases) in polarization. In sum, these observations suggest that naively adjusting the engagement level of news content using off-the-shelf LLMs and prompt-based approaches can inadvertently change an article’s overall polarization/slant. This, in turn, can potentially make the content inconsistent with the preferred editorial stance of the media firm. 



To tackle this problem, in the second part of the paper, we propose a solution that builds on open-source LLMs (e.g., Llama or Alpaca), is explicitly calibrated to reflect the firm's preferred balance of objectives through a weighting vector, and can be trained on the firm's data. Specifically, we develop a new language model policy that simultaneously makes content more engaging while aiming to achieve a preferred editorial stance (or control polarization). Our framework leverages the Multi-Objective Direct Preference Optimization algorithm \citep[MODPO;][]{zhou2024beyond}, a novel approach that combines Direct Preference Optimization \citep[DPO;][]{rafailov2024direct} with multi-objective optimization techniques \citep{van2014multi, li2020deep}, enabling the LLM to satisfy multiple, possibly conflicting, objectives in a principled and scalable way.


MODPO first employs a linear scalarization strategy to transform a multi-objective problem into a single-objective one. This can be accomplished by assigning weights to each objective and combining them into a weighted sum, allowing firms to pre-specify the emphasis to be given to each objective. By enumerating different weight vectors, we can approximate the Pareto front of optimal language models, thereby offering firms a spectrum of solutions tailored to their strategic priorities. Once the weighted objective function is defined, MODPO follows four key phases: Phase 0 – Model Pre-training, Phase 1 – Supervised Fine-Tuning, Phase 2 – Marginal Reward Modeling, and Phase 3 – Language Modeling. 
We take the pre-training stage (Phase 0) as given and use a state-of-the-art open-source LLM -- Meta's Llama-3-8b \citep{llama2024} -- as our base model. Next, in the Supervised Fine-Tuning (SFT) stage (Phase 1), we enhance the pre-trained model’s ability to follow instructions for a given task using demonstration data. In our main analysis, this task involves generating a more engaging version of each original article. To achieve this, we construct a demonstration dataset by pairing the original content with its more engaging counterpart, resulting in an \textit{engaging SFT model}. However, in our setting, since we need to account for the agent's latent preference for both engagement and polarization, we introduce the marginal reward modeling phase (Phase 2), where we learn a reward function for the polarization objective based upon a comparison dataset that captures the agent's preferences on this dimension. Lastly, in the language modeling phase, we incorporate the learned reward model from Phase 2 as an adjustment term into the single-objective DPO loss function. This gives us a new language model, denoted as the \textit{multi-objective DPO model}, designed to generate content that is more engaging without being excessively polarizing.


To ensure a robust evaluation, we split our demonstration and comparison data into training (80\%), validation (10\%), and test (10\%) sets. The test set is used for evaluation through Gemini, Claude3, and human raters. As a baseline for comparison, we also develop an \textit{engaging DPO model}, which is designed to generate content optimized solely for engagement by applying the standard DPO approach \citep{rafailov2024direct} to maximize the engagement objective without considering polarization. We then recreate all the articles in our test dataset using both the \textit{multi-objective DPO model} and the \textit{engaging DPO model}, evaluate these recreated versions on engagement and polarization, and compare their performance to the original versions. 

We find that the \textit{multi-objective DPO model} effectively increases engagement without significantly increasing polarization, i.e., it adheres closely to the editorial stance of the firm per the original articles. In contrast, the \textit{engaging DPO model} generates articles that are more engaging, but also more polarizing. This suggests that aligning an LLM to a single objective (i.e., enhance engagement) using DPO can lead to unintended consequences, similar to those observed in naive prompt-based approaches. Overall, our results suggest that media firms can effectively leverage LLMs to generate content that is more engaging/appealing to users while minimizing unintended negative consequences such as higher polarization. However, achieving this balance requires a thoughtful approach that can benefit from combining open-source LLMs with multi-objective alignment methods to ensure development of a language policy that accommodates the multiple objectives. 

We also examine two possible levers to modify the extent to which the \textit{multi-objective DPO model} balances the two objectives -- (1) the weight vector, which represents the relative importance assigned to engagement and polarization in Phase 3, and (2) the baseline task used for the SFT model created in Phase 1. As the relative weight on the polarization objective increases, the language model policy changes -- it starts producing content that is less polarizing (and less engaging). Similarly, when changing the reference policy from the \textit{engaging SFT model} to a \textit{less polarizing SFT model}, the final language model shifts towards a policy that is inherently less polarizing. Thus, if a media firm wants to adopt a more neutral political stance, it can do so by increasing the weight on the polarization objective or by using this objective directly for supervised fine-tuning. However, given the tension between the two objectives, reducing polarization/ideological slant typically comes at the cost of some engagement. Therefore, depending on the extent to which a firm cares about adhering to a preferred editorial stance vs. increasing engagement, it can strategically use these two levers to calibrate the final language model.

In the third part of the paper, we examine \emph{how} the \textit{multi-objective DPO model} enhances engagement while controlling polarization. We focus on 12 theory-driven content strategies and find that our language model achieves this balance by adjusting specific content features that greatly impact polarization but that have a relatively smaller effect on engagement, such as \textit{provocative language} and \textit{balanced perspective}. Hence, in the process of aligning multiple conflicting objectives, it seems that the language model learns to identify and adjust content structures and language elements that can simultaneously and effectively balance both objectives.

Finally, we present an extensive set of robustness checks and extensions to validate our approach and findings. These include: (1) comparing the results of our proposed model to those from a prompt-engineering approach that uses ChatGPT, (2) using an alternative open-source LLM (Alpaca) as the base model, (3) ruling out hallucination as a concern, (4) validating the results with a real-world news dataset where original articles are classified by their political stance, and (5) extending our main analysis 
to train a \textit{multi-objective DPO model} that prioritizes left-leaning perspectives while accounting for right-leaning views; this enables flexible control over the degree of left- or right-leaning slant. Across all these robustness checks and extensions, we find consistent support for our findings and approach.

\subsection{Contributions}
Our paper makes several key contributions to the literature. First, from a substantive perspective, we establish that human-written news articles in our dataset exhibit a positive relationship between engagement and polarization. Specifically, the linguistic choices that enhance engagement, such as the selection of words, tone, and emphasis on (or omission of) specific details, also tend to increase polarization. Furthermore, we show that naive prompt-based approaches to boost engagement using LLMs can inadvertently intensify polarization and cause the recreated content to deviate from the firm's intended editorial priorities. Second, from a methodological perspective, we provide a constructive solution to this problem. We leverage the recently developed MODPO algorithm and demonstrate how it can be applied to news media settings to effectively enhance engagement while maintaining a preferred level of slant. Our approach is built on open-source LLMs, is relatively light on data and computational resources, and can flexibly accommodate multiple objectives and varying preferences over these objectives. Third, from a managerial perspective, our work provides a practical framework for media firms seeking to create content that is both engaging and aligned with their editorial priorities. We further note that the framework is not only applicable to news media but also extends to a variety of business contexts where firms use LLMs for content creation to achieve multiple objectives. For instance, it can inform the design of advertising copy that balances brand voice and audience engagement or social media posts that depict brand sponsorship as well as retain creator authenticity and style. Fourth, we contribute to the understanding of how the \textit{multi-objective DPO model} balances multiple objectives by demonstrating how it leverages content characteristics that strongly influence one objective but have a relatively smaller impact on the other objectives.

Finally, from a public policy perspective, our research has significant implications for the impact of generative AI and LLMs on media polarization. A growing body of literature shows that media polarization has been rising over the past few decades \citep{skelley_fuong_2022} and that consumer trust in the media industry is declining \citep{trust2024}. Our research suggests that the use of generative AI without proper controls to limit polarization can further exacerbate these concerns. We offer a viable solution to this predicament -- showing how the thoughtful use of LLMs with careful alignment can help policymakers, news aggregators, and media firms limit polarization while enhancing engagement. 

\section{Related Literature}
\label{sec:related_lit}
First, our work contributes to the growing body of research on the role of LLMs in marketing applications. Early studies on LLMs in marketing primarily focused on simulating human behavior and examining their potential in market research \citep{li2024frontiers, gui2023challenge, brand2023using, wang2024large}. More recently, researchers have begun to examine the potential of LLMs for content creation and experimentation. \cite{ye2024lola} introduce LOLA, an LLM-Assisted Online Learning Algorithm framework that integrates LLMs with adaptive experimentation to optimize content delivery. \citet{angelopouloscausal2024} fine-tune a language model to generate new email creatives for marketing communication and evaluate their performance through a large-scale field experiment. \citet{ellickson2024using} combine the contextual embeddings of LLMs with customer demographics and prior treatment outcomes to model subsequent heterogeneous treatments of content creatives. We add to this literature by introducing a multi-objective LLM content alignment approach, with an application to balancing content engagement and polarization/editorial slant in news media settings.

Second, our work also contributes to the literature on how language shapes consumer behavior and market outcomes. Much of this research has focused on understanding which aspects of language drive engagement and content diffusion \citep{toubia2017idea, toubia2021quantifying, netzer2019words, berger2021makes, berger2023holds}. Recent studies, such as \citet{banerjee2024language} and \citet{batista2024words}, have examined hundreds of cognitive, motivational, affective, and grammatical cues to understand their impact on engagement with news headlines. While some linguistic constructs, such as emotional intensity \citep{andersen2004experimental}, are found to consistently enhance engagement, others yield mixed findings. Building on this literature, our work not only investigates how language affects the appeal of news content but also examines a critically underexplored outcome: polarization. Additionally, we analyze the relative importance of various content features and explore how our algorithm leverages these features to balance the dual objectives of fostering engagement and managing polarization to ensure alignment with the firm's editorial stance.

Finally, our research relates to the literature on media bias and polarization \citep{xiang2007news, bondi2023privacy, mosaffa2025}. Although a large body of work has documented the existence of polarization in media text and political speech using Natural Language Processing (NLP) techniques \citep{groseclose_milyo_2005, gentzkow2010drives, martin2017bias, gentzkow_etal_2019}, to the best of our knowledge we are the first to document that language shapes both engagement and polarization, and that polarization can be an unintended consequence of media firms' desire to offer more engaging content. We further show that naively using LLMs to increase engagement can make news content more polarizing, thus exacerbating the overall polarization level of the news landscape. We then provide a constructive solution based on multi-objective alignment that can be used to boost engagement while managing polarization using LLMs.

\section{Setting and Data}
\label{sec:data}

Our data consists of $13,508$ widely shared articles from \textit{The New York Times} published between January 1, 2019, and May 30, 2021. The dataset was initially assembled by \citet{yoganarasimhan2024feeds} to study users' sharing patterns of news articles in emails and on Facebook. This corpus serves as an excellent setting for our research since it covers a wide variety of news articles over a sufficiently long period of time. Below, we provide a brief description of the data and refer interested readers to the earlier study for details of the data collection process. 

For each article in the data, we have information on its metadata (which includes the headline, publication date, abstract, and section) as well as the full text of the article. Further, we perform an LDA analysis and obtain the distribution of topics for each article. Figure \ref{fig:topic} in Web Appendix $\S$\ref{app:topics_nytimes} shows the 40 most common topics across all articles in the corpus. The most prevalent topics are \textit{Family}, \textit{Politics}, and \textit{Emotions and Feelings}, while the least prevalent are \textit{Judaism}, \textit{Pets and Animals}, and \textit{Horse Racing and Farms}. 

Next, we obtain the engagement and polarization scores for each of these articles using ``GPT-4o mini", a LLM developed by OpenAI, known for its strong performance across diverse language tasks \citep{gpt4o2024}. We ask the model to score each article on a scale of 1 to 10 on both engagement and polarization, with higher values indicating the article is more engaging or polarizing, respectively. To ensure output consistency, we set the model's ``temperature" parameter to zero and score each article on both dimensions once.\footnote{The GPT model is stochastic and can generate different responses to the same prompt over different iterations. Setting the temperature to zero ensures stable and consistent responses.} For details of the prompts used, please refer to Web Appendix $\S$\ref{app:prompts_syntheticdata}. The engagement score captures the extent to which an article attracts the readers' attention and keeps them interested, while the polarization score captures the extent to which an article is considered to be politically divisive.\footnote{Given the source (\textit{The New York Times}), articles in our corpus tend to be left-leaning to neutral. So, when we ask GPT for a polarization score, we are effectively asking it to score the article on the extent of leftward slant. We also consider an alternative scoring approach, where we define polarization as the extent of extreme views/left-leaning bias in the article and repeat the scoring procedure. The results are largely consistent with those shown here.} Table \ref{tab:originaldata} shows the summary statistics for the engagement and polarization scores of the articles in our corpus.

\begin{table}[t]
\caption{\small Summary Statistics of the Engagement and Polarization in the Original Corpus}
\centering
\small
\begin{tabular}{c|c|c}
\hline
      & Engagement  & Polarization  \\ \hline
Median  & 8.00       & 6.00     \\
Mean  & 7.68       & 5.53   \\
Std   & 0.82       & 2.37 \\
N & 13,508.00        & 13,508.00 \\\hline
\end{tabular}
\label{tab:originaldata}
\end{table}

We expect the scores obtained from the ``GPT-4o mini" model to be realistic and sufficiently accurate representations of human user ratings for several reasons. First, LLM models have been shown to possess remarkable zero-shot learning proficiency, which allows them to understand and respond to queries they were not explicitly trained for \citep{xie2021explanation}. This adaptability enables them to handle structured questions on possibly unfamiliar tasks, such as scoring news articles on engagement and polarization. Second, prior research supports the reliability of LLMs in the context of news content. For example, the GPT-3 technical report \citep{brown2020language} demonstrates that LLMs can generate news articles that are often indistinguishable from authentic human-written content, as indicated by human evaluations. These capabilities have further improved with the development of more recent LLM models. Third, recent studies provide direct evidence that LLMs are capable of understanding polarization. For example, \citet{yoganarasimhan2024feeds} compares ``GPT-3.5-turbo" ratings and human ratings for polarization scores on this dataset and shows that GPT ratings align well with those of humans, on average. More recently, \citet{fang2024shifting} use OpenAI’s ``text-embedding-ada-002" model to evaluate the review text and unravel review polarization. Finally, to further validate the reliability of the scores in our case, we also collect human ratings for engagement and polarization on a random subset of approximately 1,000 articles in the data; see Web Appendix $\S$\ref{app:humancoding_eval}. We find a strong correlation between user ratings and GPT-4o mini's ratings, with correlation coefficients of 0.72 for polarization and 0.65 for engagement.\footnote{Prior research has shown that even among human raters, the agreement rate is only about 70\% in most annotation tasks \citep{ouyang2022training, liu2020learning}. Thus, the agreement we find between GPT and human ratings is similar to that between human annotators. } 

\section{Relationship between Engagement and Polarization}
\label{sec:eng_pol}
We now examine the relationship between an article's engagement and polarization scores. First, we regress the polarization score of an article on its engagement score and present the results in column (1) of Table \ref{tab:allarticles}. The results suggest a strong and significant positive relationship -- more engaging articles in our corpus tend to be more polarizing ($\beta$ = 1.092, p $<$ 0.001). However, this positive relationship may simply reflect the fact that certain topics are both more engaging and more polarizing. Therefore, in column (2) of Table \ref{tab:allarticles}, we present the regression results after controlling for the topic distribution of each article (obtained from the LDA analysis discussed earlier). We see that the positive relationship between engagement and polarization remains robust.\footnote{The positive relationship between polarization and engagement is also found in the human ratings we collect ($\beta$ = 0.728, $p <$ 0.001). Similarly, human ratings continue to show a significant positive relationship between polarization and engagement when controlling for an article's topics ($\beta$ = 0.724, $p <$ 0.001). See Web Appendix $\S$\ref{app:humancoding_eval} for details.} This suggests that certain aspects of language may simultaneously influence a content's engagement and polarization levels. 
\begin{table}[htp!]
    \centering
    \caption{\small Relationship between Polarization and Engagement in the Original Data}
    {
\def\sym#1{\ifmmode^{#1}\else\(^{#1}\)\fi}
\small
\begin{tabular}{l*{2}{cc}}
\hline\hline
            &\multicolumn{2}{c}{(1)}           &\multicolumn{2}{c}{(2)}           \\
            &\multicolumn{2}{c}{Polarization}&\multicolumn{2}{c}{Polarization}\\
\hline
Engagement&       1.092\sym{***}&    (0.0227)&       1.070\sym{***}&    (0.0177)\\
Constant      &      -2.852\sym{***}&     (0.175)&      -0.508         &     (0.386)\\
\hline
Topics       &       No         &            &       Yes         &            \\
\hline
\(N\)       &       13,508         &            &       13,508         &            \\
\(R^{2}\)   &      0.1427         &            &      0.5725         &            \\
\hline\hline
\multicolumn{5}{l}{\scriptsize Standard errors in parentheses}\\ [-3pt]
\multicolumn{5}{l}{\scriptsize \sym{*} \(p<0.05\), \sym{**} \(p<0.01\), \sym{***} \(p<0.001\)}\\
\end{tabular}
}
    \label{tab:allarticles}
\end{table}

To further explore the interplay between engagement and polarization and how they might co-evolve, we consider the perspective of an editor whose goal is to make news content more engaging. As discussed in $\S$\ref{sec:intro}, an increasingly common approach to this problem is to use LLMs to enhance engagement by providing baseline news content and prompting it to generate a ``more engaging” version of the content. To examine the ability of simple prompt-based approaches to accomplish this task, we use ``GPT-4o-mini" to create a more engaging version for each article in the corpus. Details of the prompts used are shown in Web Appendix $\S$\ref{app:prompts_syntheticdata}. Next, we obtain the engagement and polarization scores for the LLM-generated ``more-engaging" version of each article using the scoring approach discussed earlier. The results are summarized in the respective left column of each panel in Table \ref{tab:summary_stats}. As expected, these LLM-generated versions are indeed rated as more engaging, with the median engagement score increasing to 9, compared to 8 in the original dataset (per Table \ref{tab:originaldata}). However, this improvement in engagement comes with an unintended consequence -- greater polarization. The median polarization score of the ``more-engaging" versions increases to 7, up from 6 in the original dataset (per Table \ref{tab:originaldata}). Two-sample t-tests indicate that both differences are significant. Essentially, the LLM-generated versions have moved away from the editorial stance in the original corpus and have become more polarized. 



\begin{table}[htp!]
    \centering
    \caption{\small Summary Statistics of Engagement and Polarization for More and Less Engaging Versions}
    \small
\def\sym#1{\ifmmode^{#1}\else\(^{#1}\)\fi}
\begin{tabular}{ccc|cc}
\hline
      & \multicolumn{2}{c|}{Engagement}               & \multicolumn{2}{c}{Polarization}              \\ \hline
      & \begin{tabular}[c]{@{}c@{}}More Engaging \\ Version\end{tabular} & \begin{tabular}[c]{@{}c@{}}Less Engaging \\ Version\end{tabular} & \begin{tabular}[c]{@{}c@{}}More Engaging \\ Version\end{tabular} & \begin{tabular}[c]{@{}c@{}}Less Engaging \\ Version\end{tabular} \\ \hline
Median  & 9.00              & 3.00              & 7.00              & 3.00             \\
Mean  & 8.99             & 3.10             & 5.89             & 3.09             \\
Std   & 0.11              & 0.33              & 2.46              & 0.95              \\
N & 13,508          & 13,508          & 13,508         & 13,508          \\
\hline
\multicolumn{5}{l}{\scriptsize \sym{*} \(p<0.05\), \sym{**} \(p<0.01\), \sym{***} \(p<0.001\)}\\
\end{tabular}
\label{tab:summary_stats}
\vspace{-10pt}
\end{table}

If increasing engagement tends to increase polarization, a natural question follows: does reducing engagement lead to lower polarization? To explore this, we prompt the LLM to generate a less engaging version of each article. The details of these prompts are provided in Web Appendix $\S$\ref{app:prompts_syntheticdata}. We then evaluate the engagement and polarization levels of these ``less-engaging" versions, and present the results in the respective right column of each panel in Table \ref{tab:summary_stats}. As expected, the engagement scores for the less engaging versions are significantly lower than those of the original articles, with a median of 3 compared to 8 for the original articles (per Table \ref{tab:originaldata}). Further, we see that polarization also decreases, with a median score of 3, compared to 6 in the original corpus. Two-sample t-tests confirm that these differences are statistically significant as well. See Web Appendix $\S$\ref{app:example_synthetic} for an example of the three versions of an article from our dataset. 



\begin{figure}[t] 
  \subcaptionbox*{\small $\Delta \textrm{Engagement}$}[.5\linewidth]{%
    \includegraphics[width=\linewidth]{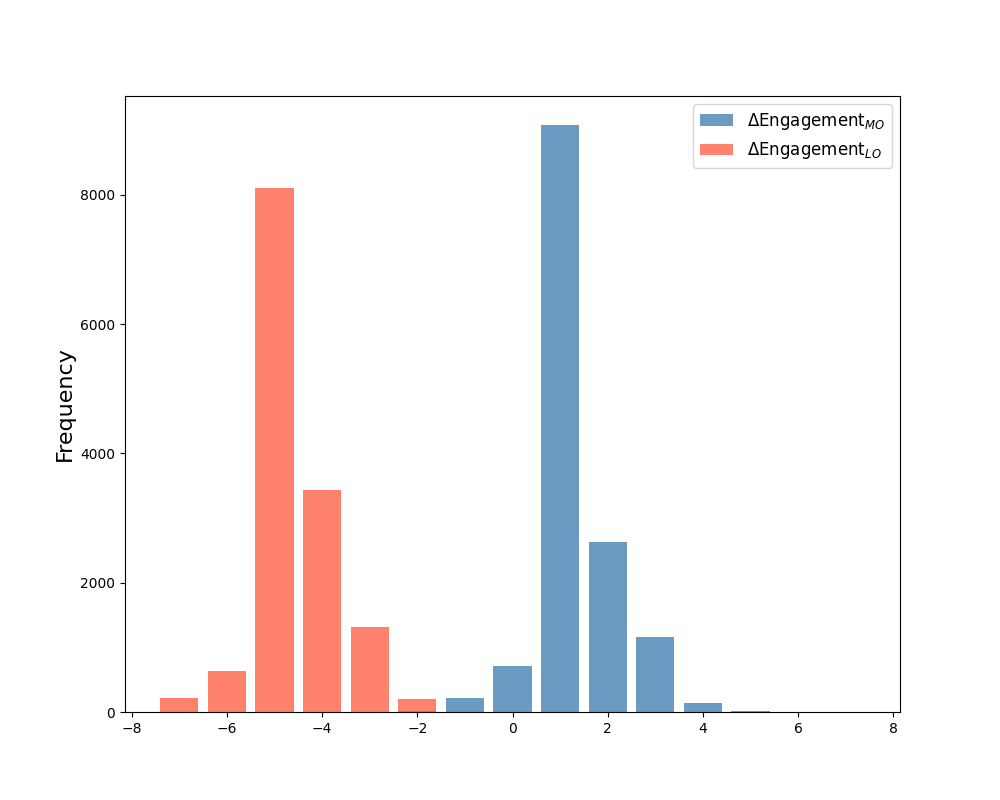}%
  }%
  \subcaptionbox*{\small $\Delta \textrm{Polarization}$}[.5\linewidth]{%
    \includegraphics[width=\linewidth]{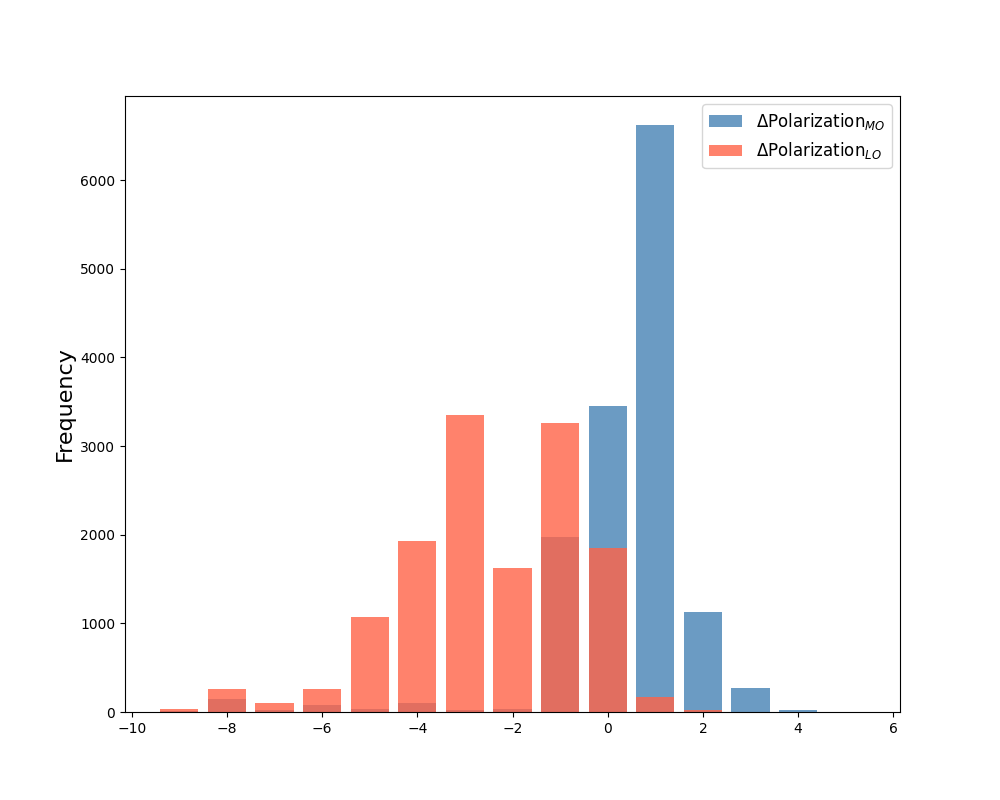}%
  }
  \caption{\small Engagement/Polarization Difference between LLM-Generated Versions and Original Articles}
  \label{fig:engagement_polarization_diff}
\end{figure}

So far, we have summarized how using an LLM to generate content shifts the engagement and polarization scores at the corpus level. We now examine these changes at the article level. Recall that we now have three versions of each article -- (1) the original version, (2) the more engaging version, and (3) the less engaging version. To quantify how the engagement and polarization scores of the LLM-generated versions change relative to the original articles, we define four measures: (1) $\Delta \textrm{Engagement}_{MO}$ -- the difference in engagement scores between the more engaging version and the original version, (2) $\Delta \textrm{Engagement}_{LO}$ -- the difference in engagement scores between the less engaging version and the original version, (3) $\Delta \textrm{Polarization}_{MO}$ --  the difference in polarization scores between the more engaging version and the original version, and (4) $\Delta \textrm{Polarization}_{LO}$ --  the difference in polarization scores between the less engaging version and the original version. Figure \ref{fig:engagement_polarization_diff} illustrates the distributions of these four difference measures. We see that the distributions of $\Delta \textrm{Engagement}_{MO}$ and $\Delta \textrm{Polarization}_{MO}$ are right-skewed, whereas the distributions of $\Delta \textrm{Engagement}_{LO}$ and $\Delta \textrm{Polarization}_{LO}$ are left-skewed; together these findings confirm the results from the previous analysis.

Next, we examine whether the {\it degree} to which an article’s engagement increases (or decreases) after it is modified to be more (or less) engaging affects the {\it extent} to which it becomes more (or less) polarizing.  Table \ref{tab:diff_diff} presents the results of two regressions: (1) $\Delta \textrm{Engagement}_{MO}$ regressed on $\Delta \textrm{Polarization}_{MO}$, and (2) $\Delta \textrm{Engagement}_{LO}$ regressed on $\Delta \textrm{Polarization}_{LO}$. In both cases, we observe a strong positive relationship. Specifically, a one-unit increase in engagement is associated with an almost 0.6-unit increase in polarization in the more engaging version (relative to the original), while a one-unit decrease in engagement is associated with an approximately 0.75-unit decrease in polarization for the less engaging version (relative to the original version). These results suggest that articles that experience larger increases (or decreases) in engagement also exhibit proportional increases (or decreases) in polarization. In other words, when LLMs are prompted to adjust an article’s engagement level, they simultaneously induce a proportional shift in its polarization level.

\begin{table}[t]
    \centering
    \caption{\small Difference in Polarization = f (Difference in Engagement) }
    \small
    {
\def\sym#1{\ifmmode^{#1}\else\(^{#1}\)\fi}
\begin{tabular}{l*{2}{c}}
\hline\hline
            &\multicolumn{1}{c}{(1)}&\multicolumn{1}{c}{(2)}\\
            &\multicolumn{1}{c}{$\Delta \textrm{Polarization}_{MO}$}&\multicolumn{1}{c}{$\Delta \textrm{Polarization}_{LO}$}\\
\hline
$\Delta \textrm{Engagement}_{MO}$&       0.559\sym{***}&                     \\
            &    (0.0161)         &                     \\
[1em]
$\Delta \textrm{Engagement}_{LO}$ &                     &       0.748\sym{***}\\
            &                     &    (0.0157)         \\
[1em]
Constant      &     0.00638         &      -0.326         \\
            &     (0.321)         &     (0.343)         \\
\hline
Topics      &       Yes         &       Yes         \\

\hline
\(N\)       &       13,508         &       13,508         \\
\(R^{2}\)   &      0.1612         &      0.3907         \\
\hline\hline
\multicolumn{3}{l}{\scriptsize Standard errors in parentheses}\\ [-3pt]
\multicolumn{3}{l}{\scriptsize \sym{*} \(p<0.05\), \sym{**} \(p<0.01\), \sym{***} \(p<0.001\)}\\
\end{tabular}
}   
\vspace{-12 pt}
\label{tab:diff_diff}
\end{table}
In sum, our analysis in this section highlights three key takeaways. First, we find that human-written news articles in our dataset exhibit a positive correlation between engagement and polarization, even after controlling for their topic coverage. This suggests that the choice of words, emphasis on (or omission of) specific details, and tone elements that make content more engaging may also make it more polarizing. Second, we demonstrate that modern LLMs can effectively enhance engagement, offering media firms a powerful tool to increase the appeal of their content. However, this comes with an unintended consequence: naively using LLMs to enhance engagement with news content can proportionately amplify polarization. This presents a trade-off for news editors and outlets -- while efforts to boost engagement may be attractive to readers, they may also inadvertently intensify polarization and make the content inconsistent with the firm's editorial stance. 
Recognizing this challenge, in the next section, we examine how news outlets and editors can leverage LLMs to enhance engagement while controlling for unintended consequences, such as increased polarization.



\section{Problem Formulation}
\label{sec:problem}

Consider a media firm that has dual goals -- (1) make content more engaging, and (2) maintain a specific editorial stance (e.g., a preferred degree of left/right-leaning slant). 
\squishlist 
\item \textbf{Enhancing engagement:} The goal of increasing news content engagement is directly tied to revenue generation. Consumers who find content engaging are more likely to spend additional time on the platform, return in the future, and interact with more advertisements. As a result, higher engagement with news content is likely to translate to greater readership (eyeballs), increased advertising revenue, and potentially more subscription revenues \citep{iizuka2021effect, goldminead2024}.  

\item \textbf{Maintaining a preferred editorial stance:} A media firm's commitment to a specific editorial stance can stem from three motivations. First, this editorial stance may reflect the ideological preferences of its owners, journalists, or editors, using the news outlet as a vehicle to promote their viewpoints \citep{baron_2006}. For instance, in February 2025, Jeff Bezos directed \textit{The Washington Post}’s opinion section to focus exclusively on ``personal liberties and free markets,” explicitly stating that ``viewpoints opposing those pillars will be left to be published by others” \citep{bezos2025}. Similarly, \textit{Fox News} has long upheld a conservative editorial stance, shaped by its founder, Rupert Murdoch, with coverage frequently aligning with right-leaning perspectives on issues such as gun rights, immigration, and economic policy \citep{foxnews2024}. Second, firms may wish to maintain a particular level of left/right leaning slant to align with the preferences of their readership because consumers tend to favor news that conforms to their existing beliefs \citep{mullainathan_shleifer_2005}. Finally, firms may adopt an editorial stance consistent with the views of their readership due to reputational concerns \citep{gentzkow2006media}.\footnote{\citet{gentzkow2006media} show that firms may slant news content to be consistent with the views of their readership because consumers often perceive news content to be of higher quality and more trustworthy when it conforms to their prior expectations.}
\squishend



The firm's goal is thus to develop a content-generation policy that effectively balances the dual objectives of enhancing engagement and maintaining a desired editorial stance. We emphasize that while our discussion focuses on these two objectives, the framework and solution can extend to any context where a firm aims to achieve multiple objectives in its content creation process.\footnote{In practice, firms may pursue objectives beyond engagement and editorial stance. For instance, they might seek to adopt a specific stylistic tone (e.g., serious, authoritative, or humorous) or ensure the content is accurate and credible (e.g., by minimizing misinformation). These additional objectives may also come into tension with each other (or with engagement). Although our paper does not explicitly model these objectives, the proposed framework and solution approach remain applicable in these broader settings.}

A natural starting point to address this problem is to employ prompt engineering with an off-the-shelf LLM, such as GPT or Gemini, by asking the LLM to consider both objectives when generating text. However, prompt-based approaches face several challenges in accomplishing this task. These include limited and imprecise control over how the model navigates trade-offs between different objectives, inability to leverage the full extent of the firm's historical data, poor scalability when incorporating multiple goals, and concerns related to data privacy, transparency, and interpretability. Later, in $\S$\ref{ssec:comparison_prompt}, we will empirically show that a prompt-based approach underperforms our proposed approach. Therefore, we focus on a scalable solution that uses open-source LLMs trained on proprietary data that is explicitly calibrated to reflect the firm’s preferred balance between objectives via a tunable weight vector.

With the above in mind, we formalize the firm's problem and start by defining a few relevant concepts. We first introduce the concept of a language model policy, denoted by $\pi$, which governs how content is generated. 
\begin{definition}
The language model policy $\pi(y|x)$ represents a probability distribution that assigns probabilities to sequences of words $y$, given an input sequence $x$. It is typically parameterized by $\theta$, which denotes the model parameters that the (large) language model aims to learn. 
\end{definition}
The policy $\pi(y|x)$ represents how a language model generates content for a specific task. For example, if the task is question-answering, $x$ is the input question, $y$ denotes the generated answer, and $\pi$ denotes the language model policy used to generate $y$ based upon input $x$. Similarly, in a summarization task, $x$ denotes the original content, and $y$ represents the generated summary. In our setting, the task is to take original news content ($x$) and generate a modified version ($y$) that aligns with a set of $K$ objectives. To formalize each objective $k$, we introduce an objective-specific reward function \( r_k: \mathcal{X} \times \mathcal{Y} \to \mathbb{R} \), where \( \mathcal{X} \) represents the original input (e.g., a draft article) and \( \mathcal{Y} \) represents the model-generated version. This definition captures a wide range of reward structures that may depend on both the original input and the generated output. Building on this, we can further define the optimal language model for a specific objective $k$ as follows:
\begin{definition}
The optimal language model $\pi_{r_k}^*$ for objective $k$ is the policy that maximizes the expected reward for that objective, such that $\pi_{r_k}^* = \arg\max_{\pi}~\mathbb{E}_{x,~y \sim \pi(y|x)} \left[r_k(x, y)\right]$, where x is drawn from a fixed dataset.
\end{definition}
The definitions above provide the foundation for developing an optimal language model $\pi_{r_k}^*$ for a single objective $k$. In the single-objective case, optimizing the policy is straightforward: responses can be rank-ordered based on the reward function $r_k(x,y)$, allowing for the development of an optimal policy. For example, if the firm's sole objective is to increase engagement ($k=1$), it can develop a policy $\pi^*_{r_1}$ that maximizes the expected reward for engagement ($\mathbb{E}[r_1(x,y)]$) given input $x$. 

However, in a multi-objective setting, the firm's goal is to develop an optimal language model $\pi^*$ that aligns with multiple objectives ($K$). This complicates the development of the optimal policy $\pi^*$ because there can be inherent trade-offs across objectives. For example, if the firm also has a secondary objective ($k=2$), where the goal is to reduce polarization, a dilemma arises. As discussed in $\S$\ref{sec:eng_pol}, a policy that maximizes engagement does not necessarily reduce polarization; in fact, it may increase polarization. One solution is to develop a new policy $\pi^*_{r_2}$ that minimizes the expected reward for polarization given input $x$ (i.e., $\mathbb{E}[r_2(x,y)]$). This approach, however, presents two key challenges. First, the two policies are likely to be misaligned. Specifically, if $\pi^*_{r_1}$ outperforms $\pi^*_{r_2}$ on engagement ($\mathbb{E}_{\pi^*_{r_1}}[r_1(x,y)] > \mathbb{E}_{\pi^*_{r_2}}[r_1(x,y)]$) but underperforms on polarization control ($\mathbb{E}_{\pi^*_{r_1}}[r_2(x,y)] < \mathbb{E}_{\pi^*_{r_2}}[r_2(x,y)]$), it is unclear which language model is preferable, as each outperforms the other on one of the objectives. Second, this produces two separate policies that accommodate different objectives, whereas, in practice, the firm needs one optimal policy that effectively balances both objectives -- maximizing engagement while controlling polarization. A natural solution to this problem is to leverage the concept of Pareto optimality, where the Pareto frontier of the policy space represents the set of policies that are non-dominated by any other policy. Formally, we define our goal as: 
\begin{definition}
Consider a firm that seeks to achieve $K$ objectives simultaneously. The firm's goal is to find the Pareto front $\Pi_f$ -- a set of policies that are Pareto optimal with respect to each objective $k$, characterized by the reward function $r_k$. That is, for each $\pi \in \Pi_f$, there exists no other policy $\pi'$ such that $\mathbb{E}_{\pi'}(r_k)  > \mathbb{E}_{\pi}(r_k)$ for every objective $k$.
\end{definition}
A common approach to solving multi-objective optimization problems is linear scalarization \citep{van2013scalarized}. Given a non-negative weight vector, $\mathbf{w} = [w_1, w_2, \dots, w_K]$, where $\sum_{k=1}^K w_k = 1$, and a corresponding reward function vector, $\mathbf{r}(x,y) = [r_1(x,y), r_2(x,y), \dots, r_K(x,y)]$, we can convert the firm's multiple objectives into a single objective using a weighted sum. The weight $w_k$ captures the relative importance of the $k^{th}$ objective for the firm. See \citet{rafieian2024multiobjective} and \citet{wang2024recommending} for recent marketing papers that also use this scalarization approach for multi-objective optimization. In our setting, a firm that cares a lot about engagement and not too much about the political editorial stance may assign a lower weight to polarization, and vice-versa. The firm’s goal can thus be formally defined as follows. 
\begin{definition}
\textbf{Firm's Problem:} Given a weight vector ($\mathbf{w}=[w_1, w_2, ...w_K]$), where $\sum_{k=1}^K w_k= 1$, and a reward function vector ($\mathbf{r}(x,y)=[r_1(x,y), r_2(x,y),..., r_K(x,y)]$), the firm's goal is to identify the Pareto-optimal language model $\pi_{\mathbf{w^Tr}}^*$ by optimizing the following objective function:
\begin{align}
    \pi_{\mathbf{w^Tr}}^* = \arg\max_{\pi}~\mathbb{E}_{x, ~y \sim \pi(y|x)} \left[\mathbf{w^Tr}(x,y)\right],
    \label{eq:firmproblem_ob}
\end{align}
where x is drawn from a fixed dataset.
\end{definition}
In summary, we formulate the firm's language model alignment challenge as a multi-objective optimization problem, where each objective has a pre-specified weight that can be chosen by the firm based on its strategic priorities.\footnote{The optimal balance between the two objectives is likely news vendor specific. For example, \emph{The New York Times} could conduct research on its readers to determine how much it should control the polarization level (not stray too far left), while increasing content engagement -- by looking at metrics such as fully reading, sharing or commenting on articles -- in order to achieve the best financial outcomes. This balance might be different for another news outlet, such as \emph{The Washington Post}.} By enumerating across different weight vectors, we can approximate the Pareto front of the optimal language models, denoted as $\left\{ \pi^*_{\mathbf{w^Tr}}| \mathbf{w}\in \Omega \right\}$, thereby providing firms with a spectrum of solutions tailored to their editorial and business objectives.\footnote{Because our objective space is convex, linear scalarization methods can effectively approximate the full Pareto frontier.}

\section{Model}
\label{sec:model}
We now provide an overview of the solution concept to the firm's problem stated in Equation \eqref{eq:firmproblem_ob}). We start with a baseline case ($\S$\ref{ssec:single_obj}), where the goal is to optimize a single objective, and extend the discussion to multi-objective settings ($\S$\ref{ssec:modpo}).

\subsection{Baseline Case with a Single Objective}
\label{ssec:single_obj}
Consider a simple setting where we only have one objective that we want to maximize:
\begin{align}
    \pi_r^* = \arg\max_{\pi}~\mathbb{E}_{x,~y \sim \pi(y|x)}\left[r(x,y)\right].
    \label{eqn:obj1}
\end{align}
In $\S$\ref{sssec:rlhf}, we describe how this problem can be solved using a reinforcement learning algorithm. Then, in $\S$\ref{sssec:dpo}, we discuss some drawbacks of this approach and present the simpler DPO algorithm.
 
\subsubsection{Reinforcement Learning from Human/AI Feedback}
\label{sssec:rlhf}
Reinforcement Learning from Human/AI Feedback (RLHF/RLAIF) is a machine learning algorithm that incorporates feedback from humans/AI agents, often in the form of rankings or comparisons, to perform tasks more aligned with the agents' revealed preferences. We provide a brief outline of the RLHF/RLAIF pipeline here and refer readers to prior works for detailed discussions; see \citet{bai2022training, ouyang2022training, leerlaif, bai2022constitutional}. The pipeline usually comprises three stages: (1) supervised fine-tuning (SFT), (2) reward modeling, and (3) reinforcement learning optimization. We describe each of these stages below.

\textbf{Supervised Fine-Tuning (SFT)}: The first step of the process starts with adapting a pre-trained, general-purpose language model to a specific downstream task through supervised learning on demonstration data ($\mathcal{D}$). Formally, we define the demonstration dataset as $\mathcal{D} = \left\{x^{(i)}, y^{(i)}\right\}_{i=1}^N$, where $x$ represents the input, $y$ represents the ideal output, and $N$ denotes the number of observations in the dataset $\mathcal{D}$. This step enhances the model’s ability to follow instructions for a given task, thereby laying the foundation for subsequent RL training \citep{chu2025sft}. In our context, SFT can enhance the model's ability to produce content that performs a specific task, such as generating modified news articles based on the original articles. This process results in a fine-tuned language model, denoted as $\pi_{\text{sft}}$. 


\textbf{Reward Modeling (RM)}: In this step, we use comparison data ($\mathcal{C}$) that capture the preferences of downstream users of the language model to develop a reward model that reflects their preferences. For example, given a prompt $x$, the comparison data would consist of two responses, $y_w$ and $y_l$, reflecting the agent's (human or AI) rating or relative preference ranking. We denote this preference as $y_w \succ y_l | x$, where $y_w$ and $y_l$ represent the more- and less-preferred responses, respectively.\footnote{The subscript $w$ in $y_w$ refers to the {\it w}inning response and the subscript $l$ in $y_l$ denotes the {\it l}osing response, respectively.} The preferences in the comparison data set are assumed to be governed by a latent reward function $r^*(x, y)$. Assuming preferences follow the Bradley-Terry model \citep{bradley1952rank},\footnote{Mathematically, the Bradley-Terry (BT) model shares structural and functional similarities with the logit model \citep{ben1997modeling}, which is widely used in economics and marketing to study consumer behavior.} their distribution $p^{*}$ can be expressed as:
\begin{align}
    \label{eqn:btpreferencemodel}
    p^*(y_w \succ y_l | x) = \frac{\exp(r^*(x, y_w))}{\exp(r^*(x, y_w)) + \exp(r^*(x, y_l))} = \sigma(r^*(x, y_w) - r^*(x, y_l)),
\end{align}
where $\sigma$ denotes the sigmoid function. Given a dataset of comparisons $\mathcal{C} = \left\{x^{(i)}, y_w^{(i)}, y_l^{(i)}\right\}_{i=1}^N$ sampled from $p^*$, we can define a reward model $r_\phi(x,y)$ and estimate its parameters $\phi$ via maximum likelihood to approximate the latent reward model $r^*(x, y)$. This optimization can be framed as a binary classification problem, where the parameters are updated to minimize the loss function, $\mathcal{L}_R(r_\phi; \mathcal{C})$, as follows: 
\begin{align}
    \mathcal{L}_R(r_\phi; \mathcal{C}) = -\mathbb{E}_{(x, y_w, y_l)\sim \mathcal{C}~}[\log \left(\sigma(r_\phi(x, y_w) - r_\phi(x, y_l))\right)].
    \label{eqn:reward}
\end{align}

\textbf{Reinforcement Learning Optimization (RL)}: In the final step, we use the learned reward function $r_\phi(x,y)$ to optimize the language model further using reinforcement learning algorithms. Specifically, suppose we have a language model policy parameterized by $\theta$, defined by $\pi_{\theta}$. We can then align $\pi_{\theta}$ to the learned reward function $r_\phi(x,y)$ by solving the following optimization problem:
\begin{align}
     \pi_{r_\phi}^* = \arg\max_{\pi_\theta}~\mathbb{E}_{x \sim \mathcal{C},~y \sim \pi_\theta(y|x)}\left[r_\phi(x,y) - \beta \log \frac{\pi_\theta (y|x)}{\pi_{\text{ref}}(y|x)}\right].
    \label{eqn:obj2}
\end{align}
The first term in the above optimization problem directly corresponds to the firm's problem defined in Equation \eqref{eqn:obj1}. The second term represents a Kullback–Leibler (KL) divergence term, which serves as a penalty for moving too far from a reference policy. The parameter $\beta$ controls the degree of deviation from the reference policy $\pi_{\text{ref}}$, which is usually the initial supervised fine-tuned model $\pi_{\text{sft}}$ (from the SFT step discussed earlier). This penalty plays a critical role in fine-tuning LLMs, as it prevents the model from excessively deviating from the distribution on which the reward model is valid while preserving generative diversity. Additionally, it mitigates the risk of reward over-optimization, which can degrade the quality of the generated content \citep{gao2023scaling}. Finally, we note that the above optimization problem is typically solved using reinforcement learning algorithms, such as Proximal Policy Optimization \citep[PPO;][]{schulman2017proximal}. 

\subsubsection{Direct Preference Optimization (DPO)}
\label{sssec:dpo}
While RLHF/RLAIF has shown promise in producing high-quality models, it is not without drawbacks. In particular, the RL step can be unstable and computationally intensive due to frequent sampling from the policy during training, which entails significant computational costs. Further, RLHF/RLAIF is an {\it online} algorithm. This means that, at each step of the optimization in Equation \eqref{eqn:obj2}, the algorithm generates new text based on the updated policy, which then needs to be annotated by humans/AI to get preference ratings. This can be very time-consuming and costly. 


To address these challenges, recent research on LLM alignment has introduced a simpler and more efficient alternative that avoids the need for explicit reward modeling and online annotation. This approach, proposed by \citet{rafailov2024direct}, is known as Direct Preference Optimization (DPO). Unlike RLHF, which requires continuously generating new text and collecting human feedback (e.g., labels or preferences) during training, DPO relies solely on {\it offline} data, significantly simplifying implementation and reducing computational demands. 
Crucially, instead of relying on explicit reward modeling and reinforcement learning, DPO derives a theoretical mapping between the true reward function $r^*$ and the optimal policy $\pi_{r^*}^*$. This connection allows the preference loss to be defined directly as a function of the optimal policy. Specifically, prior works \citep{peters2007reinforcement,peng2019advantage} have shown that the optimal solution to the KL-constrained reward maximization objective function in Equation \eqref{eqn:obj2} takes the following form:\footnote{This is done by replacing the true reward function $r^*$ instead of estimated reward model $r_{\phi}$ in Equation \eqref{eqn:obj2}, taking the first-order condition with respect to $\pi_{\theta}$, and normalizing by the partition function $Z(x)$. For a detailed derivation, see Appendix A.1 of \cite{rafailov2024direct}.} 
\begin{align}
 \pi_{r^*}^*(y|x) = \frac{1}{Z(x)} \pi_{\text{ref}}(y|x) \exp (\frac{1}{\beta}r^*(x,y)),
 \label{eqn:policy}
\end{align}
where $Z(x)$ is the partition function, defined as: $\sum_{y}\pi_{\text{ref}}(y|x)\exp(\frac{1}{\beta}r^*(x,y))$.\footnote{The partition function is a mathematical construct that computes the probability of an agent’s possible trajectories. It serves as a normalization constant, ensuring that the sum of probabilities across all states is equal to one \citep{rahme2019theoretical}.} Taking the logarithm of both sides of Equation \eqref{eqn:policy} and rearranging terms allows us to express the reward function in terms of its corresponding optimal policy $\pi_{r^*}^*$ and the reference policy $\pi_{\text{ref}}$:
\begin{align}
    r^*(x, y) = \beta \log\frac{\pi_{r^*}^*(y|x)}{\pi_{\text{ref}}(y|x)} + \beta \log Z(x).
    \label{eqn:reparameterization}
\end{align}
By substituting this reparameterization of $r^*(x,y)$ given in Equation \eqref{eqn:reparameterization} into the preference model in Equation \eqref{eqn:btpreferencemodel}, the partition function $Z(x)$ cancels out. This allows expressing the agent's preference probability purely in terms of the optimal policy $\pi_{r^*}^*$ and the reference policy $\pi_{\text{ref}}$. Thus, the optimal policy $\pi_{r^*}^*$ under the Bradley-Terry model satisfies the following preference model:
\begin{align}
    \label{eqn:btpreferencemodeldpo}
    p^*(y_w \succ y_l | x) = \frac{1}{1 + \exp \left( \beta \log\frac{\pi_{r^*}^*(y_w|x)}{\pi_{\text{ref}} (y_w|x)} - \beta \log\frac{\pi_{r^*}^* (y_l|x)}{\pi_{\text{ref}} (y_l|x)} \right)}. 
\end{align}
This reformulation bypasses the need for explicit reward modeling by directly connecting human preference probabilities to the optimal policy $\pi_{r^*}^*$. Using this, we can define the maximum likelihood objective for a parameterized policy $\pi_\theta$. The resulting objective, known as the DPO loss, can be defined as follows:
\begin{align}
        \mathcal{L}_{\text{DPO}}(\pi_\theta; \pi_{\text{ref}}, \mathcal{C}) = - \mathbb{E}_{(x, y_w, y_l) \sim \mathcal{C}~}\left[\log \sigma\left(\beta \log \frac{\pi_\theta (y_w|x)}{\pi_{\text{ref}}(y_w|x)} - \beta \log \frac{\pi_\theta (y_l|x)}{\pi_{\text{ref}}(y_l|x)} \right)\right].
    \label{eqn:dpo}
\end{align}
We can minimize this loss function for the language model policy $\pi_\theta$ by directly using maximum likelihood without relying on reinforcement learning. In summary, DPO consists of two steps:
\squishlist
\item Step 1: Supervised Fine-Tuning, where we fine-tune the pre-trained language model on the demonstration dataset $\mathcal{D}$ to obtain the SFT policy $\pi_{\text{sft}}$, which serves as the reference policy.
\item Step 2: Directly optimize the language model policy $\pi_{\theta}$ using maximum likelihood estimation on the comparison dataset $\mathcal{C}$, based on the loss function in Equation \eqref{eqn:dpo}. 
\squishend
DPO avoids both the explicit reward modeling in Equation \eqref{eqn:reward} (Step 2 in RLHF) and the computationally intensive reinforcement learning process in Equation \eqref{eqn:obj2} (Step 3 in RLHF). As such, DPO offers a more stable, efficient, and practical alternative for aligning LLMs.


\subsection{Multi-Objective Direct Preference Optimization (MODPO)}
\label{ssec:modpo}

While the DPO approach described above works well, it is designed for single-objective settings. Yet, our context consists of two objectives -- (1) increasing engagement and (2) controlling polarization. To that end, we adapt the objective function in Equation \eqref{eq:firmproblem_ob} to the multi-objective setting by incorporating a KL divergence term, similar to the formulation in Equation \eqref{eqn:obj2}. With this adjustment, our objective function becomes:
\begin{align}
    \pi_{\mathbf{w^Tr}}^* = \arg\max_{\pi}~\mathbb{E}_{x,~y\sim \pi(y|x)} \left [\mathbf{w^Tr}(x,y) - \beta \log \frac{\pi(y|x)}{\pi_{\text{ref}}(y|x)}\right].
    \label{eq:modpo_eq}
\end{align}


To solve the multi-objective alignment problem described in Equation \eqref{eq:modpo_eq}, we adopt the recently proposed Multi-Objective Direct Preference Optimization (MODPO) algorithm \citep{zhou2024beyond}. This novel algorithm combines ideas from Direct Preference Optimization \citep[DPO;][]{rafailov2024direct} with multi-objective optimization techniques \citep{van2014multi, li2020deep} to align LLMs on multiple objectives. This method has mainly been used in LLM safety research in computer science, where the goal is to train LLMs to provide answers that are both helpful and harmless in response to sensitive prompts such as, ``how to commit suicide?" or ``how to make a bomb" \citep{safe-rlhf, beavertails, liu2024enhancing}. Our work extends these principles to the media domain, where firms might seek to increase engagement while maintaining a preferred editorial stance. We provide a brief summary of this approach below. 

Similar to the theoretical mapping of DPO for a single objective as shown in Equation \eqref{eqn:reparameterization}, MODPO builds on the theoretical relationship between the weighted ground-truth collective reward models $\mathbf{w^Tr^*}$ across multiple objectives and the optimal language model $\pi^*_{\mathbf{w^Tr^*}}$ as follows:
\begin{align}
    \mathbf{w^Tr^*}(x, y) = \beta \log\frac{\pi_{\mathbf{w^Tr^*}}^*(y|x)}{\pi_{\text{ref}}(y|x)} + \beta \log Z_m(x),
    \label{eqn:reparameterization_multiobj2}
\end{align}
where $Z_m(x) = \sum_{y}\pi_{\text{ref}}(y|x)\exp(\frac{1}{\beta}\mathbf{w^Tr^*}(x,y))$ is the partition function for the multi-objective alignment problem. Let $r_k^*(x, y)$ denote the true reward function for the $k^{th}$ objective. Then, as before, we can assume that the agent's preferences for each objective $k$ follow the Bradley-Terry model. Consequently, the preference distribution $p_k^{*}$ for two responses $y^k_{w}$ and $y^k_{l}$ to the same input $x$ for each objective $k$ is given by:
\begin{align}
\label{eqn:btpreferencemodel_multiobj}
    p_k^*(y^k_{w} \succ y^k_{l} | x) = \frac{\exp(r_k^*(x, y^k_{w}))}{\exp(r_k^*(x, y^k_{w})) + \exp(r_k^*(x, y^k_{l}))} = \sigma(r_k^*(x, y^k_{w}) - r_k^*(x, y^k_{l})).
\end{align}
We can then rearrange Equation \eqref{eqn:reparameterization_multiobj2} to parameterize $r_k^*$ as a function of $\pi_{\mathbf{w^Tr^*}}^*$, $\mathbf{r_{-k}^*}$, $\pi_{\text{ref}}$, and $Z_m(x)$. Here, $r^*_k$ represents element $k$ of vector $\mathbf{r}$, while $\mathbf{r^*_{-k}}$ represents all elements of vector $\mathbf{r}$ except for element $k$ and $\mathbf{w_{-k}}$ is similarly defined. The resulting parametrized function of $r_k^*$ is as follows: 
\begin{align}
    r_k^*(x, y) = \frac{1}{w_k}\left(\beta \log\frac{\pi_{\mathbf{w^Tr^*}}^*(y|x)}{\pi_{\text{ref}}(y|x)} + \beta \log Z_m(x) - \mathbf{w_{-k}^Tr_{-k}^*}(x, y)\right).
    \label{eqn:reparameterization_multiobj}
\end{align}
Substituting the parametrization for $r_k^*(x, y)$ above into the preference model in Equation \eqref{eqn:btpreferencemodel_multiobj} causes the partition function $Z_m(x)$ to cancel out, and we get:
\begin{align}
\label{eqn:btpreferencemodel_multiobj_v2}
    &p_k^*(y^k_w \succ y^k_l | x) = \\ \notag
    &\sigma \left(\frac{1}{w_k} \left( \beta \log \frac{\pi_{\mathbf{w}^T \mathbf{r}^*}^*(y^k_{w}|x)}{\pi_{\text{ref}}(y^k_{w}|x)} - \beta \log \frac{\pi_{\mathbf{w}^T \mathbf{r}^*}^*(y^k_{l}|x)}{\pi_{\text{ref}}(y^k_{l}|x)} - \mathbf{w}_{-k}^T \left( \mathbf{r}_{-k}^*(x, y^k_w) - \mathbf{r}_{-k}^*(x, y^k_l) \right) \right)\right).
\end{align}
The equation above captures the preference distribution for the $k^{th}$ objective. Notice that it is similar in spirit to Equation \eqref{eqn:btpreferencemodeldpo} in the single-objective DPO case, with one key difference -- the additional term $\mathbf{w}_{-k}^T \left (\mathbf{r_{-k}^*}(x, y^k_w) - \mathbf{r_{-k}^*}(x, y^k_l)\right)$, which represents the adjustment added to reflect the marginal impact of the other objectives on the $k^{th}$ objective. This added complexity means that, unlike in the single-objective DPO case, we cannot bypass the explicit reward modeling step. However, based on Theorem 1 from \citep{rafailov2024direct}, we can parameterize the reward model implicitly. For all objectives except $k$, we define $K-1$ reward models, $r_{\phi, i}(x,y)$, each approximating the latent reward function $r_i^*$ as follows: 
\begin{align}
    \label{eq:reward_approx}
    r_{\phi, i}(x,y) = \beta \log \frac{\pi_{\phi, i}(y|x)}{\pi_{\text{ref}}(y|x)}.
\end{align}
Note that this parameterization assumes that we have already obtained $\pi_{\phi, i}(y|x)$ for each individual objective through the single-objective DPO process illustrated in $\S$\ref{sssec:dpo}. We then construct a reward function vector for all objectives except the $k^{th}$ objective, denoted as $\mathbf{r_{\phi, -k}}(x, y)$. Next, we substitute $\mathbf{r_{-k}^*}(x, y)$ with the pre-estimated values $\mathbf{r_{\phi, -k}}(x, y)$. Based upon this substitution, we then formulate an objective for learning the target policy $\pi_{\theta_\mathbf{w}}$ using the comparison dataset for the $k^{th}$ objective, analogous to Equation \eqref{eqn:dpo}: 
\begin{flalign}
    \label{eq:modpo_obj_eqn}
    &\mathcal{L}_{\text{MODPO}}(\pi_{\theta_\mathbf{w}};\mathbf{r_{\phi,-k}}, \pi_{\text{ref}}, \mathcal{C}_k) =  \\ \notag 
    &- \mathbb{E}_{(x, y_w, y_l)\sim \mathcal{C}_k}\left[ \log  \sigma \left(\frac{1}{w_k} \left( \beta \log \frac{\pi_{\theta_w}(y^k_w|x)}{\pi_{\text{ref}}(y^k_w|x)} - \beta \log \frac{\pi_{\theta_w}(y^k_l|x)}{\pi_{\text{ref}}(y^k_l|x)} - \underset{\text{margin},~m_{\phi}(x, y^k_w, y^k_l)}{\underbrace{\mathbf{w_{-k}^T (r_{\phi, -k}}(x, y^k_w) - \mathbf{r_{\phi, -k}}(x, y^k_l) )}} \right) \right) \right].
\end{flalign}
This loss formulation ensures that the target policy $\pi_{\theta_\mathbf{w}}$ closely approximates the optimal language model $\pi_{\mathbf{w^Tr}}^*$ for a given weight vector $\mathbf{w}$, regardless of which comparison dataset $\mathcal{C}_k$ is used in the $\mathcal{L}_{\text{MODPO}}$ objective function to learn the policy.\footnote{Theoretically, the choice of comparison dataset $\mathcal{C}_k$ in the objective function $\mathcal{L}_{\text{MODPO}}$ to learn the target policy $\pi_{\theta_\mathbf{w}}$ should not affect the ability to recover the optimal language model policy $\pi_{\mathbf{w^Tr}}^*$ for a given weight vector $\mathbf{w}$. This is because, for any objective $k$ and specified weight vector $\mathbf{w}$, the corresponding reparameterized reward function $r_k^*(x, y)$ (as defined in Equation \eqref{eqn:reparameterization_multiobj}) belongs to the same equivalence class. Reward functions from the same equivalence class induce the same optimal policy under the constrained reinforcement learning problem. Formal proofs can be found in Appendix A2 of \cite{zhou2024beyond} and Appendix A5 of \cite{rafailov2024direct}. In practice, though, the choice of comparison dataset $\mathcal{C}_k$ in the objective function $\mathcal{L}_{\text{MODPO}}$ might lead to small numerical differences in the obtained language model policy $\pi_{\theta_\mathbf{w}}$.} Intuitively, $L_\text{MODPO}$ is similar in spirit to the loss function of the single-objective DPO shown in Equation \eqref{eqn:dpo}, with two main differences: (1) additional weights $w_k$ to capture the weighted objective function, and (2) a margin term to ensure the language model is guided by multiple objectives. 

In summary, MODPO consists of three main phases:
\squishlist
\item Phase 1: Similar to RLHF and DPO, the process begins with Supervised Fine-Tuning, where we fine-tune the pre-trained language model on the demonstration dataset for the primary objective to obtain the SFT policy $\pi_{\text{sft}}$, which serves as the reference policy for subsequent steps. 
\item Phase 2: Next, we move to reward modeling using a set of comparison datasets, denoted by $\left\{\mathcal{C}_1, \mathcal{C}_2, ..., \mathcal{C}_K \right\}$, where each dataset $\mathcal{C}_k = \left\{x^{(i)}, y^{k(i)}_{w}, y_{l}^{k(i)}\right\}_{i=1}^N$ is sampled from $p_k^*$ for a specific objective $k$. Specifically, for each objective $k$, we perform reward modeling for all objectives other than $k$ using $\mathbf{\mathcal{C}_{-k}}$, which comprises comparison datasets for all objectives except $k$. This results in a reward function vector $\mathbf{r_{\phi, -k}}(x, y)$ for all objectives other than $k$. 
\item Phase 3: Finally, we optimize the language model policy $\pi_{\theta_\mathbf{w}}$ for a given weight vector $\mathbf{w}$ based on the loss function shown in Equation \eqref{eq:modpo_obj_eqn}. This process builds upon the reference policy $\pi_{\text{ref}}$ from Phase 1, the learned reward models $\mathbf{r_{\phi, -k}}(x, y)$ from Phase 2, and the comparison data $\mathcal{C}_k$ for objective $k$.
\squishend

MODPO offers many advantages in our setting. First, from a managerial perspective, it is capable of accommodating diverse and potentially conflicting business objectives. By adopting a linear scalarization strategy, we are able to transform a multi-objective problem into a single-objective one by assigning weights to each objective and combining them into a weighted sum. MODPO thus allows firms and managers to specify how much they care about different objectives and develop LLM policies precisely aligned with those weights. Second, MODPO is both practically stable and computationally efficient, making it feasible for real-world implementation. Similar to DPO, it does not rely on reinforcement learning algorithms for LLM alignment. Instead, it employs a novel parameterization of the reward model, enabling it to learn optimal policies in closed form. Under this approach, the optimization problem collapses to simple classification loss, and eliminates the need for sampling from the language model during training and extensive hyperparameter tuning. Finally, from a theoretical perspective, MODPO is statistically guaranteed to recover the exact optimal solution that would be obtained through multi-objective RLHF,\footnote{Multi-objective RLHF \citep{rame2024rewarded, williams2024multi} extends RLHF to optimize multiple objectives by incorporating human input to guide the learning process. Like RLHF, it involves both reward modeling and reinforcement learning. We do not discuss multi-objective RLHF in detail here since MODPO offers greater stability and efficiency than RLHF-style approaches.} and it has been empirically demonstrated to match or surpass existing methods in safety alignment and long-form question answering \citep{zhou2024beyond}.



\subsection{Implementation}
\label{ssec:implementation}
\begin{figure}[htp!]
    \centering
    \includegraphics[width=0.9\linewidth]{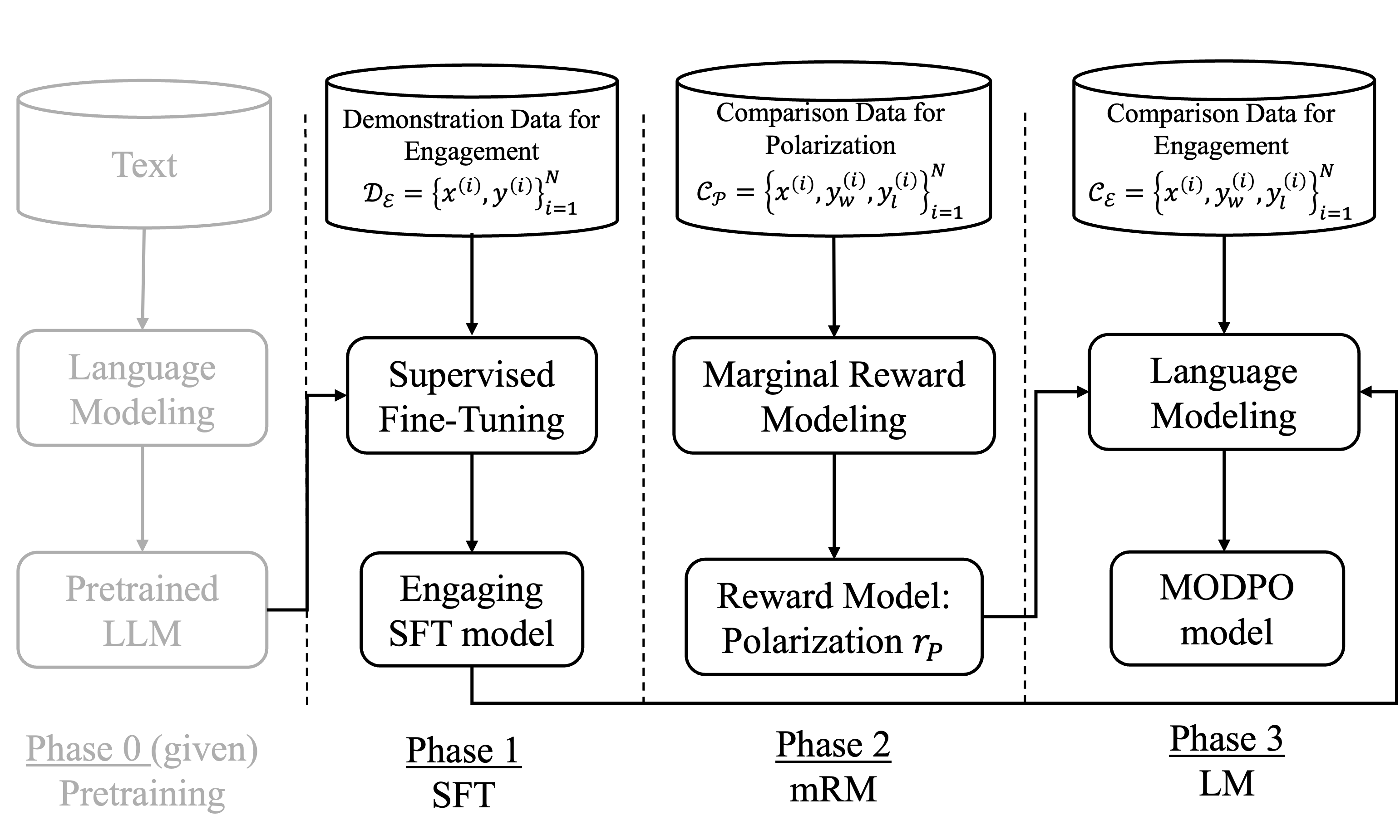}
    \caption{\small Overview of MODPO Implementation}
    \label{fig:modpo}
\end{figure}

Figure \ref{fig:modpo} illustrates the four phases of MODPO implementation: model pre-training, supervised fine-tuning, marginal reward modeling, and language modeling. Each phase builds upon the previous one, enabling the model to progressively align with multiple business objectives. We describe each in detail below. 

\subsubsection{Phase 0: Pre-Trained LLM}
Phase 0 consists of model pre-training, a foundational stage where the model learns general linguistic patterns, knowledge, and reasoning capabilities from a vast and diverse dataset. In other words, this is the stage where a general-purpose LLM is built. The standard approach in LLM alignment is to take a pre-trained LLM as given since training LLMs from scratch requires millions of dollars in resources \citep{llmcost2024}. Following this practice, we adopt a state-of-the-art open-source LLM -- Meta's Llama-3-8b \citep{llama2024} -- as our base model, taking its pre-training as given in Phase 0. To ensure that the substantive results are not idiosyncratic to the choice of the base LLM, in $\S$\ref{ssec:robustness_alpaca}, we present robustness checks with alternative pre-trained open-source LLMs, such as Alpaca.

\subsubsection{Phase 1: Supervised Fine-Tuning (SFT)}

In Phase 1, we adapt the pre-trained model to perform specific tasks or operate effectively within targeted domains using demonstration data. In our setting, this task involves generating a new version of an article from a base version.

\textbf{Demonstration Data:} To fine-tune the pre-trained model, we use a demonstration dataset, denoted as $\mathcal{D_E} = \left\{x^{(i)}, y^{(i)}\right\}_{i=1}^N$, where the original content serves as the input ($x$), and the more engaging response, described in $\S$\ref{sec:eng_pol}, serves as the preferred output ($y$). Essentially, this gives us a new language model that generates more engaging versions of original articles. Firms can also use alternative demonstration datasets for supervised fine-tuning, depending on their strategic goals.\footnote{Note that because of the KL divergence term added in Equation \eqref{eq:modpo_eq}, the final policy produced by the MODPO model will not deviate too far from the baseline/reference policy generated in this phase. As such, it is critical to choose a demonstration dataset that is aligned with the firm's primary objective of interest.} For instance, if a firm’s primary objective is to limit polarization/bias, it can instead construct a demonstration data $\mathcal{D_P}$, where the ideal output ($y$) consists of less polarized versions of the original articles. We consider this extension in $\S$\ref{sssec:managepolarization}.



\textbf{Supervised Fine-Tuning (SFT):} Building on Llama-3-8b, we use the demonstration data $\mathcal{D_E}$ to fine-tune and train a model optimized for generating engaging content, referred to hereafter as the \textit{engaging SFT model}. To fine-tune the model efficiently, we employ Low-Rank Adaptation (LoRA), a widely used Parameter-Efficient Fine-Tuning (PEFT) technique \citep{dettmers2024qlora, ye2024lola, angelopouloscausal2024}, originally developed by \citet{hu2021lora}. LoRA enables practical fine-tuning of large-scale LLMs on medium-scale hardware under the assumption that updates during model adaptation exhibit an intrinsic low-rank property. Specifically, instead of modifying all model parameters, LoRA introduces a set of low-rank trainable matrices into each layer of the Transformer model, significantly economizing on both memory usage and computational time. This vastly reduces the number of trainable parameters. Please see Web Appendix $\S$\ref{sec:lora} for additional details on the LoRA fine-tuning technique.

\subsubsection{Phase 2: Marginal Reward Modeling (mRM)}
\label{sssec:rm}

\textbf{Comparison Data for Polarization Objective, $\mathcal{C_P}$:} For reward modeling in Phase 2, we leverage comparison data to capture the agent's preferences for polarization. Formally, we define this dataset as $\mathcal{C_P} = \left\{x^{(i)}, y^{(i)}_{w}, y_{l}^{(i)}\right\}_{i=1}^N$, where each observation consists of an original content ($x$) and two modified versions ($y_w$ and $y_l$) evaluated by the agent in terms of polarization. Recall that our stated goal is to avoid excessive polarization/slant and ensure that the final output remains close to the original version of the article, which presumably aligns with \textit{The New York Times}'s current editorial stance.\footnote{In our main analysis, we assume that the original article already reflects the preferred editorial stance of the news outlet. Our goal is to enhance engagement while maintaining the polarization level of the original article. In $\S$\ref{ssec:left_right}, we relax this assumption and explore cases where the media outlet adjusts an original article’s polarization/slant to a specific preferred level.} As such, in our setting, $y_w$ corresponds to the less polarizing version, while $y_l$ represents the more polarizing version; both versions are obtained from the articles generated through the process described in $\S$\ref{sec:eng_pol}.

\textbf{Marginal Reward Modeling:} In our setting, we need to account for the firm's preference for both engagement ($k=1$) and polarization ($k=2)$. To learn the optimal policy $\pi_{\theta_{\bf w}}$ based on Equation \eqref{eq:modpo_obj_eqn}, we need to optimize on one objective $k$ and model rewards for the remaining $K-1$ objectives. In theory, we can choose any of the $k$ objectives to optimize Equation \eqref{eq:modpo_obj_eqn}, and this should yield the same optimal language model for a given weight $\mathbf{w}$, as discussed earlier. In practice, we choose polarization ($k=2$) as the objective for which we learn the reward function, while engagement serves as the objective on which we will learn the language policy based on Equation \eqref{eq:modpo_obj_eqn}. Therefore, we first learn the reward model for polarization, denoted as $r_{\phi, P}$. This process begins with a single-objective DPO procedure, as outlined in Equation \eqref{eqn:dpo}, to derive the language model policy $\pi_{\phi, P}$ based on the comparison data $\mathcal{C_P}$. We then parametrize the reward model as $r_{\phi, P}=\beta \log \frac{\pi_{\phi, P}(y|x)}{\pi_{\text{ref}}(y|x)}$, following the approximation in Equation \eqref{eq:reward_approx}. 

\subsubsection{Phase 3: Language Modeling (LM)}
\textbf{Comparison Data for Engagement Objective, $\mathcal{C_E}$:} 
For language modeling in Phase 3, we leverage comparison data to capture the agent's preferences for engagement ($\mathcal{C_E}$). Specifically, we denote this dataset as $\mathcal{C_E} = \left\{x^{(i)}, y^{(i)}_{w}, y_{l}^{(i)}\right\}_{i=1}^N$, where each observation consists of an original content piece ($x$) and two modified versions ($y_w$ and $y_l$) evaluated by the agent based on engagement. Since our objective is to enhance engagement, $y_w$ represents the more engaging version, while $y_l$ corresponds to the less engaging version; as before, both versions are obtained from the articles generated in $\S$\ref{sec:eng_pol}. 

\textbf{Language Modeling:} Here, we incorporate the reward model obtained from Phase 2 as an adjustment term into the objective function in Equation \eqref{eq:modpo_obj_eqn} to directly optimize the language model $\pi_{\theta_\mathbf{w}}$. Similar to Phase 1, the language modeling process uses the LoRA procedure for the actual optimization. 

In sum, we obtain a Pareto-optimal language model $\pi_{\theta_\mathbf{w}}$ by leveraging the capability of pre-trained models in Phase 0, supervised fine-tuning in Phase 1, marginal reward modeling in Phase 2, and language modeling in Phase 3. This approach balances the dual objectives of enhancing engagement and controlling polarization, letting firms tailor content that aligns with diverse business goals (reflected in the choice of weights $w_k$).  

\subsubsection{Other Implementation Details}
Lastly, we discuss a few remaining points related to our implementation.  

\textbf{Article Generation:} We note that all modified versions of articles in the demonstration datasets used in Phase 1, as well as the comparison datasets used in Phase 2 and Phase 3, are generated with the help of ChatGPT, as described in $\S$\ref{sec:eng_pol}. This approach aligns with the growing trend of using synthetic data generated by LLMs for model training, given their remarkable zero-shot learning proficiency and low cost in data generation \citep{liu2024visual, ubani2023zeroshotdataaug}. Notably, it has been demonstrated that LLMs can generate news articles that human evaluators often find indistinguishable from those written by journalists \citep{brown2020language}. With recent advancements, these capabilities have only improved, making synthetic data an increasingly ubiquitous and, in many cases, preferred alternative to purely human-generated data. For example, the training process of the recent ``DeepSeek-R1" model used a substantial amount of synthetic data, including 600,000 reasoning examples and 200,000 non-reasoning instances, generated using internal models and the ``DeepSeek-V3” model \citep{guo2025deepseek}. 

We wish to point out that our approach is quite general and does not inherently require AI-generated content. It can be easily adapted to settings where all the demonstration and comparison data consist of human-written articles. In $\S$\ref{ssec:AllSides}, we present an application of the MODPO approach with news articles published by various media outlets, which serve as our demonstration and comparison datasets. These articles are sourced from {\it AllSides} \citep{allsides2012}, a news aggregator that presents multiple perspectives by curating coverage from outlets across the political spectrum (left, center, and right) alongside its own factual summaries of the same news events.

\textbf{Preference Generation:} In our empirical application, the preference ratings designating which articles are more engaging and which ones are more polarizing are determined using ChatGPT, as described in $\S$\ref{sec:eng_pol} and Web Appendix $\S$\ref{app:prompts_syntheticdata}. There has been ongoing discussion about the pros and cons of using human vs. AI-generated feedback to align language models. While the early stream of work primarily relied on human feedback, collecting human feedback is often time-consuming, costly, and difficult to scale. It is also subject to inconsistencies due to individual differences and can inadvertently introduce human biases into the training process. Therefore, recent studies \citep{leerlaif, bai2022constitutional} have increasingly relied on AI-generated feedback as a viable alternative to human input. Notably, AI-generated feedback has demonstrated performance comparable to human input in tasks like summarization \citep{leerlaif}, offering advantages such as rapid scalability, greater consistency, and facilitation of large-scale training. Indeed, even the original DPO paper \citep{rafailov2024direct} employs AI-generated preference data for tasks such as controlled sentiment generation and single-turn dialogue. Naturally, if AI preferences are biased/misleading, then the language model obtained from the alignment procedure will perform poorly in the eventual content generation. As such, it is good practice to ensure that the content produced after LLM alignment using MODPO (or any other alignment procedure) is consistent with human evaluations on the task of interest. To that end, we validate our final results using human ratings (in addition to the evaluation of results by another independent LLM); see Web Appendix $\S$\ref{app:humancoding_eval} for details.

\textbf{Training, Validation, and Test Data:} We split our demonstration and comparison data into training (80\%), validation (10\%), and test sets (10\%). The test set serves as a corpus to conduct evaluations using independent LLMs (Gemini and Claude3) and human raters.\footnote{In our implementation, we exclude articles with identical polarization ratings, resulting in a comparison dataset of 11,814 articles. To determine the number of training epochs for all three phases, we use 10\% of the dataset for validation. Within the test set, which initially consists of 1,181 articles (10\% of the total), a small fraction fails to generate due to the SFT step and certain idiosyncrasies of the Llama model. As a result, the final test set contains 983 news articles. When using an alternative base LLM—Alpaca, as discussed in $\S$\ref{ssec:robustness_alpaca}—the final test set includes 1,028 news articles.} 

\textbf{Weight Selection:} For our main results in $\S$\ref{ssec:enhance_engage_modpo}, we set equal weights (i.e., $w = [0.5, 0.5]$) for engagement and polarization. We examine the impact of modifying the weight vectors on the language model policy and content generated in $\S$\ref{sssec:diff_weights}.


\section{Results}
\label{sec:results}
We present the main results from our analysis in $\S$\ref{ssec:enhance_engage_modpo} and document the model's sensitivity to the weight vector and data used for SFT in $\S$\ref{ssec:tuning}. In $\S$\ref{ssec:content_strategies}, we provide some insights into how the \textit{multi-objective DPO model} accommodates multiple objectives by examining how it leverages different types of content strategies.

\subsection{Evaluation Results for MODPO}
\label{ssec:enhance_engage_modpo}
We now examine how the \textit{multi-objective DPO model} performs on our test data on the two objectives of interest -- engagement and polarization -- using a weight vector of $w = [0.5, 0.5]$. For each article in the test data, we generate a new version of the article based on the \textit{multi-objective DPO model} outlined above. To aid comparison, we also consider two benchmark versions of each article:
\squishlist
\item Original version -- The version of the article, as originally published by \emph{The New York Times}.
\item Engaging DPO version -- This is the version of the article generated using a single-objective DPO approach that only optimizes engagement without controlling polarization.\footnote{In terms of implementation, Figure \ref{fig:modpo} simplifies to a single objective ($K=1$) without requiring the reward modeling phase. Additionally, the weight parameter is set to $w=1$, and the loss function for language modeling simplifies to Equation \eqref{eqn:dpo}.} 
\squishend

We use ``Gemini-1.5-Flash" to evaluate the engagement and polarization of all three versions of each article in the test data. Further, to ensure consistent evaluations and capture the relative differences more effectively, we provide all three versions within the same prompt and ask Gemini to rate each one in direct comparison to the others.\footnote{We report the results in terms of relative differences in engagement and polarization scores, as our primary interest lies in understanding how the MODPO and engaging DPO versions compare to the original version and each other. The raw scores assigned by ``Gemini-1.5-Flash" to the original versions are available in Web Appendix $\S$\ref{app:rawvalues}.} Please refer to Web Appendix $\S$\ref{app:gemini_eval} for details on the exact prompts used.

\begin{table}[t]
\caption{\small Gemini Evaluation Results (with \textit{Engaging SFT} as the Reference Policy)}
\vspace{-12 pt}
\begin{center}
\small
\def\sym#1{\ifmmode^{#1}\else\(^{#1}\)\fi}
\begin{tabular}{cccc|ccc}
\hline
      & \multicolumn{3}{c|}{Difference in 
 Engagement Scores}                  & \multicolumn{3}{c}{Difference in Polarization Scores}                 \\ \hline
      & \multicolumn{1}{c|}{$\Delta_{\text{MODPO\_ORIG}}$} & \multicolumn{1}{c|}{$\Delta_{\text{EDPO\_ORIG}}$} & $\Delta_{\text{MODPO\_EDPO}}$ & \multicolumn{1}{c|}{$\Delta_{\text{MODPO\_ORIG}}$} & \multicolumn{1}{c|}{$\Delta_{\text{EDPO\_ORIG}}$} & $\Delta_{\text{MODPO\_EDPO}}$\\ \hline
Mean & \multicolumn{1}{c|}{2.13\sym{***}}                      & \multicolumn{1}{c|}{2.57\sym{***}}                    & -0.44\sym{***}                       & \multicolumn{1}{c|}{0.80\sym{***}}                               & \multicolumn{1}{c|}{1.75\sym{***}}                             & -0.95\sym{***}                                \\
Std  & \multicolumn{1}{c|}{1.15}                      & \multicolumn{1}{c|}{1.58}                    & 1.28                        & \multicolumn{1}{c|}{1.66}                               & \multicolumn{1}{c|}{2.08}                             & 1.37                                 \\
Median & \multicolumn{1}{c|}{2.00}                      & \multicolumn{1}{c|}{3.00}                    & -1.00                       & \multicolumn{1}{c|}{1.00}                               & \multicolumn{1}{c|}{2.00}                             & -1.00                                \\
N  & \multicolumn{1}{c|}{983.00}           & \multicolumn{1}{c|}{983.00}              & 983.00      & \multicolumn{1}{c|}{983.00}          & \multicolumn{1}{c|}{983.00}             & 983.00      \\ \hline
\multicolumn{7}{l}{\scriptsize \sym{*} \(p<0.05\), \sym{**} \(p<0.01\), \sym{***} \(p<0.001\). } \\ [-3pt]
\multicolumn{7}{l}{\scriptsize $\Delta_{\text{MODPO\_ORIG}}$ = MODPO version - Original version; $\Delta_{\text{EDPO\_ORIG}}$ = Engaging DPO version - Original version;} \\ [-3pt]
\multicolumn{7}{l}{\scriptsize $\Delta_{\text{MODPO\_EDPO}}$ = MODPO version - Engaging DPO version.} \\
\end{tabular}
\vspace{-15pt}
\end{center}
\label{tab:gemini_evaluation_eng}
\end{table}
Table \ref{tab:gemini_evaluation_eng} shows how the engagement and polarization scores of the MODPO and engaging DPO versions change compared to the original version, as well as how they relate to each other.\footnote{We report the results in terms of relative differences in engagement and polarization scores, as our primary interest lies in understanding how the MODPO and engaging DPO versions compare to the original version and each other. The raw scores assigned by ``Gemini-1.5-Flash" to the original versions are available in Web Appendix $\S$\ref{app:rawvalues}.} We find that the articles recreated using both the \textit{engaging DPO model} and \textit{multi-objective DPO model} are significantly more engaging than the original articles. ($\Delta_{\text{EDPO\_ORIG}} = 2.57$, $p < 0.001$; $\Delta_{\text{MODPO\_ORIG}} = 2.13$, $p < 0.001$). Interestingly, articles generated by the \textit{multi-objective DPO model} exhibit only slightly lower engagement than those generated by the \textit{engaging DPO model}, which suggests that moving from the single-objective setting (of solely maximizing engagement) to the multi-objective setting does not severely harm performance on engagement. Further, we observe that the \textit{multi-objective DPO model} is more effective in limiting excessive polarization (unlike the \textit{engaging DPO model}, which exhibits the highest level of polarization) and adheres more closely to the editorial stance in the original version ($\Delta_{\text{MODPO\_ORIG}} = 0.80$, $p < 0.001$; $\Delta_{\text{EDPO\_ORIG}} = 1.75$, $p < 0.001$; $\Delta_{\text{MODPO\_EDPO}} = -0.95$, $p < 0.001$). Nonetheless, it is not able to fully avoid an increase in polarization, which is understandable given the desire to boost engagement. In sum, our results demonstrate that the \textit{multi-objective DPO model} effectively balances engagement while not deviating too far from the preferred editorial stance of the firm. As such, it allows content creators to mindfully leverage open-source LLMs in combination with multi-objective alignment to successfully generate content that is engaging for users while minimizing unintended negative consequences, such as excessive polarization. 

Finally, we make note of a few points regarding the evaluation process. As mentioned earlier, we use the ``Gemini-1.5-Flash" model to score all articles on engagement and polarization for the results shown in the main text. This choice is guided by two reasons. First, ``Gemini-1.5-Flash" has demonstrated superior performance in various benchmarks such as enhanced Massive Multitask Language Understanding (MMLU-Pro), challenging math problems (MATH), language translation (WMT23), and multimodal college-level understanding and reasoning tasks (MMMU) \citep{geminiflash2024}. Second, we intentionally avoid using GPT for evaluation because both the demonstration data and comparison data were generated by GPT. Prior research suggests that LLMs can exhibit self-preference, meaning they may favor their own output over those of other models or humans, even when human evaluators rate these as having equally high quality \citep{panickssery2024llm, xu2024pride}. Using a different LLM for evaluation ensures that our results remain independent of the data generation process for model training. 

To further validate our results and ensure that they are not specific to Gemini, we also consider three alternative evaluation approaches to score engagement and polarization for the test data.  First, we use an alternative LLM, ``claude-3-5-sonnet-20240620," a top-ranked and widely used model known for its strong performance on major NLP benchmarks. The results from this evaluation are reported in Web Appendix $\S$\ref{app:claude3_eval}. Second, we conduct a Qualtrics experiment to gather human evaluations on a subset of articles; see Web Appendix $\S$\ref{app:qualtrics_eval}. Finally, we use a human rater to score all the articles in the full test dataset; details are provided in Web Appendix $\S$\ref{app:humancoding_eval}. Our key findings remain consistent across all the evaluation methods.

\subsection{Impact of Firm Preferences for Maximizing Engagement vs. Managing Polarization}
\label{ssec:tuning}
We now examine how different baseline preferences for engagement vs. polarization affect the language model's content creation process. In the results presented so far, the weight vector for the relative importance of engagement and polarization was set to $\mathbf{w}=\{0.5, 0.5\}$, and the baseline reference policy $\pi_{\text{ref}}$ was developed to maximize engagement. In $\S$\ref{sssec:diff_weights}, we examine how changing the relative weights on engagement vs. polarization affects the behavior of the language model. Subsequently, in $\S$\ref{sssec:managepolarization}, we examine how changing the reference policy used in the first MODPO phase affects the language model.

\subsubsection{Effect of Modifying the Weight Vector in MODPO}
\label{sssec:diff_weights}

Media firms can have differing priorities when balancing the objective of increasing engagement with the need to adhere to a preferred editorial stance or limit polarization. We now examine how adjusting the weight vector influences the language model’s output. This exercise is valuable for two reasons. First, it allows us to assess the sensitivity of the \textit{multi-objective DPO model} to different input weights, helping us calibrate the extent to which the language model’s policy shifts in response to these weight changes. 
Second, from a managerial perspective, it presents a natural way for firms to align their editorial strategy with the language model employed. For example, a firm that cares significantly about engagement and only marginally about its editorial stance may choose a lower weight on polarization and vice-versa. This provides firms with a flexible framework to select content creation solutions that best align with their strategic priorities, enabling a more tailored and adaptive approach.


We vary the polarization weight over a range of values ($w_p \in [0.1, 0.2, \dots, 0.9]$), adjusting the engagement weight accordingly ($1-w_p \in [0.9, 0.8, \dots, 0.1]$), and fine-tune nine versions of the \textit{multi-objective DPO model}. Using each model, we then generate nine recreated versions for each original article in the test set. We then evaluate each recreated version on both engagement and polarization using the ``Gemini-1.5-Flash". Next, we regress the polarization and engagement scores of the recreated articles on the corresponding weight $w_p$ used to fine-tune the language model. As shown in columns (1) and (2) of Table \ref{tab:polarization_weight}, we find that the coefficient for the polarization weight ($w_p$) is negative and significant for both polarization ($b=-0.754$, $p<0.001$) and engagement ($b=-1.428$, $p<0.001$) scores.\footnote{These results remain robust when controlling for article fixed effects.} 
This indicates that increasing the polarization weight in the \textit{multi-objective DPO model} reduces not only the polarization score but also the engagement score of the generated articles (consistent with our earlier results reported in $\S$\ref{sec:eng_pol}). Note that to ensure that our findings are not driven by baseline differences in the original articles, we control for the polarization and engagement scores of the original versions. The coefficient for the original article’s polarization score is positive and significant ($b_{\text{Polarization}_\text{ORIG}} = 0.260$, $p < 0.001$), suggesting that articles with higher initial polarization levels tend to remain relatively more polarizing, even after regenerating the article using the MODPO process to balance engagement and polarization.



\begin{table}[t]
    \centering
    \caption{\small Polarization and Engagement Level of MODPO Version as a Function of Polarization Weight}
    \small
    {
\def\sym#1{\ifmmode^{#1}\else\(^{#1}\)\fi}
\begin{tabular}{l*{2}{c}}
\hline\hline
            &\multicolumn{1}{c}{(1)}&\multicolumn{1}{c}{(2)}\\
            &\multicolumn{1}{c}{$\text{Polarization}_\text{MODPO}$}&\multicolumn{1}{c}{$\text{Engagement}_\text{MODPO}$}\\
\hline
Weight ($w_p$)        &      -0.757\sym{***}&      -1.424\sym{***}\\
            &    (0.0660)         &    (0.0716)         \\
[1em]
$\text{Polarization}_\text{ORIG}$&       0.260\sym{***}&                     \\
            &   (0.00807)         &                     \\
[1em]
$\text{Engagement}_\text{ORIG}$&                     &     0.00586         \\
            &                     &    (0.0230)         \\
[1em]
Constant     &       5.168\sym{***}&       7.957\sym{***}\\
            &    (0.0617)         &     (0.182)         \\
\hline
\(N\)       &        8927         &        8927         \\
\(R^{2}\)   &      0.1158         &      0.0424         \\
\hline\hline
\multicolumn{3}{l}{\scriptsize Standard errors in parentheses}\\[-3pt]
\multicolumn{3}{l}{\scriptsize \sym{*} \(p<0.05\), \sym{**} \(p<0.01\), \sym{***} \(p<0.001\)}\\
\end{tabular}
}
\label{tab:polarization_weight}
\end{table}

\subsubsection{Changing the Reference Policy to Focus on Limiting Polarization}
\label{sssec:managepolarization}

Recall that the first step of MODPO consists of supervised fine-tuning to generate a reference policy that focuses on a single objective, denoted as $\pi_{\text{ref}}$. In our main analysis, we use engagement as the objective on which this reference policy is fine-tuned. Intuitively, the reference policy $\pi_{\text{ref}}$ reflects the firm's primary objective, while additional objectives are incorporated in later phases (Phase 2 and Phase 3). Furthermore, due to the KL divergence term in Equation \eqref{eq:modpo_obj_eqn}, the final policy generated by \textit{multi-objective DPO model} cannot move too far from the reference policy. As such, the choice of reference policy plays a crucial role in the final policy. 

We now examine how changing the reference policy from one that enhances engagement to one that minimizes polarization affects the language model. To implement this, we change the demonstration data from $\mathcal{D_E}$ to $\mathcal{D_P}$, training a \textit{less polarizing SFT model} as the new reference policy $\pi_{\text{ref}}$. In the demonstration data ($\mathcal{D_P}$), the original content serves as input ($x$), while a less polarized version of the content is used as the ideal output ($y$). If the firm’s sole objective is to reduce polarization, we can further refine this model by using comparison data that capture a preference for lower polarization levels. Employing DPO on this data yields a \textit{less polarizing DPO model}, which recreates articles optimized specifically to reduce polarization.


If the firm also values improving engagement as a secondary objective (in addition to controlling for polarization), we can use \textit{multi-objective DPO} with the \textit{less polarizing SFT model} as the reference policy. In this case, we first use the comparison data $\mathcal{C_{E}}$, where the agent ranks different content versions based upon engagement, as described in $\S$\ref{sssec:rm}, to train a reward model for the second objective—enhancing engagement in Phase 2. This reward model is then incorporated as an adjustment term in the objective function in Phase 3 of the \textit{multi-objective DPO model}, which optimizes Equation \eqref{eq:modpo_obj_eqn} using $\mathcal{C_P}$. Finally, to evaluate the outcomes, we score the engagement and polarization levels of all three versions of each article from the test data using Gemini as before; see Web Appendix $\S$\ref{app:gemini_reference_change} for details.

\begin{table}[t]
\caption{\small Evaluation Results using \textit{Less polarized SFT} as the Reference Policy}
\vspace{-10 pt}
\begin{center}
\small
\def\sym#1{\ifmmode^{#1}\else\(^{#1}\)\fi}
\begin{tabular}{cccc|ccc}
\hline
      & \multicolumn{3}{c|}{Difference in Engagement Scores}                  & \multicolumn{3}{c}{Difference in Polarization Scores}                 \\ \hline
      & \multicolumn{1}{c|}{$\Delta_{\text{MODPO\_ORIG}}$} & \multicolumn{1}{c|}{$\Delta_{\text{LDPO\_ORIG}}$} & $\Delta_{\text{MODPO\_LDPO}}$ & \multicolumn{1}{c|}{$\Delta_{\text{MODPO\_ORIG}}$} & \multicolumn{1}{c|}{$\Delta_{\text{LDPO\_ORIG}}$} & $\Delta_{\text{MODPO\_LDPO}}$\\ \hline
Mean      & \multicolumn{1}{c|}{-0.05}                     & \multicolumn{1}{c|}{-3.24\sym{***}}                   & 3.19\sym{***}                        & \multicolumn{1}{c|}{-0.78\sym{***}}                              & \multicolumn{1}{c|}{-3.29\sym{***}}                            & 2.51\sym{***}                                 \\
Std       & \multicolumn{1}{c|}{2.73}                      & \multicolumn{1}{c|}{2.15}                    & 2.12                        & \multicolumn{1}{c|}{2.70}                               & \multicolumn{1}{c|}{2.65}                             & 2.39                                 \\
Median     & \multicolumn{1}{c|}{1.00}                      & \multicolumn{1}{c|}{-3.00}                   & 3.00                        & \multicolumn{1}{c|}{-2.00}                              & \multicolumn{1}{c|}{-4.00}                            & 3.00                                 \\
N  & \multicolumn{1}{c|}{1154.00}           & \multicolumn{1}{c|}{1154.00}              & 1154.00      & \multicolumn{1}{c|}{1154.00}          & \multicolumn{1}{c|}{1154.00}             & 1154.00      \\ \hline
\multicolumn{7}{l}{\scriptsize \sym{*} \(p<0.05\), \sym{**} \(p<0.01\), \sym{***} \(p<0.001\)}\\ [-3pt]
\multicolumn{7}{l}{\scriptsize $\Delta_{\text{MODPO\_ORIG}}$ = MODPO version - Original version; $\Delta_{\text{LDPO\_ORIG}}$ = Less Polarizing DPO version - Original version;} \\ [-3pt]
\multicolumn{7}{l}{\scriptsize $\Delta_{\text{MODPO\_LDPO}}$ = MODPO version - Less Polarizing DPO version.} 
\end{tabular}
\end{center}
\vspace{-18pt}
\label{tab:gemini_evaluation_pol}
\end{table}

The evaluation results are shown in Table \ref{tab:gemini_evaluation_pol}. Compared to the original articles, recreated versions generated using the \textit{less polarizing DPO} and the \textit{multi-objective DPO} are significantly less polarizing ($\Delta_{\text{LDPO\_ORIG}} = -3.29$, $p < 0.001$; $\Delta_{\text{MODPO\_ORIG}} = -0.78$, $p < 0.001$). In other words, both models make the content more neutral than the original version and significantly reduce the left-leaning bias. However, MODPO version is able to avoid a significant drop in engagement ($\Delta_{\text{MODPO\_ORIG}} = -0.05$, $p > 0.1$), while the less polarizing DPO version concedes significant ground on the engagement objective ($\Delta_{\text{LDPO\_ORIG}} = -3.24$, $p < 0.001$). This suggests that media firms can successfully prioritize their primary goal by fine-tuning the \textit{SFT model}, and achieve secondary goals by incorporating them as marginal rewards during the fine-tuning process of the \textit{multi-objective DPO}.

\subsection{Uncovering Content Strategies for Enhancing Engagement and Managing Polarization}
\label{ssec:content_strategies}

We now provide some insights into how the \textit{multi-objective DPO model} is able to enhance engagement while simultaneously avoiding excessive polarization. 

\subsubsection{Theory-Driven Content Strategies}
\label{sssec:content_theories}
We identify three sets of theory-driven features that have been shown to affect polarization and engagement based on prior literature: (1) content structure and flow, (2) emotional and persuasive content elements, and (3) informational and factual content elements. 

\textbf{Content Structure and Flow.} Prior research has shown that three key aspects of content structure can affect engagement and polarization -- \textit{strong opening}, \textit{narrative structure}, and \textit{framing issues as questions}. A \textit{strong opening}, also referred to as a hook, is designed to capture the audience's attention and set the tone for the content that follows \citep{ting2020hooks, tiktok2024}. Although widely recognized for their effectiveness in driving engagement, strong openings may also have implications for polarization. For example, hooks tailored to resonate with specific audiences might inadvertently reinforce pre-existing beliefs, potentially heightening polarization \citep{smith2024polarization}. Beyond a strong opening, \textit{narrative structure} can also play a critical role in driving engagement. This facet refers to how the story is organized and presented, including its flow and themes. Past research highlights storytelling elements such as the speed of plot progression, the build-up of cognitive tension, and the circuitous nature of a story as effective tools for keeping audiences engaged \citep{toubia2021quantifying, laurino2022speed}. Another important content feature, particularly in news articles, is \textit{framing issues as questions}, where topics or analysis are presented through thought-provoking or guiding questions. However, the impact of this characteristic on engagement remains inconclusive. For example, \cite{scacco2020curiosity} find that news articles with question headlines tend to generate less intention to engage, while \cite{lai2014makes} observe that such headlines often spark greater readership interest. In addition, the way questions are framed is critical. When questions emphasize division or stereotypes, they may reinforce existing biases and deepen polarization, whereas those framed inclusively can encourage diverse perspectives, foster empathy, and reduce polarization \citep{polarizationquestion2019}. 

\textbf{Emotional and Persuasive Techniques.} We explore six content features along this dimension -- \textit{emotion level}, \textit{dramatic metaphors}, \textit{provocative language}, \textit{definitive language}, \textit{call to action}, and \textit{urgency level}. \textit{Emotion level} refers to the degree to which the content elicits emotional responses, such as joy, anger, fear, or hope. Highly emotional content often sparks increased sharing and commenting, reflecting a higher level of engagement \citep{berger2012makes}. However, it may also reinforce echo chambers by amplifying preexisting biases and encouraging users to interact primarily with like-minded individuals \citep{garrett2009echo}.  \textit{Dramatic metaphors} use vivid and figurative language to frame events or issues, making complex situations more relatable \citep{lakoff2008metaphors}. Although this can improve engagement, it can also deepen polarization by aligning individuals with groups that share similar identities \citep{thibodeau2019role}. Similarly, both \textit{provocative language} and \textit{definitive language} may influence how engaging the content is \citep{banerjee2024language}. \textit{Provocative language} evokes strong visceral reactions and thoughts, while \textit{definitive language} projects authority and certainty. However, they may also polarize audiences by leaving little room for alternative interpretations, thus discouraging opposing views. \textit{Call to action} -- explicit or implicit instructions or suggestions, compelling the readers to act, can significantly influence engagement by creating stimulation or aligning with audience motivations. Yet, calls to action, particularly those encouraging sharing, may amplify the dissemination of content consistent with users' preexisting beliefs, thereby fostering echo chambers and reinforcing polarized opinions \citep{bakshy2015exposure}. \textit{Urgency level}, which emphasizes immediacy and the need to act quickly, has been found to drive higher engagement \citep{perez2016measuring}, though its impact on polarization has not been documented before. 

\textbf{Informational and Factual Elements.} We consider three content features along these dimensions -- \textit{fact level}, \textit{relevant statistics}, \textit{balanced perspective}. \textit{Fact level} refers to specific and verifiable information included in an article that serves as the foundational building block of its narrative. This feature helps distinguish factual claims from opinions, interpretations, or commentary, providing clarity about what is being presented as a fact, and potentially reduces polarization. However, prior research has shown that it is possible to cherry-pick which factual sources are cited in order to present a slanted view of events \citep{groseclose_milyo_2005}. Complementing this feature, \textit{relevant statistics} provide numerical data to support the factuality of the article, enabling readers to assess the magnitude or significance of the presented information. Together, these two features can increase content credibility and trustworthiness, often driving higher engagement \citep{van2020reconstructive}. \textit{Balanced perspective} is characterized by fair, objective, and impartial presentation of diverse information. By appealing to readers who value objectivity, this element can help mitigate polarization \citep{gentzkow2006media, allen2020evaluating}. A balanced perspective may be particularly effective in reducing confirmation bias that often fuels divisive discourse \citep{stroud2011niche}. 

\subsubsection{Illustrative Example of Language Model Content Strategies}
\label{sssec:example}
\begin{figure}[htp!] 
    \centering
    \includegraphics[width=0.9\linewidth]{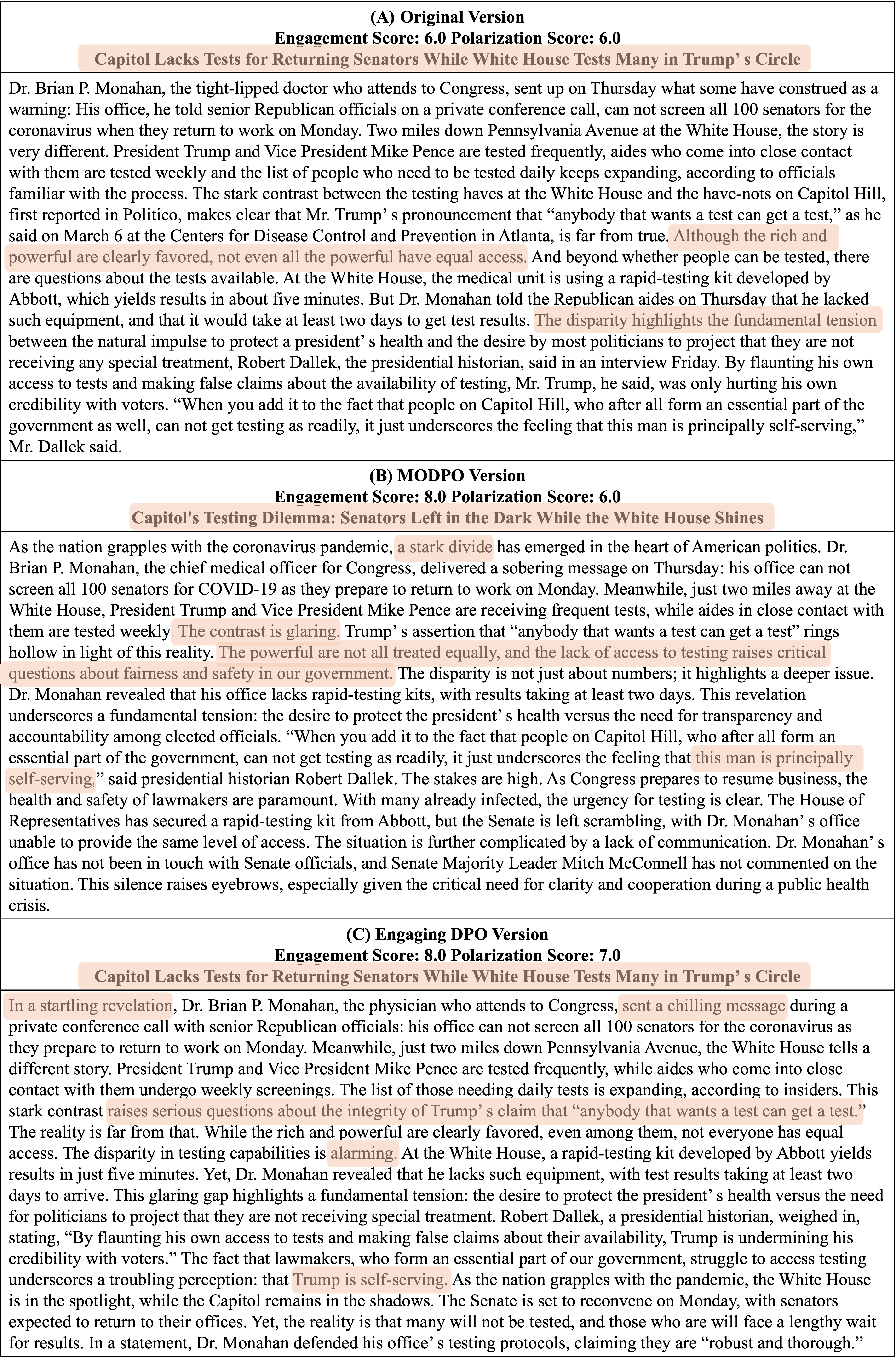}
    \caption{\small Three Versions of an Example Article}
    \label{fig:example1}
\end{figure}

We now highlight how different language models, specifically \textit{the multi-objective DPO model} vs. \textit{the engaging DPO model}, leverage the afore-discussed content strategies using an illustrative example from the test data. Due to space constraints, we only provide one example in Figure \ref{fig:example1} and refer readers to Web Appendix $\S$\ref{app:examples_content_strategies} for additional examples.
In general, all three versions describe the disparity in access to COVID-19 testing between the White House, where rapid and frequent testing was prioritized, and Capitol Hill, where senators faced limited availability. One can glean from Figure \ref{fig:example1} that both MODPO and engaging DPO versions enhance engagement by having a strong opening, although they employ distinct approaches. The MODPO version begins with a striking contrast in its title, “Senators Left in the Dark While the White House Shines,” which immediately grabs attention through vivid imagery. Meanwhile, the engaging DPO version opens with dramatic phrases like ``startling revelation” and “chilling message” to evoke an emotional response. By contrast, the original article starts more descriptively, which, while factual, lacks the same allure. The narrative structure also varies significantly across the versions. The original article is linear, fact-driven, and dense, offering comprehensive detail but lacking the same immediate appeal. In comparison, the MODPO version is more logically organized, guiding readers through the content with greater clarity and flow. The engaging DPO, on the other hand, prioritizes emotional impact and critique.

Polarization management differs across the versions as well. The MODPO version employs moderately provocative language, such as “stark divide” and “glaring contrast,” which effectively highlight inequities without excessive dramatization. By comparison, the engaging DPO version uses more emotive terms, such as “startling,” “chilling,” and “self-serving,” alongside dramatic metaphors like “The Capitol remains in the shadows.” While this approach evokes a strong sense of injustice, it also intensifies polarization. For its part, the original article remains relatively neutral, using mild language to point out disparities, such as, “Although the rich and powerful are clearly favored, not even all the powerful have equal access.” Finally, the tone and perspective also set the versions apart. The original article maintains a relatively factual tone, though its implicit critique of inequality and the Trump administration shows partisanship and leans left. The MODPO version also achieves a somewhat balanced perspective by focusing on systemic issues, as reflected in statements like, ``The powerful are not all treated equally, and the lack of access to testing raises critical questions about fairness and safety in our government." Conversely, the engaging DPO version is the most polarizing, with direct criticism of Trump. For example, it asserts, “This stark contrast raises serious questions about the integrity of Trump's claim that ‘anybody that wants a test can get a test'." 

\subsubsection{Relationship between Content Strategies and Engagement/Polarization in the Original Version}
\label{sssec:content_strat_engpol}
While the example discussed gives some qualitative insights into the content strategies of different language models, we now perform a more systematic evaluation. Specifically, we use Gemini to assess the use of the 12 theory-driven strategies for each news article in the test set. As before, we provide all three versions within the same prompt, and use ``Gemini-1.5-Flash" to compare and rate the content strategies for each version. Details on the exact prompts used are provided in Web Appendix $\S$\ref{app:prompts_contentstrategies}.  
We now examine how much of the variance in the two measures of interest -- engagement and polarization -- can be explained by the content strategies discussed above. To that end, we specify the following regression:
\begin{align}
    Y_i = \alpha + \sum_{j=1}^{12} \beta_i X_i +  \varepsilon_i,
\end{align}
where $Y_i$ denotes the polarization or engagement score of original article $i$ in the test data and $X_i$ represents the 12 content features summarized in $\S$\ref{sssec:content_theories}. The regression results are shown in Table \ref{tab:engagement_polarization_features} in Web Appendix $\S$\ref{app:regression_content_eng_pol}. We find that content strategies explain a larger proportion of the variance in polarization scores than in engagement scores (61.44\% vs. 15.54\%, respectively). Further, we find that \textit{strong openings} and \textit{narrative structure} emerge as important drivers of engagement, exhibiting a significant positive relationship with engagement scores. In contrast, \textit{provocative language} and \textit{urgency level} are positively associated with polarization while a \textit{balanced perspective} is negatively associated with polarization.


\begin{figure}[t] 
  \subcaptionbox*{\small Engagement (Original Version)}[.5\linewidth]{%
    \includegraphics[width=\linewidth]{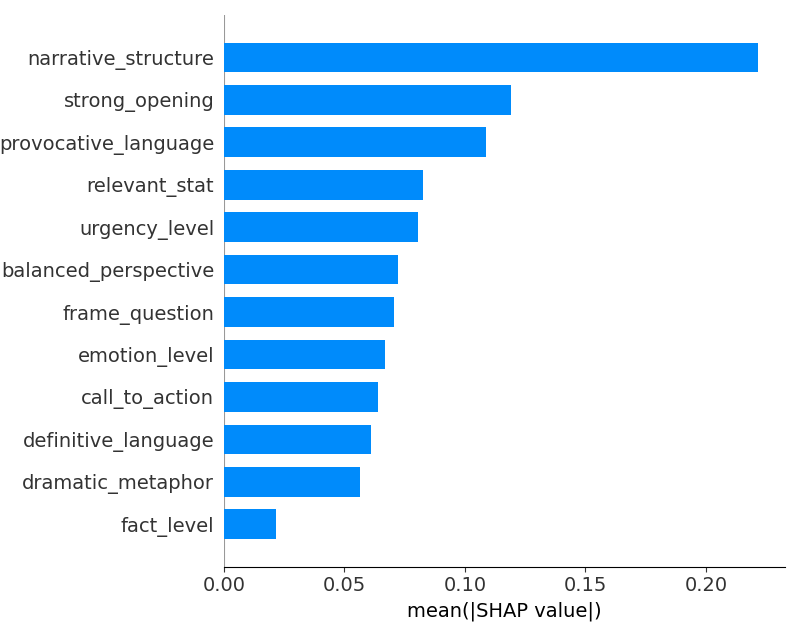}%
  }%
  \subcaptionbox*{\small Polarization (Original Version)}[.5\linewidth]{%
    \includegraphics[width=\linewidth]{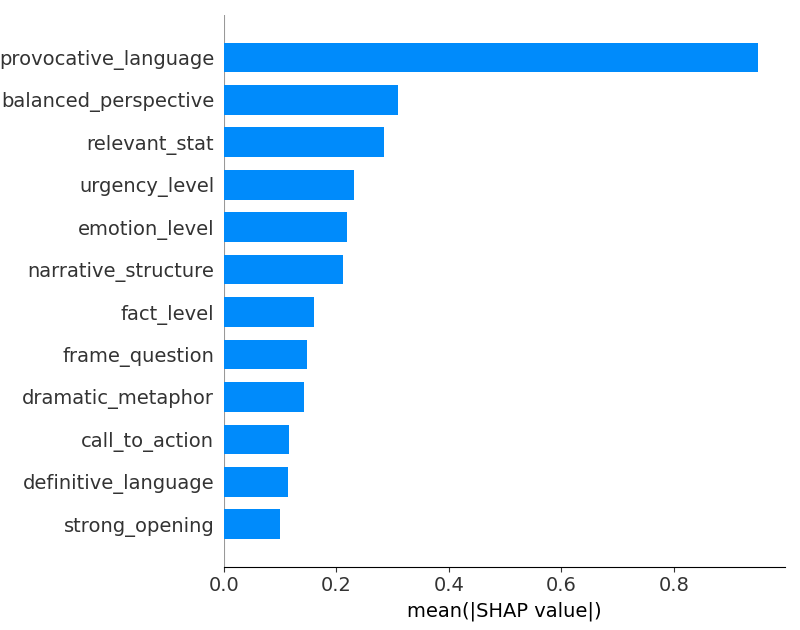}
  }
  \caption{\small The Importance of Content Strategies on Engagement and Polarization (SHAP Value)}
  \label{fig:shap}
\end{figure}
To gain further insights, we apply SHAP \citep{shap2017}, which can be used to quantify the importance of each factor in explaining engagement and polarization. As illustrated in Figure \ref{fig:shap}, \textit{narrative structure} and \textit{strong opening} are the two most influential content features driving engagement, while \textit{provocative language} and \textit{balanced perspective} are the top two features that explain content polarization. These findings align closely with the regression results.

\subsubsection{MODPO Content Strategies}

\begin{figure}[t] 
    \centering
\includegraphics[width=0.85\linewidth]{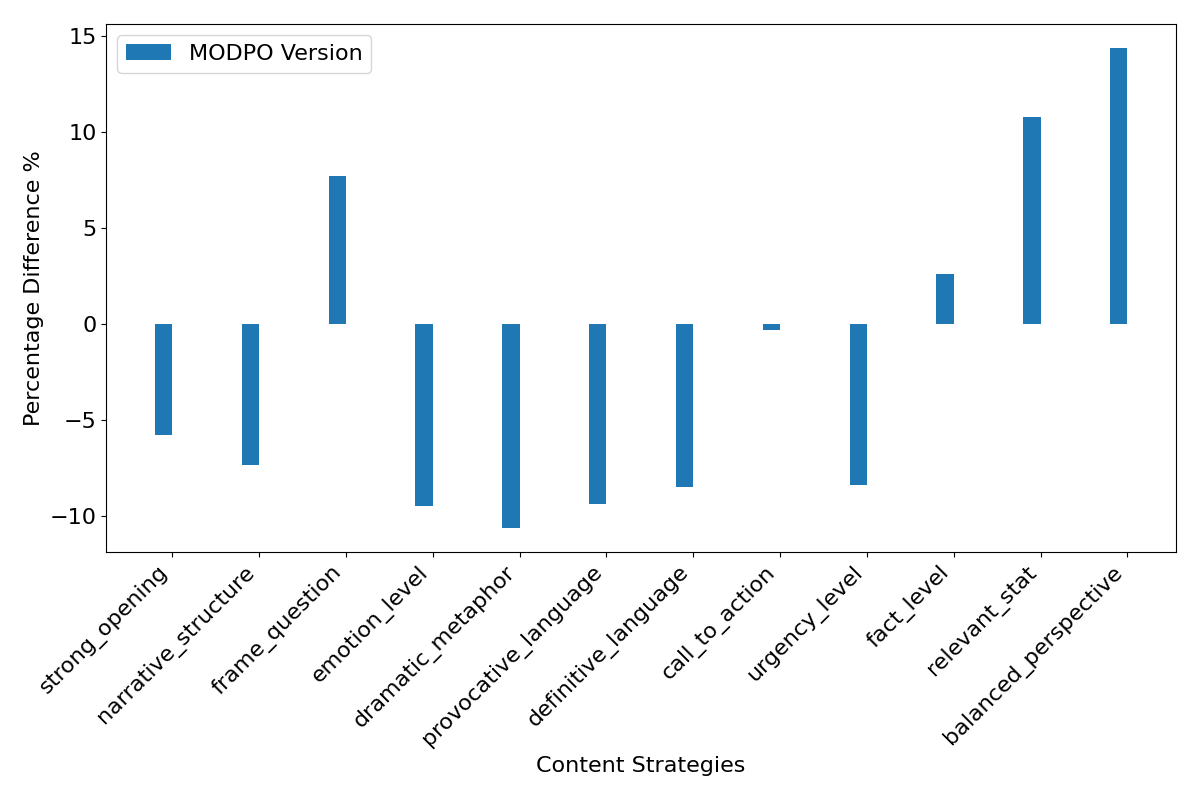}
    \caption{\small Relative Differences in Content Strategies Between the MODPO and Engaging DPO Versions}
    \captionsetup{justification=justified, singlelinecheck=false}
    \caption*{\small \textit{Note:} This figure illustrates the percentage difference in each content strategy between articles generated by the \textit{multi-objective DPO model} and those generated by the \textit{engaging DPO model}.}
    \label{fig:barchange}
\end{figure}

The \textit{multi-objective DPO model} seeks to recreate content that is more engaging without exacerbating polarization. Consequently, it should leverage strategies that strongly influence polarization but have a relatively smaller impact on engagement. To examine this conjecture, we focus on two content characteristics, \textit{provocative language} and \textit{balanced perspective}, which are relatively more important for explaining polarization than engagement (based on the SHAP analysis in Figure \ref{fig:shap}).\footnote{This observation is further supported by the Pearson correlation coefficients reported in Web Appendix $\S$\ref{app:correlation_coefs_content}, which reflect how each of these content strategies correlates with the polarization and engagement scores. Specifically, both \textit{provocative language} and \textit{balanced perspective} exhibit strong correlations with polarization but only modest correlations with engagement (\emph{provocative language}: $p_{eng}$ = 0.24, $p_{pol}$ = 0.68; \emph{balanced perspective}: $p_{eng}$ = -0.11, $p_{pol}$ = -0.58).} 
To understand whether and how the \textit{multi-objective DPO model} takes advantage of these two content characteristics, we compare the scores of these two features in the article versions generated by this model relative to the versions generated by the single-objective \textit{engaging DPO model}. Figure \ref{fig:barchange} presents the average percentage difference in each content strategy across the versions generated by the two models. 
We see that the \textit{multi-objective DPO model} seems to have learned that it should reduce the use of \textit{provocative language} and enhance the use of \textit{balanced perspectives} if it wants to generate content that is engaging without markedly increasing polarization.


Similarly, because both {\it narrative structure} and \textit{strong opening} greatly impact engagement, the {\it engaging DPO model} tends to pursue these content strategies more aggressively. However, because these same characteristics are also associated with greater polarization, the \textit{multi-objective DPO model} acts on them more conservatively. This differential approach can be gleaned from Figure \ref{fig:barchange}, where the relative percentage difference for these characteristics is negative when comparing the two models. More broadly, there are significant correlations between all content strategies (see Figure \ref{fig:correlation} in the Web Appendix $\S$\ref{app:correlation_coefs_content}). This interdependence implies that adjusting one strategy may inevitably affect others. The \textit{multi-objective DPO model} thus tries to balance different content strategies to simultaneously achieve its multiple goals.

\section{Robustness Checks and Extensions}
\label{sec:robustness}
We now present a series of robustness checks and extensions for our approach and findings.

\subsection{Comparison with Prompt-based Approaches}
\label{ssec:comparison_prompt}
A simplistic approach to address the problem of achieving multiple objectives is to use prompt engineering with off-the-shelf LLMs such as GPT or Gemini by asking the LLM to consider both objectives when generating text. However, prompt-based approaches tend to face several challenges in executing this task effectively. First, they offer limited and often imprecise control over how the LLM navigates trade-offs between different objectives, especially at scale. In Web Appendix $\S$\ref{app:prompt_eval_results}, we present an exercise where we use ChatGPT to recreate content for our multi-objective setting with different weights. We find that irrespective of the weight vector specified, the output remains largely unchanged. Thus, it is not possible for a firm/manager to provide relative preferences for different objectives and obtain results consistent with those objectives. 
This is due to the fundamental difference between prompt engineering approaches vs. language modeling approaches. In prompt-based approaches, the language model $\pi$ is fixed, and the firm/manager tries to shift the outcome by changing the context or input provided to the LLM. In contrast, in language modeling approaches (such as MODPO), we are directly modifying the language model policy $\pi$ to make it consistent with the firm's preferences. Therefore, prompt-based approaches usually exhibit poor performance on complex or domain-specific tasks compared to fine-tuned LLMs that directly change the language model policy trained on domain-specific data \citep{zhao2021calibrate, wei2022chain, ye2024lola}. 



In addition, prompt-based approaches are hard to scale and replicate across instances since GPT/Gemini models are often embedded in a black-box system with limited transparency. Finally, as discussed earlier, prompt-based approaches suffer from privacy and security concerns, especially for organizations handling proprietary or sensitive business data. Inputs provided to GPT models may be used as training data for future GPT model versions, which often conflicts with a company’s values and data protection standards. For example, \textit{The New York Times} has sued OpenAI and Microsoft for the unpermitted use of news articles to train GPT models in December 2023 \citep{nytopenai2024}. 



\subsection{An Alternative Base LLM -- Alpaca}
\label{ssec:robustness_alpaca}
In the main analysis, we use Llama 3-8b as the base model on which we perform the multi-objective alignment. However, we can use any open-source LLMs in  Phase 0 of Figure \ref{fig:modpo}. To test whether our results are robust to the use of alternative base LLMs, we implement the MODPO approach (with the same data) using Alpaca in Phase 0. Alpaca is a fine-tuned model based on the Llama 2-7B model, trained on 52,000 instruction-following demonstrations. Despite its smaller size, it outperforms the base Llama model in various instruction-following tasks, including summarization, text generation, and question answering \citep{taori2023alpaca}. Additionally, Alpaca is often used as the foundation model in the LLM safety alignment area, where the goal is to optimize both helpfulness and harmlessness \citep{safe-rlhf}. We find that the results using Alpaca as the base model are similar to those with Llama 3-8B. For detailed results, please refer to Web Appendix $\S$\ref{app:evalresults_alpaca}.


\subsection{Hallucination}
\label{ssec:hallucination}

One concern when using LLMs to rewrite content is hallucination, which refers to cases where the generated content is nonsensical or unfaithful to the source material \citep{filippova2020controlled}. 
To see if this is an issue in our context, we examine the extent to which our exercise suffers from factual hallucinations \citep{ji2023survey} by manually reviewing all the 983 generated news articles in our evaluation set.
We find that only 3\% of the generated articles exhibit signs of hallucination, which suggests that hallucination is not a significant problem in our setting. Please see Web Appendix $\S$\ref{app:hallucination} for a detailed discussion of this analysis. 


\subsection{Validation Using a Real-World Dataset}
\label{ssec:AllSides}
We now examine how our approach performs when using alternative data sources for training the MODPO model. To that end, we collect news articles from {\it AllSides} \citep{allsides2012}, a news aggregator that presents multiple perspectives by featuring coverage from outlets with varying political stances (i.e., left, center, and right), alongside its own factual summaries of the same news events. For example, for a given news event, {\it AllSides} may present news articles from {\it Fox News} (right), {\it CNN} (left), and {\it Reuters} (center). More details about the {\it AllSides} data collection process and an example news event from its website are provided in $\S$\ref{ssec:left_right} and Web Appendix $\S$\ref{app:AllSides}. We use this {\it AllSides} dataset to create the demonstration and comparison datasets to train the \textit{engaging DPO model} and the \textit{multi-objective DPO model} for this setting, following the same procedure described in $\S$\ref{ssec:enhance_engage_modpo}. We find that the results in this setting, where we use real-world news articles from multiple sources, are consistent with those obtained in our \textit{The New York Times} setting, where we used synthetic data (from GPT) for training. See Web Appendix $\S$\ref{app:evalresults_AllSides} for details. 




\subsection{A Specific Level of Slant: The Case of Targeting Left-Leaning Audiences} 
\label{ssec:left_right}
As noted earlier, different media organizations often have distinct editorial goals regarding the type and extent of political slant in their news content. In our main analysis, we effectively assume that the original news articles already reflected the preferred editorial stance of the news outlet, and the goal was to enhance engagement while maintaining that stance. However, our framework can easily extend to cases where the firm seeks to recreate content to achieve a desired level of left- or right-leaning polarization. To show how this can be achieved, we again use the real-world articles from the news aggregator {\it AllSides} described above. Specifically, we collect data on 9,982 news events from {\it AllSides} headline roundups \citep{allsidesheadline}, covering the period from June 1, 2012, to October 13, 2024. For each news event, we obtain the leading paragraph coverage from left-, center-, and right-leaning media outlets.\footnote{As outlined in Web Appendix $\S$\ref{app:AllSides}, {\it AllSides} \citep{allsides2012} provides media bias ratings for news outlets, which we validate using polarization scores generated by ChatGPT.}  To ensure completeness of the final sample, we retain 5,656 events where coverage from all three perspectives is available. 




Consider a firm that has amassed neutral content but now seeks to align it with a moderately left-leaning ideological stance.\footnote{This scenario has been selected for illustrative purposes; yet our framework is flexible and general. For example, it can be used to generate ``right-leaning” versions of articles if the firm prefers a conservative slant. In practice, media firms may be uncertain about the polarization/slant of their existing content. To achieve a desired level of slant, they can first assess the original ideological slant of the content and then adjust it using the approach proposed here to align with their preferred editorial stance. Our approach can easily accommodate this full workflow.} One way to accomplish this is to make the content left-leaning and incorporate some right-leaning language to adjust to the particular level desired by the firm. The news articles from {\it AllSides}-- which classifies articles as left, center, or right leaning -- serve as a real-world benchmark for constructing demonstration and comparison datasets. Specifically, to train a model that reflects the firm’s editorial preferences, we first construct a demonstration dataset, denoted as $\mathcal{D_L} = \left\{x^{(i)}, y^{(i)}\right\}_{i=1}^N$. Here, the content from the center outlet serves as the input ($x$), and the corresponding coverage from a left-leaning outlet is used as the ideal output ($y$). If the firm seeks to reinforce its left-leaning stance, we can construct a comparison dataset that captures its preference for left-leaning content ($\mathcal{C_L} = \left\{x^{(i)}, y^{L(i)}_{w}, y_{l}^{L(i)}\right\}_{i=1}^N$), where each observation consists of an article from the center outlet ($x$), a corresponding left-leaning piece ($y_w$), and a corresponding right-leaning content piece ($y_{l}$). Alternatively, we also consider another comparison dataset to capture preference for right-leaning content ($\mathcal{C_R} = \left\{x^{(i)}, y_{w}^{R(i)}, y_{l}^{R(i)}\right\}_{i=1}^N$), where each observation consists of an article from the center outlet ($x$), a corresponding right-leaning piece ($y_w$), and a corresponding left-leaning piece ($y_{l}$). 

Next, using these different datasets, we construct two language models and generate three versions of each article in the test data:
\squishlist
    \item Center version: The original version of the article from the center outlet. 
    \item Left-leaning DPO version: This version is generated using a single-objective DPO approach that only optimizes for left-leaning perspectives without incorporating right-leaning viewpoints. Specifically, we first fine-tune a \textit{SFT model} using $\mathcal{D_L}$, and then train a \textit{left-leaning DPO model} using $\mathcal{C_L}$.
    \item MODPO version: This version is generated using the MODPO approach, which reflects left-leaning perspectives while also accounting for right-leaning perspectives. Specifically, we first fine-tune a \textit{SFT model} using $\mathcal{D_L}$, then use $\mathcal{C_R}$ for the marginal reward modeling step (as illustrated in Figure \ref{fig:modpo}), and finally use $\mathcal{C_L}$ for the language modeling step. We set the weight vector for the relative importance of left- and right-leaning preferences to $\mathbf{w}=\{0.5, 0.5\}$.
\squishend

\begin{table}[t]
\caption{\small Gemini Evaluation Results -- Difference in Scores (with \textit{Left-leaning SFT} as the Reference Policy)}
\vspace{-12pt}
\begin{center}
\small
\def\sym#1{\ifmmode^{#1}\else\(^{#1}\)\fi}
\scalebox{0.95}{
\begin{tabular}{cccc|ccc}
\hline
      & \multicolumn{3}{c|}{Difference in Left-leaning Polarization Scores}                  & \multicolumn{3}{c}{Difference in Engagement Scores}                 \\ \hline
      & \multicolumn{1}{c|}{$\Delta_{\text{MODPO\_CENTER}}$} & \multicolumn{1}{c|}{$\Delta_{\text{LFDPO\_CENTER}}$} & $\Delta_{\text{MODPO\_LFDPO}}$ & \multicolumn{1}{c|}{$\Delta_{\text{MODPO\_CENTER}}$} & \multicolumn{1}{c|}{$\Delta_{\text{LFDPO\_CENTER}}$} & $\Delta_{\text{MODPO\_LFDPO}}$\\ \hline
Mean      & \multicolumn{1}{c|}{0.16\sym{*}}                                                        & \multicolumn{1}{c|}{1.29\sym{***}}                                                           & -1.12\sym{***}        & \multicolumn{1}{c|}{0.62\sym{***}}             & \multicolumn{1}{c|}{1.50\sym{***}}                & -0.88\sym{***}        \\
Std       & \multicolumn{1}{c|}{1.28}                                                        & \multicolumn{1}{c|}{1.86}                                                           & 2.00         & \multicolumn{1}{c|}{1.60}             & \multicolumn{1}{c|}{2.64}                & 2.73         \\
Median     & \multicolumn{1}{c|}{0.00}                                                        & \multicolumn{1}{c|}{1.00}                                                           & -1.00        & \multicolumn{1}{c|}{1.00}             & \multicolumn{1}{c|}{2.00}                & -2.00        \\
N  & \multicolumn{1}{c|}{534.00}           & \multicolumn{1}{c|}{534.00}              & 534.00      & \multicolumn{1}{c|}{534.00}          & \multicolumn{1}{c|}{534.00}             & 534.00      \\ \hline
\multicolumn{7}{l}{\scriptsize \sym{*} \(p<0.05\), \sym{**} \(p<0.01\), \sym{***} \(p<0.001\)}\\ [-3pt]
\multicolumn{7}{l}{\scriptsize $\Delta_{\text{MODPO\_CENTER}}$ = MODPO version - Center version; $\Delta_{\text{LFDPO\_CENTER}}$ = Left-leaning DPO version - Center version;} \\ [-3pt]
\multicolumn{7}{l}{\scriptsize $\Delta_{\text{MODPO\_LFDPO}}$ = MODPO version - Left-leaning DPO version.} 
\end{tabular}
}
\end{center}
\label{tab:gemini_diffval_leftright}
\vspace{-20pt}
\end{table}

Similar to our main analysis, we use ``Gemini-1.5-Flash" to evaluate the left-leaning polarization level and engagement of all three versions of each article in the test data. The left panel of Table \ref{tab:gemini_diffval_leftright} shows how the polarization of the different models compares to each other. Our findings suggest that both the \textit{left-leaning DPO} and \textit{multi-objective DPO} models generate articles that are significantly more left-leaning than the original articles ($\Delta_{\text{LFDPO\_CENTER}} = 1.29$, $p < 0.001$; $\Delta_{\text{MODPO\_CENTER}} = 0.16$, $p < 0.05$). However, the \textit{multi-objective DPO model}, which incorporates right-leaning perspectives, exhibits a lower left-leaning polarization level compared to the pure \textit{left-leaning DPO model} ($\Delta_{\text{MODPO\_LFDPO}} = -1.12$, $p < 0.001$). Notably, by adjusting the relative weights on left- and right-leaning preferences, we can achieve any specific desired level of left-leaning polarization. In contrast, such control is not feasible with a single-objective DPO, as we typically don't have preference data for a specific level of polarization. 


Interestingly, while our fine-tuning process only targets different types of polarization (left and right), we also observe shifts in engagement scores (see the right panel of Table \ref{tab:gemini_diffval_leftright}). The left-leaning DPO version, which exhibits the highest level of left-leaning polarization, also has the highest engagement rating compared to the original content from center outlets. In contrast, the MODPO version, which is moderately left-leaning, shows higher engagement compared to the center version. This pattern re-affirms our earlier finding (from $\S$\ref{sec:eng_pol}) that engagement and polarization tend to move together, and modifying one of these without directly controlling for the other can inadvertently shift the other metric as well.

\section{Conclusion}
\label{sec:conclusion}
In conclusion, this paper studies the relationship between content engagement and polarization and explores how media firms can leverage and adapt LLMs to create content that balances these two objectives. Using a large dataset from \emph{The New York Times}, we first show that more engaging articles also tend to be more polarizing. Further, we show that using naive prompt-based or single-objective alignment approaches (e.g., DPO) to improve engagement can inadvertently heighten the level of polarization.

We present a constructive solution to this problem by adapting the Multi-Objective Direct Preference Optimization algorithm, a newly proposed language alignment approach in the LLM safety literature that combines Direct Preference Optimization with multi-objective optimization techniques, to our setting and data. We further show that the content created by the newly aligned language model effectively balances the trade-off between enhancing engagement and adhering to the firm's preferred editorial policy. We further show that the approach is flexible, and the language model policy is responsive to the weight vector specified by the firm and the demonstration data used for supervised fine-tuning. We also provide some insights into how the \textit{multi-objective DPO model} leverages different content strategies to balance between the two objectives. Lastly, using an alternative data source from {\it AllSides}, we show that the model can be easily extended to accommodate a different set of conflicting objectives and alternative data sources.

Our proposed framework and findings have important managerial implications for media firms and news providers. In particular, since more engaging content tends to be more appealing to consumers, with studies showing a direct correlation between such greater engagement and repeated site visitation as well as receptivity to ads, media companies have a strong incentive to present news content in the most engaging way possible. However, as noted, naively boosting engagement tends to render the content more polarizing, which, in turn, can mitigate or negate the positive impact of greater engagement. This is because when content is misaligned with the attitudes and opinions of the target audience, e.g., much more left- or right-leaning, readers are inclined to view such content as less credible, reputable and trustworthy \citep{gentzkow2006media} leading to lower loyalty (e.g., canceling subscriptions) or less openness to ads (e.g., reduced click through rates).\footnote{Studies show that a major factor in ad effectiveness in news media is readers' degree of trust in the news source, which makes them more receptive to the content of accompanying ads \citep{goldminead2024, kellogginsight2022}.} Hence, as media outlets increasingly leverage LLMs to generate news content, the ability to balance engagement and polarization, as proposed here, is crucial for their monetization prospects. 

From a policy perspective, our research has implications for the impact of generative AI and LLMs on media polarization. We show that the widespread use of LLMs in news media, without safeguards to limit ideological slant, can exacerbate media polarization. As such, policymakers may wish to monitor how media firms employ AI-based solutions as part of their news generation workflow and offer appropriate guidelines.



We note that our general framework can be adapted to a variety of other business applications where LLM-based approaches are increasingly being used for content creation. For example, advertising agencies now use LLMs to craft ad copies \citep{forbes2023}, e-commerce platforms use them to automate product descriptions \citep{productdescription2023}, and social media influencers utilize LLMs to generate posts \citep{lyu2024preliminary}. In all these settings, content creators and firms often seek to balance multiple objectives simultaneously. For instance, Jasper, a leading artificial intelligence (AI) marketing content platform, strives to ensure its generated content remains true to a brand’s voice while also effectively engaging targeted audiences \citep{jasper2024}. Social media creators seek to produce posts that promote the brands they are collaborating with while staying true to their style and maintaining authenticity with followers \citep{forbes2023}. E-commerce companies seek to craft product descriptions that not only correspond to brand values but also convey accurate product information. We expect our approach to be relevant and effective across all these marketing applications.

\section*{Funding and Competing Interests Declaration}
\noindent Author(s) have no competing interests to declare. 

\newpage

\begin{appendices}
\setcounter{table}{0}
\setcounter{figure}{0}
\setcounter{equation}{0}
\setcounter{page}{0}
\renewcommand{\thetable}{A\arabic{table}}
\renewcommand{\thefigure}{A\arabic{figure}}
\renewcommand{\theequation}{A\arabic{equation}}
\renewcommand{\thepage}{\roman{page}}
\pagenumbering{roman}
\section{Prevalence of Topics in \textit{The New York Times} Articles}
\label{app:topics_nytimes}
\begin{figure}[htb!]
    \centering
    \includegraphics[width=0.8\linewidth]{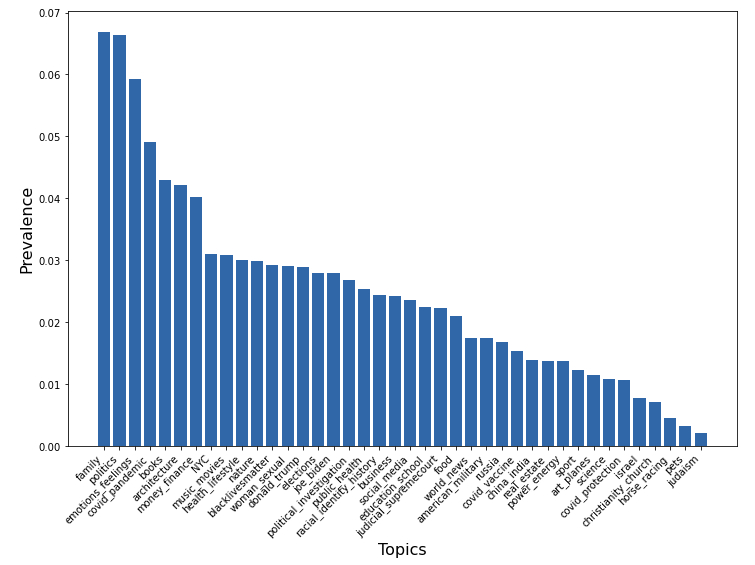}
    \caption{Topic Prevalence on the New York Times Corpus}
    \label{fig:topic}
\end{figure}

\noindent Figure \ref{fig:topic} shows the prevalence of the 40 most popular topics across all articles in the corpus from the LDA analysis, in decreasing order of popularity. We see that the most prevalent topics in our corpus are \textit{Family}, \textit{Politics}, and \textit{Emotions and Feelings}, while the least prevalent include \textit{Judaism}, \textit{Pets and Animals}, and \textit{Horse Racing and Farms}. For further details of the LDA analysis and the keywords recovered for each topic, please see \citet{yoganarasimhan2024feeds}.

\section{Prompting Procedure to Generate Synthetic Data}
\label{app:prompts_syntheticdata}
\noindent In our study, we use the ``GPT-4o mini” model to generate responses in a zero-shot manner, enabling us to 1) evaluate the polarization and engagement levels of the original \emph{The New York Times} articles, 2) create both a more engaging and a less engaging version of each article, and 3) assess the engagement and polarization levels of these generated versions. 
The ``GPT-4o mini” model has been optimized for chat using the Chat Completion API and can produce responses based on the provided chat history. Therefore, for consistency, we use the same prompt below as the input whenever we run the model on a new article.\footnote{We do not explicitly define engagement and polarization in the LLM prompt used for our main analysis. However, we perform a robustness check by including the same definitions provided to human raters in Web Appendix $\S$\ref{app:humancoding_eval}, where ``engaging" is defined as ``a news article is considered engaging if it attracts your attention and keeps you interested from start to finish," and ``polarizing" is defined as ``a news article is considered politically polarizing if it contains extreme views from either Democrats or Republicans." We find that the ratings generated by the GPT-4o mini model remain consistent regardless of whether these definitions are included.} Building on recent works \citep{zhang2021datasetgan, he2023targeted} that leverage LLM-generated synthetic data for model training, our prompting procedure to produce synthetic data is as follows:
\begin{itemize}
    \item Step 1: Evaluating the polarization and engagement levels of the original article.
    
    \textit{
    User: I will provide you with some text from a news article. Please rate from 1 to 10 how polarizing this article is. Let’s call this article as ``original version":*article text*}

    \textit{
    User: I will provide you with some text from a news article. Please rate from 1 to 10 how engaging this article is. Let’s call this article as ``original version":*article text*}

    \item Step 2: Generating two new versions of the original article by making it more engaging or less engaging.
    
    \textit{
    User: Could you help make the text of this article more engaging and attract more attention? Let's call this new version ``version A".}
    
    \textit{
    User: Could you help make the text of this article less engaging and attract less attention? Let's call this new version ``version B".}

    \item Step 3: Evaluating the engagement levels of the two generated versions.
    
    \textit{
    User: Please rate from 1 to 10 how engaging ``version A" is and ``version B" is, and explain which version of the article is more engaging and likely to attract more attention: \\
    ``version A": *more engaging version generated from Step 2* \\
    ``version B": *less engaging version generated from Step 2*.}

    \item Step 4: Evaluating the polarization levels of the two generated versions.
    
    \textit{
    User: Please rate from 1 to 10 how polarizing ``version A" is and ``version B" is, and explain which version of the article is less polarizing: \\
    ``version A": *more engaging version generated from Step 2* \\
    ``version B": *less engaging version generated from Step 2*}
\end{itemize}

\section{Example of an Article generated by GPT}
\label{app:example_synthetic}
We provide an example from \textit{The New York Times}, published in May 2020, which reports on so-called ``coronavirus parties” as a potential means of gaining immunity. The original article presents the issue with a cautious tone:
\begin{center}
\begin{minipage}{0.8\columnwidth}
\vspace{12 pt}
\label{h1_old}
\textit{\centering ``Covid-19 Parties" Probably Didn’t Involve Intentional Spread \\ 
Amid growing impatience over stay-at-home orders and rising unemployment, public health experts have worried that some people may try to expose themselves to the coronavirus in a risky bid to gain immunity. One fear is the prospect of ``coronavirus parties," much like the chickenpox parties of the past that preceded the development of a chickenpox vaccine, designed to deliberately spread infection.
County health officials in southeastern Washington State reported this week that they had evidence that one or more such gatherings had been linked to at least two new coronavirus cases. But on Wednesday night, the officials retracted those comments and said the so-called Covid-19 parties may have been more innocent gatherings. Meghan DeBolt, the director of community health for Walla Walla County, said county officials were learning more about the cases that have emerged from the recent social gatherings. She said they were still hearing reports of parties where infected people were present but do not have evidence that the people who became ill after the gatherings had attended out of a desire to be exposed.  
The county had said in a news release on Monday that the authorities were “receiving reports of Covid-19 parties occurring in our community, where non-infected people mingle with an infected person in an effort to catch the virus.” Officials later elaborated on those reports in interviews, saying the parties were discovered after tracing the paths of people who were found to be infected after the gatherings. Officials in Walla Walla have been working in recent weeks to contain a large outbreak at a meat processing facility in the area. Ms. DeBolt said the county was not close to halting the rise in infections, and health officials are concerned that more people are engaging in ill-advised social interactions despite stay-at-home requirements. “We know that people are exhausted from isolation and quarantine,” Ms. DeBolt said. “We want to be able to reopen, too. We want to be able to go to restaurants and socialize with friends and family members. We need our community’s help to be diligent for a little bit longer so that we can get ahead of this.” The prospect of infection parties for people who wish to quickly contract the disease in the hope of gaining immunity has been a fear among some health experts because the country has a long history of people choosing purposeful infection. In the case of chickenpox parties, the Centers for Disease Control and Prevention “strongly recommends against hosting or participating in these events. }
\end{minipage}
\end{center}
In contrast, the LLM-generated ``more engaging” version adopts more emotionally charged language and creates a heightened sense of urgency. It frames the ``Covid-19 parties" as a major public threat, which aligns with Democrats' political narratives advocating for strong government intervention to protect public health based on expert recommendations:
\begin{center}
\begin{minipage}{0.8\columnwidth}
\vspace{12 pt}
\label{h1_old}
\textit{\centering The Shocking Rise of ``Covid-19 Parties": A Dangerous Gamble for Immunity? \\ 
As frustration mounts over relentless stay-at-home orders and skyrocketing unemployment, a troubling trend is emerging: the rise of ``Covid-19 parties." In a desperate bid for immunity, some individuals are reportedly risking it all to expose themselves to the coronavirus, reminiscent of the infamous chickenpox parties from years past. Imagine a gathering where the goal is not to celebrate, but to contract a virus. Public health experts are sounding the alarm as reports surface of these reckless events, where non-infected individuals mingle with those who have tested positive for COVID-19, hoping to catch the virus and gain immunity. Recently, health officials in southeastern Washington State revealed they had evidence linking such gatherings to at least two new cases of the virus. However, just as quickly, they walked back those statements, suggesting that these so-called “Covid-19 parties” might not be as sinister as initially thought. Meghan DeBolt, the director of community health for Walla Walla County, explained, “We’re still piecing together what’s happening. While we’ve heard reports of gatherings with infected individuals, we don’t have clear evidence that attendees are intentionally trying to get sick.” The situation is further complicated by a significant outbreak at a local meat processing facility, prompting officials to urge the community to remain vigilant. “We know that people are exhausted from isolation and quarantine,” DeBolt continued. “We want to reopen, too. But we need everyone’s help to stay diligent just a little longer. The stakes are too high.” The concept of infection parties raises serious concerns among health experts, who recall a troubling history of purposeful infections. The Centers for Disease Control and Prevention has strongly advised against such gatherings, warning of the potential consequences that could arise from this dangerous trend. As the pandemic drags on, the question looms: Are we witnessing a desperate attempt to reclaim normalcy, or are these gatherings a reckless gamble with lives at stake? Only time will tell, but one thing is clear: the fight against COVID-19 is far from over, and the choices we make today could shape the future of our communities. } 
\end{minipage}
\end{center}

In contrast, the LLM-generated “less engaging” version is more neutral and factual:
\begin{center}
\begin{minipage}{0.8\columnwidth}
\vspace{12 pt}
\label{h1_old}
\textit{\centering Reports of Gatherings During COVID-19 \\ 
Some individuals have been reported to be attending gatherings during the COVID-19 pandemic, which are sometimes referred to as “Covid-19 parties.” These events are thought to involve people trying to expose themselves to the coronavirus in order to gain immunity, similar to past chickenpox parties. Recently, health officials in southeastern Washington State mentioned that there may be a connection between these gatherings and a few new coronavirus cases. However, they later clarified that the gatherings might not have been as intentional as previously suggested. Meghan DeBolt, the director of community health for Walla Walla County, stated that officials are still investigating the situation. She noted that while there have been reports of parties with infected individuals, there is no definitive evidence that attendees are deliberately trying to contract the virus. The county is currently dealing with an outbreak at a local meat processing facility, and officials are concerned about the potential for more infections due to social interactions that go against stay-at-home orders. DeBolt mentioned that the community needs to continue following health guidelines to help control the spread of the virus. Health experts have expressed concerns about the idea of infection parties, given the history of similar practices. The Centers for Disease Control and Prevention advises against hosting or participating in such events. In summary, while there are reports of gatherings during the pandemic, the motivations behind them are still unclear, and health officials continue to monitor the situation. } 
\end{minipage}
\end{center}

\section{Details of the LoRA Fine Tuning Technique}
\label{sec:lora}
\begin{figure}[H]
    \centering
    \includegraphics[width=0.6\linewidth]{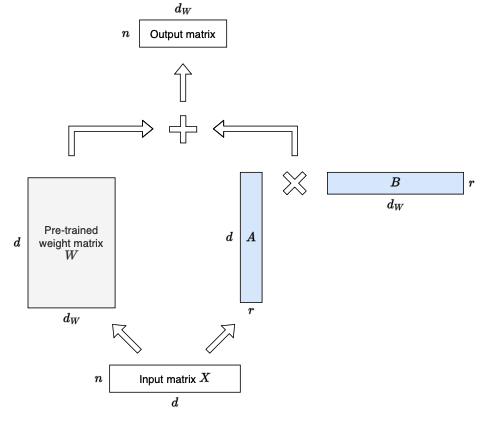}
    \caption{Illustration of LoRA Fine-Tuning}
    \label{fig:lora}
\end{figure}

\noindent In Figure \ref{fig:lora}, we illustrate LoRA \citep{hu2021lora}, a Parameter-Efficient Fine-Tuning (PEFT) method that enables fine-tuning on hardware with limited GPU memory while maintaining competitive model performance. Let $W$ represent a pre-trained weight matrix, such as $W_Q$, $W_K$, or $W_V$, which represent the query, key, and value vectors in the architecture of the Transformer’s attention mechanism.  These matrices store most of the information and knowledge learned by a LLM, and play a crucial role in calculating the attention scores and subsequently determining the relevance of different tokens in the input sequence. 
 We use $d \times d_W$ to represent the dimension of the weight matrix and $r$ to represent the rank of the low-rank matrices. LoRA modifies the weight matrix $W$ during fine-tuning as follows:

\begin{align}
    W' = W + \Delta W, \text{where \quad} \Delta W = AB, \text{\quad with\quad}A\in R^{d \times r,\quad} B\in R^{r \times d_w}.
\end{align}
 
Essentially, LoRA introduces low-rank matrices A and B to model the weight matrix change $\Delta W$ during fine-tuning, while keeping the original weight matrix $W$ frozen. This strategy significantly reduces the number of trainable parameters from $d \times d_W$ to $r \times (d+ d_W)$, as the rank $r$ is typically set to a small value, such as 4 or 8, to balance the trade-off between model capacity and computational efficiency.

\section{Evaluation Ratings for Polarization and Engagement for the Main Model}
In this section, we provide details of four approaches for obtaining article evaluation ratings on engagement and polarization levels: (a) Gemini, (b) Claude, (c) human coders, and (d) lab experiments on Qualtrics.

\subsection{Prompts for Gemini Evaluation}
\label{app:gemini_eval}
\noindent We use Gemini to assess the polarization and engagement levels of the articles in $\S$\ref{ssec:enhance_engage_modpo}. To evaluate polarization, we use the following prompt: \\
\textit{``I will provide you with three versions of lead paragraphs from a news article. Please rate each one (``Version A", ``Version B", and ``Version C") on a scale of 1 to 10 for how polarizing it is. Afterward, explain which version you find the most polarizing, which version you find the least polarizing, or if you feel all versions are equally polarizing: \\        
``Version A": the original article\\
``Version B": the article generated by the multi-objective DPO model\\
``Version C": the article generated by the engaging DPO model\\} 

\noindent Similarly, to evaluate engagement, we use the following prompt:\\
\textit{``I will provide you with three versions of lead paragraphs from a news article. Please rate each one (``Version A", ``Version B", and ``Version C") on a scale of 1 to 10 for how engaging it is. Afterward, explain which version you find the most engaging, which version you find the least engaging, or if you feel all versions are equally engaging:\\       
``Version A": the original article\\
``Version B": the article generated by the multi-objective DPO model\\
``Version C": the article generated by the engaging DPO model\\}

\subsection{Ratings from Claude Evaluation}
\label{app:claude3_eval}
We also use an alternative LLM -- ``claude-3-5-sonnet-20240620," a top-ranked and widely used model known for its strong performance on major NLP benchmarks. To ensure consistency, we use the same prompt as in the Gemini evaluation when assessing polarization with Claude, and the results are reported in Table \ref{tab:claude3_eval}.




Table \ref{tab:claude3_eval} shows how the engagement and polarization scores of the MODPO and engaging DPO versions differ from the original version, as well as how they compare to each other. We find that the articles modified using both \textit{the engaging DPO} and \textit{multi-objective DPO} are significantly more engaging than the original articles ($\Delta_{\text{EDPO\_ORIG}} = 1.49$, $p < 0.001$; $\Delta_{\text{MODPO\_ORIG}} = 1.82$, $p < 0.001$). Interestingly, articles generated by \textit{the multi-objective DPO model} exhibit even higher engagement than those generated by \textit{the engaging DPO}, which suggests that moving from a single-objective setting (focused solely on maximizing engagement) to a multi-objective setting does not necessarily compromise engagement, which tends to be the primary metric of interest for firms. Further, \textit{the multi-objective DPO model} is effective in controlling polarization compared to \textit{the engaging DPO model} ($\Delta_{\text{MODPO\_EDPO}} = -0.27$, $p < 0.001$). However, it still leads to some increase in polarization compared to the original version ($\Delta_{\text{MODPO\_ORIG}} = 0.88$, $p < 0.001$), indicating that it is not able to fully avoid any increase in polarization, which is understandable given the desire to simultaneously boost engagement. In sum, our results demonstrate that \textit{the multi-objective DPO model} effectively balances engagement and polarization as discussed in the main paper.

\begin{table}[htp!]
\caption{Claude Evaluation Results (with \textit{Engaging SFT} as the Reference Policy)}
\def\sym#1{\ifmmode^{#1}\else\(^{#1}\)\fi}
\scalebox{0.95}{
\begin{tabular}{cccc|ccc}
\hline
      & \multicolumn{3}{c|}{Difference in Engagement Scores}                  & \multicolumn{3}{c}{Difference in Polarization Scores}                 \\ \hline
      & \multicolumn{1}{c|}{$\Delta_{\text{MODPO\_ORIG}}$} & \multicolumn{1}{c|}{$\Delta_{\text{EDPO\_ORIG}}$} & $\Delta_{\text{MODPO\_EDPO}}$ & \multicolumn{1}{c|}{$\Delta_{\text{MODPO\_ORIG}}$} & \multicolumn{1}{c|}{$\Delta_{\text{EDPO\_ORIG}}$} & $\Delta_{\text{MODPO\_EDPO}}$\\ \hline
Mean      & \multicolumn{1}{c|}{1.82\sym{***}}                    & \multicolumn{1}{c|}{1.49\sym{***}}                  & 0.32\sym{**}                      & \multicolumn{1}{c|}{0.88\sym{***}}                             & \multicolumn{1}{c|}{1.15\sym{***}}                           & -0.27\sym{***}                              \\
Std       & \multicolumn{1}{c|}{1.34}                    & \multicolumn{1}{c|}{1.27}                  & 1.11                      & \multicolumn{1}{c|}{2.10}                             & \multicolumn{1}{c|}{1.96}                           & 1.24                               \\
Min       & \multicolumn{1}{c|}{-4.00}                   & \multicolumn{1}{c|}{-4.00}                 & -3.00                     & \multicolumn{1}{c|}{-6.00}                            & \multicolumn{1}{c|}{-6.00}                          & -5.00                              \\
25\%      & \multicolumn{1}{c|}{2.00}                    & \multicolumn{1}{c|}{1.00}                  & -1.00                     & \multicolumn{1}{c|}{1.00}                             & \multicolumn{1}{c|}{1.00}                           & -1.00                              \\
50\%      & \multicolumn{1}{c|}{2.00}                    & \multicolumn{1}{c|}{1.00}                  & 1.00                      & \multicolumn{1}{c|}{1.00}                             & \multicolumn{1}{c|}{2.00}                           & -1.00                              \\
75\%      & \multicolumn{1}{c|}{3.00}                    & \multicolumn{1}{c|}{2.00}                  & 1.00                      & \multicolumn{1}{c|}{2.00}                             & \multicolumn{1}{c|}{2.00}                           & 1.00                               \\
Max       & \multicolumn{1}{c|}{5.00}                    & \multicolumn{1}{c|}{5.00}                  & 4.00                      & \multicolumn{1}{c|}{5.00}                             & \multicolumn{1}{c|}{5.00}                           & 2.00                               \\ 
N  & \multicolumn{1}{c|}{983.00}           & \multicolumn{1}{c|}{983.00}              & 983.00      & \multicolumn{1}{c|}{983.00}          & \multicolumn{1}{c|}{983.00}             & 983.00      \\ \hline
\multicolumn{7}{l}{\scriptsize \sym{*} \(p<0.05\), \sym{**} \(p<0.01\), \sym{***} \(p<0.001\)}\\ [-3pt]
\multicolumn{7}{l}{\scriptsize $\Delta_{\text{MODPO\_ORIG}}$ = MODPO version - Original version; $\Delta_{\text{EDPO\_ORIG}}$ = Engaging DPO version - Original version;} \\ [-3pt]
\multicolumn{7}{l}{\scriptsize $\Delta_{\text{MODPO\_EDPO}}$ = MODPO version - Engaging DPO version.} \\
\end{tabular}
}
\label{tab:claude3_eval}
\end{table}

\subsection{Ratings from Qualtrics Experiments}
\label{app:qualtrics_eval}

\begin{figure}[htp!] 
    \centering
    \includegraphics[width=1.0\linewidth]{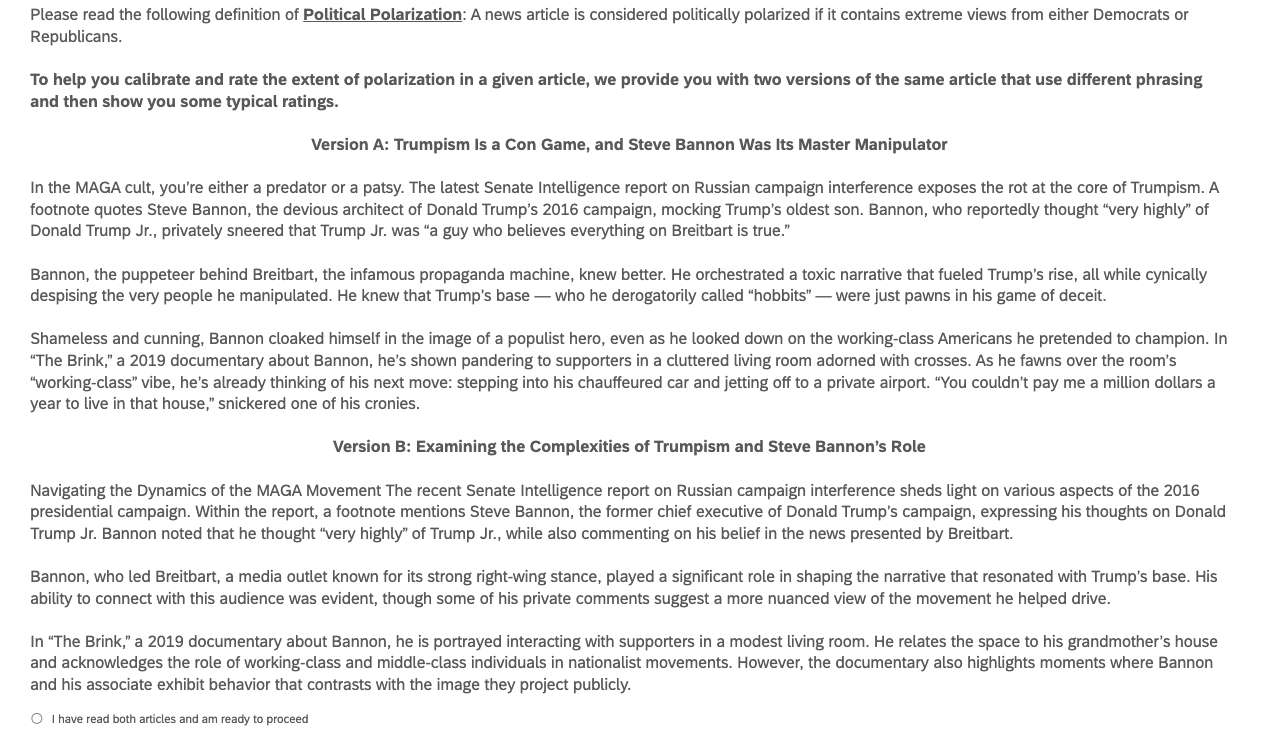}
    \caption{Definition and Illustrative Example of Polarization Rating Task}
    \label{fig:labpolarization}
\end{figure}

\begin{figure}[htp!] 
    \centering
    \includegraphics[width=1.0\linewidth]{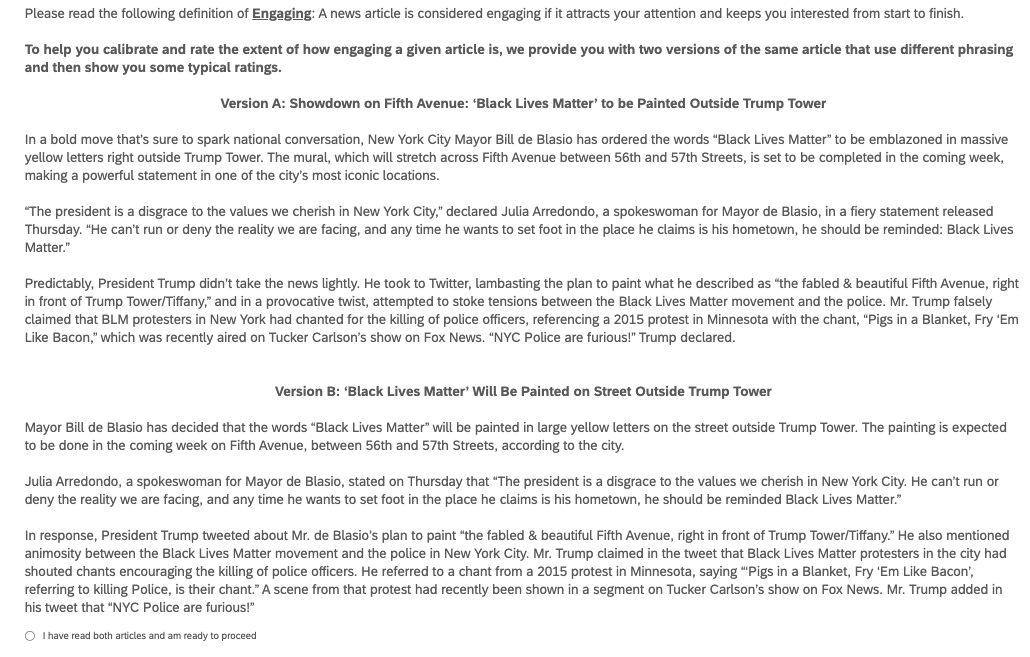}
    \caption{Definition and Illustrative Example of Engagement Rating Task}
    \label{fig:labengagement}
\end{figure}

\begin{figure}[htp!] 
    \centering
    \includegraphics[width=1.0\linewidth]{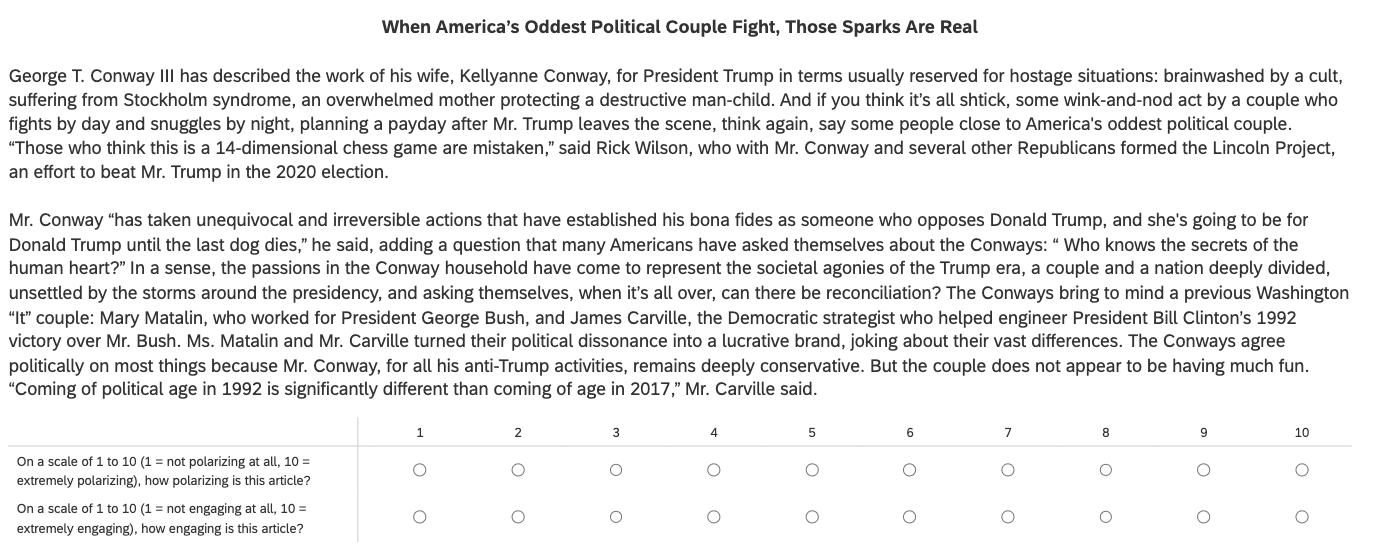}
    \caption{User Interface for Article Rating}
    \label{fig:labexample}
\end{figure}

We sought to understand how people rate news articles in terms of polarization and engagement. To achieve this, we conducted a lab experiment with 230 undergraduate students from a major public U.S. university,\footnote{Of these 230 students, 180 provided valid responses; the remaining students indicated that they did not consider their responses to be of good quality and chose not to be included in the study. On average, each article received two to three valid responses.} using a total of 70 selected news articles from our evaluation set. Each participant was randomly assigned to review three different articles presented in a randomized order: one original article, one generated using the multi-objective DPO model, and one generated using the engaging DPO model. Participants were asked to review snippets of these articles and evaluate them on two dimensions: (1) how engaging or interesting they found the article, and (2) how politically polarizing it appeared. To ensure thoughtful evaluations, we first provided participants with clear definitions of both political polarization and engagement, along with illustrative examples that demonstrated high and low levels of each dimension. Later, they were required to spend a minimum of 30 seconds reviewing and rating each article. Figures \ref{fig:labpolarization} and \ref{fig:labengagement} provide the definitions and illustrative examples shown to participants for polarization and engagement, respectively. The user interface used for article rating is presented in Figure \ref{fig:labexample}.

The polarization and engagement ratings from the Qualtrics experiment are summarized in Table \ref{apptab:qualtrics_evaluation_eng}. The results show that articles generated using \textit{the engaging DPO} model or \textit{the multi-objective DPO} model were rated as significantly more engaging compared to the original articles ($\Delta_{\text{EDPO\_ORIG}} = 0.62$, $p < 0.001$; $\Delta_{\text{MODPO\_ORIG}} = 0.62$, $p < 0.001$). Notably, while both methods enhance engagement, articles generated using \textit{the multi-objective DPO model} exhibit a substantial reduction in polarization compared to \textit{the engaging DPO model} 
($\Delta_{\text{MODPO\_EDPO}} = -0.43$, $p < 0.001$).

\begin{table}[htp!]
\caption{Ratings from Qualtrics Experiment (with \textit{Engaging SFT} as the Reference Policy)}
\def\sym#1{\ifmmode^{#1}\else\(^{#1}\)\fi}
\scalebox{0.95}{
\begin{tabular}{cccc|ccc}
\hline
      & \multicolumn{3}{c|}{Difference in Engagement Scores}                  & \multicolumn{3}{c}{Difference in Polarization Scores}                 \\ \hline
      & \multicolumn{1}{c|}{$\Delta_{\text{MODPO\_ORIG}}$} & \multicolumn{1}{c|}{$\Delta_{\text{EDPO\_ORIG}}$} & $\Delta_{\text{MODPO\_EDPO}}$ & \multicolumn{1}{c|}{$\Delta_{\text{MODPO\_ORIG}}$} & \multicolumn{1}{c|}{$\Delta_{\text{EDPO\_ORIG}}$} & $\Delta_{\text{MODPO\_EDPO}}$\\ \hline
Mean & \multicolumn{1}{c|}{0.62\sym{***}}                      & \multicolumn{1}{c|}{0.62\sym{***}}                    & -0.01                       & \multicolumn{1}{c|}{-0.20\sym{***}}                              & \multicolumn{1}{c|}{0.23\sym{***}}                             & -0.43\sym{***}                                \\
Std  & \multicolumn{1}{c|}{2.72}                      & \multicolumn{1}{c|}{2.55}                    & 2.49                        & \multicolumn{1}{c|}{2.73}                               & \multicolumn{1}{c|}{2.93}                             & 2.62                                 \\
Min  & \multicolumn{1}{c|}{-7.00}                     & \multicolumn{1}{c|}{-6.00}                   & -7.00                       & \multicolumn{1}{c|}{-9.00}                              & \multicolumn{1}{c|}{-9.00}                            & -6.00                                \\
25\% & \multicolumn{1}{c|}{-1.00}                     & \multicolumn{1}{c|}{-1.00}                   & -1.25                       & \multicolumn{1}{c|}{-2.00}                              & \multicolumn{1}{c|}{-1.00}                            & -2.00                                \\
50\% & \multicolumn{1}{c|}{1.00}                      & \multicolumn{1}{c|}{1.00}                    & 0.00                        & \multicolumn{1}{c|}{0.00}                               & \multicolumn{1}{c|}{0.00}                             & -0.50                                \\
75\% & \multicolumn{1}{c|}{2.00}                      & \multicolumn{1}{c|}{2.00}                    & 1.25                        & \multicolumn{1}{c|}{1.00}                               & \multicolumn{1}{c|}{2.00}                             & 1.00                                 \\
Max  & \multicolumn{1}{c|}{8.00}                      & \multicolumn{1}{c|}{6.00}                    & 8.00                        & \multicolumn{1}{c|}{9.00}                               & \multicolumn{1}{c|}{8.00}                             & 9.00                       \\
N  & \multicolumn{1}{c|}{180.00}           & \multicolumn{1}{c|}{180.00}              & 180.00      & \multicolumn{1}{c|}{180.00}          & \multicolumn{1}{c|}{180.00}             & 180.00      \\ \hline
\multicolumn{7}{l}{\scriptsize \sym{*} \(p<0.05\), \sym{**} \(p<0.01\), \sym{***} \(p<0.001\)}\\ [-3pt]
\multicolumn{7}{l}{\scriptsize $\Delta_{\text{MODPO\_ORIG}}$ = MODPO version - Original version; $\Delta_{\text{EDPO\_ORIG}}$ = Engaging DPO version - Original version;} \\ [-3pt]
\multicolumn{7}{l}{\scriptsize $\Delta_{\text{MODPO\_EDPO}}$ = MODPO version - Engaging DPO version.} \\
\end{tabular}
}
\label{apptab:qualtrics_evaluation_eng}
\end{table}
\subsection{Ratings from Human Rater}
\label{app:humancoding_eval}

We hired a human rater to evaluate news articles on polarization and engagement. To ensure consistency, transparency, and reliability in this annotation process, we developed a detailed annotation rubric. In the rubric, political polarization is defined as the degree to which a news article contains extreme views from either Democrats or Republicans. Additionally, it also includes some specific cues to guide raters in evaluating polarization:

\begin{enumerate}
    \item Framing of Issues
    \begin{itemize}
        \item 1-3: The article lacks a clear framing of polarization issues or pays minimal attention to them. 
        \item 4-6: Issues are framed in a basic manner, with mild divisive undertones.
        \item 7-8: The framing acknowledges multiple perspectives but creates divisions and makes assertive claims. 
        \item 9-10: The framing strongly highlights clear divisions or creates larger divisions. 
    \end{itemize}
    \item Amount of Perspectives Included
    \begin{itemize}
        \item 1-3: The article actively presents different perspectives equally, maintaining an informational tone without promoting a specific agenda. 
        \item 4-6: Diverse perspectives are integrated, demonstrating an effort toward inclusion.
        \item 7-8: Some perspectives are included, but the article subtly reinforces a particular agenda or framing.
        \item 9-10: Some key voices are marginalized, or significant perspectives are omitted entirely. 
    \end{itemize}
    \item Language and Tone
    \begin{itemize}
        \item 1-3: The language is neutral, factual, and unbiased.
        \item 4-6: The tone is slightly subjective, reflecting underlying opinions without overt bias; speculative language may be present.
        \item 7-8: Clear biases emerge, either positively or negatively, with a more definitive tone.
        \item 9-10: The language is heavily charged and intentionally inflammatory or manipulative. 
    \end{itemize}
    \item Informativeness
    \begin{itemize}
        \item 1-3: The article is factually rich, comprehensive, and balanced. 
        \item 4-6: It is well-researched and substantial but may lack key details or alternative viewpoints.
        \item 7-8: While including relevant facts, the content lacks depth or clarity.
        \item 9-10: Factual information is very limited.
    \end{itemize}
\end{enumerate}

In the rubric, engagement is defined as the extent to which an article attracts reader attention and keeps the reader interested from beginning to end. The rubric also provides some specific cues to guide raters in assessing engagement: 

\begin{enumerate}
    \item Emotional Resonance
    \begin{itemize}
        \item 1-3: Lacks emotional appeal or connection. 
        \item 4-6: Exhibits basic emotional appeal with limited resonance. 
        \item 7-8: Demonstrates moderate emotional engagement, resonating with some readers.
        \item 9-10: Creates a strong emotional connection, deeply resonating with a broad audience. 
    \end{itemize}
    \item Call to Action or Reflection
    \begin{itemize}
        \item 1-3: Lacks a clear takeaway or purpose.
        \item 4-6: Provides minimal call to action or a vague reflection.
        \item 7-8: Provides clear call to action or a modest reflection.
        \item 9-10: Offers a strong, compelling call to action or encourages deep reflection, effectively motivating the audience. 
    \end{itemize}
    \item Opening
    \begin{itemize}
    \item 1-3: Starts weakly, offering little context or interest.
	\item 4-6: Moderately engaging, though it lacks a strong hook.
	\item 7-8: Opens with a compelling question, statement, or scenario that grabs attention.
	\item 9-10: Delivers an exceptionally compelling opening, immediately drawing the readers in.
    \end{itemize}
    \item Irrelevance
    \begin{itemize}
        \item 1-3: Overly lengthy with significant irrelevant content.
        \item 4-6: Contains sections that feel too long or irrelevant but offer some value. 
        \item 7-8:  Includes minor irrelevant parts that detract from the main message. 
        \item 9-10: Concise and focused, staying relevant and to the point throughout. 
    \end{itemize}
    \item Storytelling
    \begin{itemize}
        \item 1-3: Lacks coherence and a clear flow. 
        \item 4-6: Features a basic narrative structure but may lack fluidity or strong transitions.
        \item 7-8: Provides a well-balanced narrative with good integration of storytelling and information. 
        \item 9-10: Delivers a compelling narrative with a strong structure, smoothly guiding the reader through the key points.
    \end{itemize}
\end{enumerate}

We present two sets of results based on the ratings provided by the human annotator. First, we examine how the rater evaluates the polarization and engagement of the original versions of the articles. Table \ref{apptab:human_evaluation_eng_relationship} reports the correlation between these two scores. This table is analogous to Table \ref{tab:allarticles} in the main text, which showed the corresponding relationship based on GPT ratings. Consistent with the GPT ratings, we find a significant positive relationship between polarization and engagement, regardless of whether we control for the topic distribution of the articles (column (1): $\beta$ = 0.728, $p <$ 0.001; column (2): $\beta$ = 0.724, $p <$ 0.001). 

Second, we discuss how the human rater evaluates the three versions of each article -- namely, the original version, the MODPO version, and the engaging DPO version. The results are shown in Table \ref{apptab:human_evaluation_eng}, which complements Table \ref{tab:gemini_evaluation_eng} in the main text, where we showed similar ratings using the Gemini model. Overall, we find that the human rater's evaluation is largely consistent with the Gemini evaluations. The results show that articles generated using \textit{the engaging DPO model} or \textit{the multi-objective DPO model} are significantly more engaging compared to the original articles ($\Delta_{\text{EDPO\_ORIG}} = 1.39$, $p < 0.001$; $\Delta_{\text{MODPO\_ORIG}} = 1.31$, $p < 0.001$), and there is no significant difference in engagement between the two methods ($\Delta_{\text{MODPO\_EDPO}} = -0.07$, $p > 0.1$). Notably, while both methods enhance engagement, articles generated using \textit{the multi-objective DPO} exhibit a substantial reduction in polarization ($\Delta_{\text{MODPO\_EDPO}} = -0.46$, $p < 0.001$) relative to \textit{the engaging DPO}.

\begin{table}[htp!]
    \centering
    \caption{Relationship between Polarization and Engagement in Human Annotation Data}
    {
\def\sym#1{\ifmmode^{#1}\else\(^{#1}\)\fi}
\begin{tabular}{l*{2}{cc}}
\hline\hline
            &\multicolumn{2}{c}{(1)}           &\multicolumn{2}{c}{(2)}           \\
            &\multicolumn{2}{c}{Polarization}&\multicolumn{2}{c}{Polarization}\\
\hline
Engagement& 0.728\sym{***}&    (0.0250)&       0.724\sym{***}&    (0.0258)\\
Constant      &       1.217\sym{***}&    (0.0795)&       1.835\sym{**} &     (0.704)\\
\hline
Topics       &       No         &            &       Yes         &            \\
\(N\)       &         983         &            &         983         &            \\
\hline\hline
\multicolumn{5}{l}{\scriptsize Standard errors in parentheses}\\ [-3pt]
\multicolumn{5}{l}{\scriptsize \sym{*} \(p<0.05\), \sym{**} \(p<0.01\), \sym{***} \(p<0.001\)}\\
\end{tabular}
}
    \label{apptab:human_evaluation_eng_relationship}
\end{table}

\begin{table}[htp!]
\caption{Human Annotation Evaluation Results (with \textit{Engaging SFT} as the Reference Policy)}
\def\sym#1{\ifmmode^{#1}\else\(^{#1}\)\fi}
\scalebox{0.95}{
\begin{tabular}{cccc|ccc}
\hline
      & \multicolumn{3}{c|}{Difference in Engagement Scores}                  & \multicolumn{3}{c}{Difference in Polarization Scores}                 \\ \hline
      & \multicolumn{1}{c|}{$\Delta_{\text{MODPO\_ORIG}}$} & \multicolumn{1}{c|}{$\Delta_{\text{EDPO\_ORIG}}$} & $\Delta_{\text{MODPO\_EDPO}}$ & \multicolumn{1}{c|}{$\Delta_{\text{MODPO\_ORIG}}$} & \multicolumn{1}{c|}{$\Delta_{\text{EDPO\_ORIG}}$} & $\Delta_{\text{MODPO\_EDPO}}$\\ \hline
Mean      & \multicolumn{1}{c|}{1.31\sym{***}}                    & \multicolumn{1}{c|}{1.39\sym{***}}                  & -0.07                     & \multicolumn{1}{c|}{1.42\sym{***}}                             & \multicolumn{1}{c|}{1.89\sym{***}}                           & -0.46\sym{***}                              \\
Std       & \multicolumn{1}{c|}{1.13}                    & \multicolumn{1}{c|}{1.24}                  & 1.24                      & \multicolumn{1}{c|}{1.34}                             & \multicolumn{1}{c|}{1.61}                           & 1.57                               \\
Min       & \multicolumn{1}{c|}{-2.00}                   & \multicolumn{1}{c|}{-3.00}                 & -3.00                     & \multicolumn{1}{c|}{-4.00}                            & \multicolumn{1}{c|}{-5.00}                          & -4.00                              \\
25\%      & \multicolumn{1}{c|}{1.00}                    & \multicolumn{1}{c|}{1.00}                  & -1.00                     & \multicolumn{1}{c|}{1.00}                             & \multicolumn{1}{c|}{1.00}                           & -2.00                              \\
50\%      & \multicolumn{1}{c|}{1.00}                    & \multicolumn{1}{c|}{1.00}                  & 0.00                      & \multicolumn{1}{c|}{1.00}                             & \multicolumn{1}{c|}{2.00}                           & -1.00                              \\
75\%      & \multicolumn{1}{c|}{2.00}                    & \multicolumn{1}{c|}{2.00}                  & 1.00                      & \multicolumn{1}{c|}{2.00}                             & \multicolumn{1}{c|}{3.00}                           & 1.00                               \\
Max       & \multicolumn{1}{c|}{5.00}                    & \multicolumn{1}{c|}{6.00}                  & 4.00                      & \multicolumn{1}{c|}{6.00}                             & \multicolumn{1}{c|}{6.00}                           & 5.00                               \\ 
N  & \multicolumn{1}{c|}{983.00}           & \multicolumn{1}{c|}{983.00}              & 983.00      & \multicolumn{1}{c|}{983.00}          & \multicolumn{1}{c|}{983.00}             & 983.00      \\ \hline
\multicolumn{7}{l}{\scriptsize \sym{*} \(p<0.05\), \sym{**} \(p<0.01\), \sym{***} \(p<0.001\)}\\ [-3pt]
\multicolumn{7}{l}{\scriptsize $\Delta_{\text{MODPO\_ORIG}}$ = MODPO version - Original version; $\Delta_{\text{EDPO\_ORIG}}$ = Engaging DPO version - Original version;} \\ [-3pt]
\multicolumn{7}{l}{\scriptsize $\Delta_{\text{MODPO\_EDPO}}$ = MODPO version - Engaging DPO version.} \\
\end{tabular}
}
\label{apptab:human_evaluation_eng}
\end{table}

\newpage
\section{Raw Values for the Evaluation Results}
\label{app:rawvalues}
Table \ref{tab:gemini_evaluation_eng_raw} presents the raw engagement and polarization levels obtained from ``Gemini-1.5-Flash" for the original article, the MODPO version, and the engaging DPO version we describe in $\S$\ref{ssec:enhance_engage_modpo}.
\begin{table}[htp!]
\caption{Evaluation Results Using \textit{Engaging SFT} as the Reference Policy}
\vspace{-15pt}
\begin{center}
\scalebox{1.0}{
\begin{tabular}{cccc|ccc}
\hline
     & \multicolumn{3}{c|}{Engagement}                                                                 & \multicolumn{3}{c}{Polarization}                                                               \\ \hline
     & \multicolumn{1}{c|}{$\text{Eng}_{\text{ORIG}}$} & \multicolumn{1}{c|}{$\text{Eng}_{\text{MODPO}}$} & $\text{Eng}_{\text{EDPO}}$ & \multicolumn{1}{c|}{$\text{Pol}_{\text{ORIG}}$} & \multicolumn{1}{c|}{$\text{Pol}_{\text{MODPO}}$} & $\text{Pol}_{\text{EDPO}}$ \\ \hline
Mean & \multicolumn{1}{c|}{5.37}           & \multicolumn{1}{c|}{7.39}              & 7.77       & \multicolumn{1}{c|}{5.64}           & \multicolumn{1}{c|}{6.10}              & 7.09       \\ 
Std  & \multicolumn{1}{c|}{1.34}           & \multicolumn{1}{c|}{1.45}              & 1.65       & \multicolumn{1}{c|}{2.20}           & \multicolumn{1}{c|}{1.67}              & 1.67       \\ 
Min  & \multicolumn{1}{c|}{1.00}           & \multicolumn{1}{c|}{1.00}              & 1.00       & \multicolumn{1}{c|}{1.00}           & \multicolumn{1}{c|}{1.00}              & 1.00       \\ 
25\% & \multicolumn{1}{c|}{5.00}           & \multicolumn{1}{c|}{7.00}              & 7.00       & \multicolumn{1}{c|}{4.00}           & \multicolumn{1}{c|}{5.00}              & 6.00       \\ 
50\% & \multicolumn{1}{c|}{5.00}           & \multicolumn{1}{c|}{8.00}              & 8.00       & \multicolumn{1}{c|}{6.00}           & \multicolumn{1}{c|}{6.00}              & 7.00       \\ 
75\% & \multicolumn{1}{c|}{6.00}           & \multicolumn{1}{c|}{8.00}              & 9.00       & \multicolumn{1}{c|}{7.00}           & \multicolumn{1}{c|}{7.00}              & 8.00       \\ 
Max  & \multicolumn{1}{c|}{9.00}           & \multicolumn{1}{c|}{9.00}              & 10.00      & \multicolumn{1}{c|}{19.00}          & \multicolumn{1}{c|}{19.00}             & 19.00      \\ 
N  & \multicolumn{1}{c|}{998.00}           & \multicolumn{1}{c|}{998.00}              & 998.00      & \multicolumn{1}{c|}{998.00}          & \multicolumn{1}{c|}{998.00}             & 998.00      \\ 
\hline
\multicolumn{7}{l}{\scriptsize ORIG = Original version; MODPO = MODPO version; EDPO = Engaging DPO version.} \\
\end{tabular}
}
\end{center}
\label{tab:gemini_evaluation_eng_raw}
\end{table}

Table \ref{tab:gemini_evaluation_pol_raw} presents the raw engagement and polarization levels obtained from ``Gemini-1.5-Flash" for the original article, the MODPO version, and the less polarizing DPO version we describe in $\S$\ref{sssec:managepolarization}.

\begin{table}[htp!]
\caption{Evaluation Results Using \textit{Less-polarizing SFT} as the Reference Policy}
\vspace{-15pt}
\begin{center}
\begin{tabular}{cccc|ccc}
\hline
     & \multicolumn{3}{c|}{Engagement}                                                                 & \multicolumn{3}{c}{Polarization}                                                               \\ \hline
     & \multicolumn{1}{c|}{$\text{Eng}_{\text{ORIG}}$} & \multicolumn{1}{c|}{$\text{Eng}_{\text{MODPO}}$} & $\text{Eng}_{\text{LDPO}}$ & \multicolumn{1}{c|}{$\text{Pol}_{\text{ORIG}}$} & \multicolumn{1}{c|}{$\text{Pol}_{\text{MODPO}}$} & $\text{Pol}_{\text{LDPO}}$ \\ \hline
Mean & \multicolumn{1}{c|}{6.81}           & \multicolumn{1}{c|}{6.77}              & 3.58             & \multicolumn{1}{c|}{6.40}           & \multicolumn{1}{c|}{5.62}              & 3.11             \\ 
Std  & \multicolumn{1}{c|}{1.45}           & \multicolumn{1}{c|}{1.78}              & 1.35             & \multicolumn{1}{c|}{2.09}           & \multicolumn{1}{c|}{1.46}              & 1.83             \\ 
Min  & \multicolumn{1}{c|}{1.00}           & \multicolumn{1}{c|}{1.00}              & 1.00             & \multicolumn{1}{c|}{1.00}           & \multicolumn{1}{c|}{1.00}              & 1.00             \\ 
25\% & \multicolumn{1}{c|}{6.00}           & \multicolumn{1}{c|}{5.00}              & 3.00             & \multicolumn{1}{c|}{5.00}           & \multicolumn{1}{c|}{5.00}              & 2.00             \\
50\% & \multicolumn{1}{c|}{7.00}           & \multicolumn{1}{c|}{7.00}              & 3.00             & \multicolumn{1}{c|}{7.00}           & \multicolumn{1}{c|}{6.00}              & 3.00             \\
75\% & \multicolumn{1}{c|}{8.00}           & \multicolumn{1}{c|}{8.00}              & 4.00             & \multicolumn{1}{c|}{8.00}           & \multicolumn{1}{c|}{7.00}              & 4.00             \\ 
Max  & \multicolumn{1}{c|}{10.00}          & \multicolumn{1}{c|}{9.00}              & 9.00             & \multicolumn{1}{c|}{10.00}          & \multicolumn{1}{c|}{9.00}              & 10.00            \\ 
N  & \multicolumn{1}{c|}{998.00}           & \multicolumn{1}{c|}{998.00}              & 998.00      & \multicolumn{1}{c|}{998.00}          & \multicolumn{1}{c|}{998.00}             & 998.00      \\ \hline
\multicolumn{7}{l}{\scriptsize ORIG = Original version; MODPO = MODPO version; LDPO = Less Polarizing DPO version.} \\
\end{tabular}
\end{center}
\label{tab:gemini_evaluation_pol_raw}
\end{table}

Table \ref{tab:gemini_rawval_leftright} presents the raw left-leaning polarization levels and engagement scores obtained from ``Gemini-1.5-Flash" for the original article, the MODPO version, and the left-leaning DPO version we illustrate in $\S$\ref{ssec:left_right}. 

\begin{table}[htp!]
\caption{Evaluation Results Using \textit{Left-leaning SFT} as the Reference Policy}
\vspace{-15pt}
\small
\begin{center}
\begin{tabular}{cccc|ccc}
\hline
      & \multicolumn{3}{c|}{Left-leaning Polarization}                  & \multicolumn{3}{c}{Engagement}                 \\ \hline
      & \multicolumn{1}{c|}{$\text{Pol}_{\text{ORIG}}$} & \multicolumn{1}{c|}{$\text{Pol}_{\text{MODPO}}$} & $\text{Pol}_{\text{LFDPO}}$ & \multicolumn{1}{c|}{$\text{Eng}_{\text{ORIG}}$} & \multicolumn{1}{c|}{$\text{Eng}_{\text{MODPO}}$} & $\text{Eng}_{\text{LFDPO}}$\\ \hline
Mean      & \multicolumn{1}{c|}{4.30}                                                        & \multicolumn{1}{c|}{4.46}                                                           & 5.59         & \multicolumn{1}{c|}{6.29}             & \multicolumn{1}{c|}{6.91}                & 7.79         \\
Std       & \multicolumn{1}{c|}{1.02}                                                        & \multicolumn{1}{c|}{1.29}                                                           & 1.69         & \multicolumn{1}{c|}{1.02}             & \multicolumn{1}{c|}{1.05}                & 2.33         \\
Min       & \multicolumn{1}{c|}{1.00}                                                        & \multicolumn{1}{c|}{1.00}                                                           & 1.00         & \multicolumn{1}{c|}{2.00}             & \multicolumn{1}{c|}{3.00}                & 1.00         \\
25\%      & \multicolumn{1}{c|}{4.00}                                                        & \multicolumn{1}{c|}{4.00}                                                           & 5.00         & \multicolumn{1}{c|}{6.00}             & \multicolumn{1}{c|}{6.00}                & 8.00         \\
50\%     & \multicolumn{1}{c|}{4.00}                                                        & \multicolumn{1}{c|}{5.00}                                                           & 6.00         & \multicolumn{1}{c|}{6.00}             & \multicolumn{1}{c|}{7.00}                & 9.00         \\
75\%      & \multicolumn{1}{c|}{5.00}                                                        & \multicolumn{1}{c|}{5.00}                                                           & 7.00         & \multicolumn{1}{c|}{7.00}             & \multicolumn{1}{c|}{8.00}                & 9.00         \\
Max       & \multicolumn{1}{c|}{8.00}                                                        & \multicolumn{1}{c|}{8.00}                                                           & 9.00         & \multicolumn{1}{c|}{8.00}             & \multicolumn{1}{c|}{9.00}                & 10.00        \\ 
N  & \multicolumn{1}{c|}{534.00}           & \multicolumn{1}{c|}{534.00}              & 534.00      & \multicolumn{1}{c|}{534.00}          & \multicolumn{1}{c|}{534.00}             & 534.00      \\ \hline
\multicolumn{7}{l}{\scriptsize ORIG = Original version; MODPO = MODPO version; LFDPO = Left-leaning DPO version.} \\
\end{tabular}
\end{center}
\label{tab:gemini_rawval_leftright}
\end{table}

\newpage
\section{Prompts for Polarization and Engagement Evaluation when Changing the Reference Policy}
\label{app:gemini_reference_change}

\noindent We use Gemini to assess the polarization and engagement levels of the articles in $\S$\ref{sssec:managepolarization}. To evaluate polarization, we use the following prompt: \\
\textit{``I will provide you with three versions of lead paragraphs from a news article. Please rate each one (``Version A", ``Version B", and ``Version C") on a scale of 1 to 10 for how polarizing it is. Afterward, explain which version you find the most polarizing, which version you find the least polarizing, or if you feel all versions are equally polarizing: \\        
``Version A": *the original article*\\
``Version B": *the article generated by the multi-objective DPO model*\\
``Version C": *the article generated by the less polarizing DPO model*\\} 

\noindent Similarly, to evaluate engagement, we used the following prompt:\\
\textit{``I will provide you with three versions of lead paragraphs from a news article. Please rate each one (``Version A", ``Version B", and ``Version C") on a scale of 1 to 10 for how engaging it is. Afterward, explain which version you find the most engaging, which version you find the least engaging, or if you feel all versions are equally engaging:\\       
``Version A": *the original article*\\
``Version B": *the article generated by the multi-objective DPO model*\\
``Version C": *the article generated by the less polarizing DPO model*\\}

\section{Appendix for Content Strategies}
\subsection{More Examples of Content Strategies}
\label{app:examples_content_strategies}
We provide two additional examples to illustrate the content strategies across the three article versions.
\begin{figure}[htp!] 
    \centering
\includegraphics[width=1.0\linewidth]{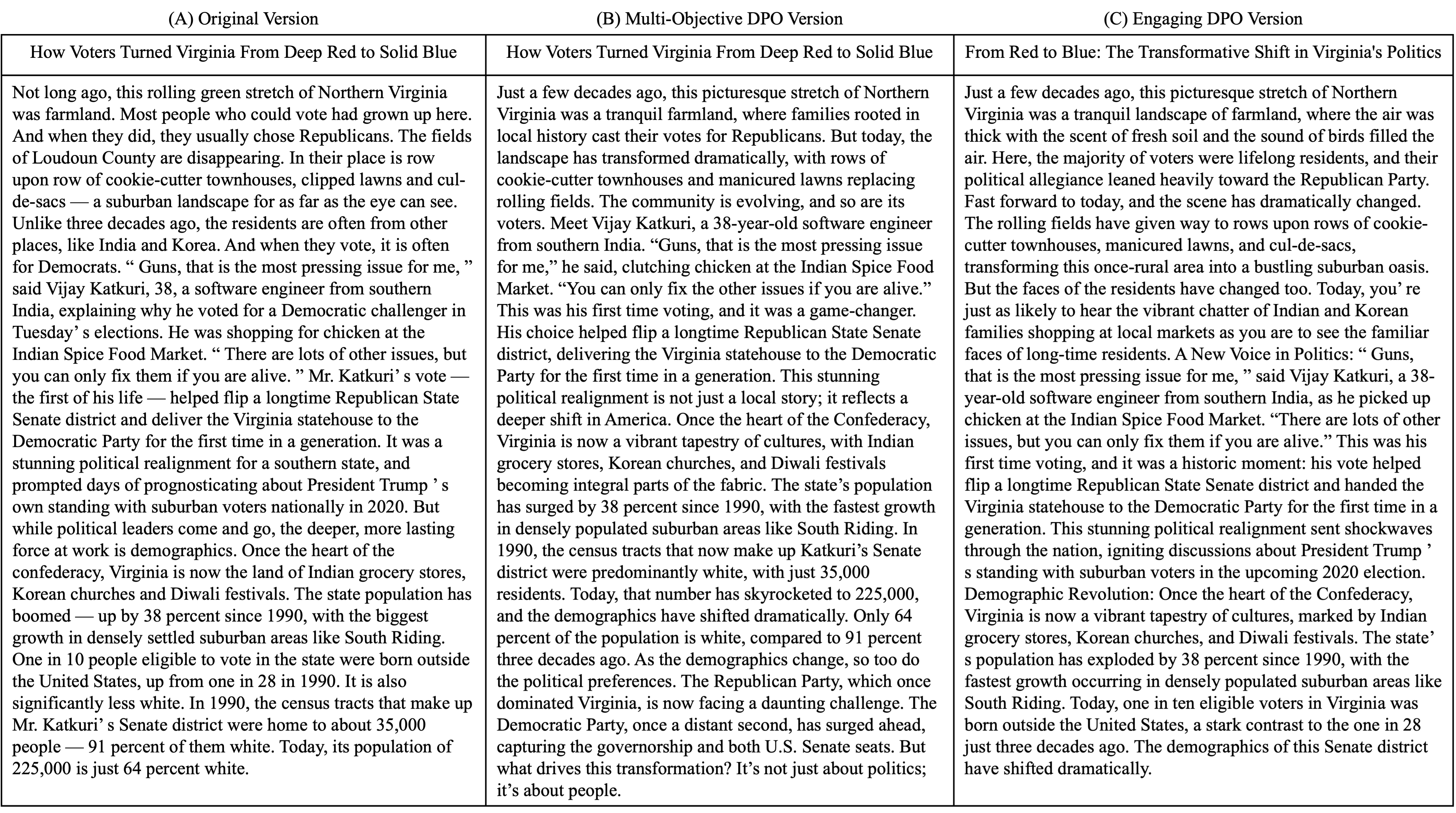}
    \caption{Example 1}
    \label{fig:example2}
\end{figure}
Figure \ref{fig:example2} presents the first example of the three versions, where version A is the original article, version B represents the article generated using the \textit{engaging DPO model}, and version C shows the article generated using the \textit{multi-objective DPO model}. According to Claude 3 evaluations, both the engaging DPO and MODPO versions are more engaging than the original article, as reflected in their engagement scores (original: 7.0, MODPO: 9.0, engaging DPO: 8.0). Additionally, the MODPO version achieves a lower polarization score compared to both the original and engaging DPO versions  (original: 5.0, MODPO: 4.0, engaging DPO: 6.0).

The versions generated using the \textit{multi-objective DPO model} and the \textit{engaging DPO model} enhance engagement by vividly portraying Virginia’s transformation. Both versions open with a compelling contrast between past tranquility and present vibrancy. This juxtaposition of farmland and suburban growth effectively hooks readers by drawing them into the story. By contrast, the original article takes a more direct and factual approach, focusing immediately on demographic and political changes. The narrative structure also varies significantly across the versions. The original article follows a linear, fact-driven flow that prioritizes information over storytelling. In comparison, the MODPO version integrates personal anecdotes, such as Vijay Katkuri’s perspective, with broader societal implications, creating a more cohesive and engaging narrative. Similarly, the engaging DPO version adopts this strategy but includes additional details, which might slow the pacing for some readers.

When it comes to managing polarization, the MODPO version contains less \textit{provocative language}, fostering a more inclusive tone compared to the original version. For instance, the original article states, “Once the heart of the Confederacy, Virginia is now the land of Indian grocery stores, Korean churches, and Diwali festivals,” a phrasing that, while impactful, might provoke strong reactions or come across as divisive. For its part, the MODPO version uses metaphors like “a vibrant tapestry of cultures,” effectively dramatizing the transformation in a way that feels neutral and inclusive. The engaging DPO version, while similarly rich in metaphors, opts for more dramatic and evocative language, describing “a tranquil landscape of farmland” and “a bustling suburban oasis.” The vivid descriptions enhance engagement, but their judgmental tone, e.g., when stating that ``The rolling fields have given way to rows upon rows of cookie-cutter townhouses",  may heighten polarization. Finally, the tone and perspective vary significantly across the versions. The original article provides a wealth of factual information but focuses heavily on demographic and political shifts, which may come across as one-sided. In contrast, the MODPO version strikes a balance, weaving cultural and political commentary with personal stories to create a narrative that feels more neutral. The engaging DPO version emphasizes the dramatic impact of the shifts, delivering a compelling but potentially more emotionally charged narrative.

\begin{figure}[htp!] 
    \centering
    \includegraphics[width=1.0\linewidth]{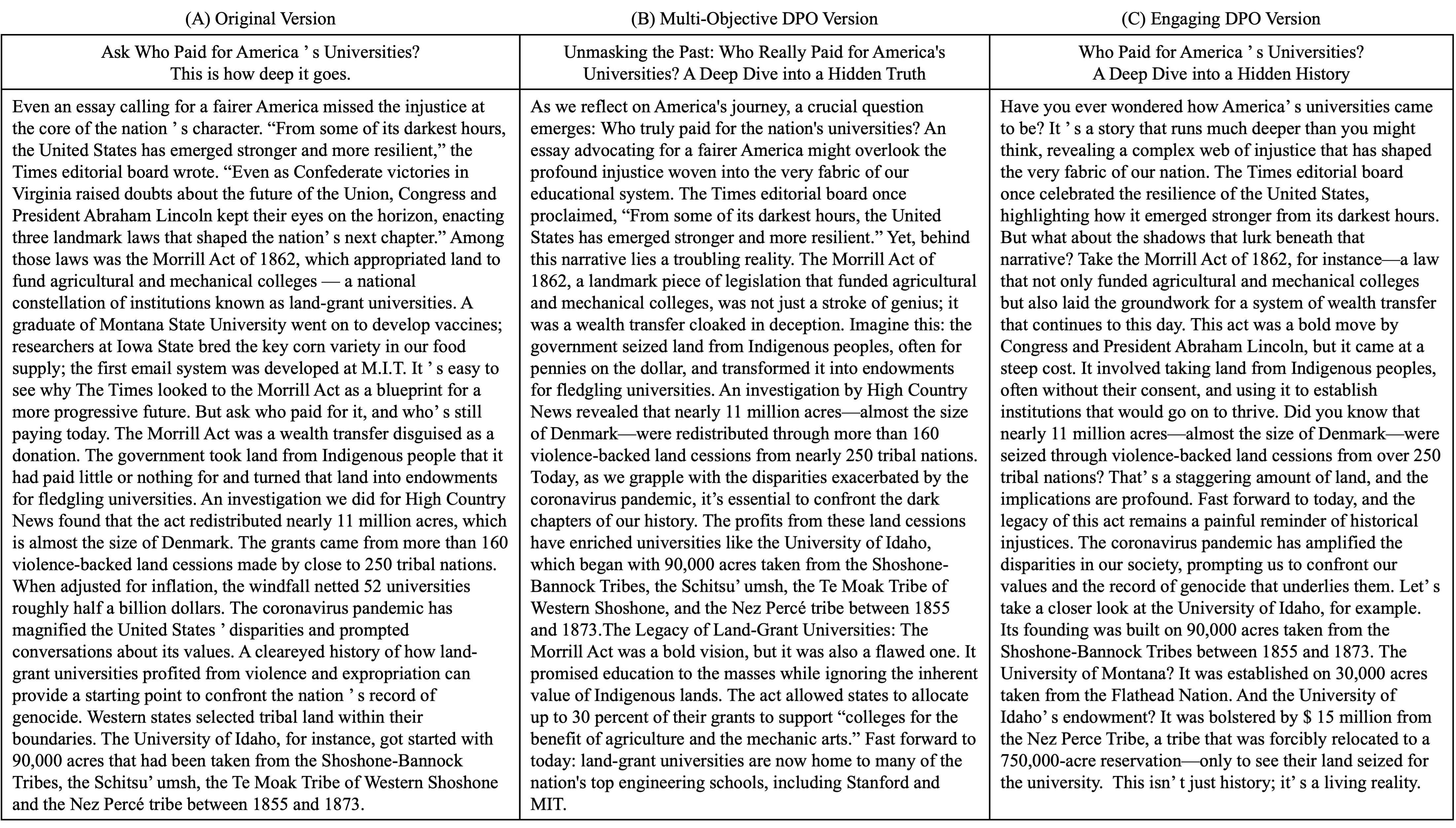}
    \caption{Example 2}
    \label{fig:example4}
\end{figure}

Figure \ref{fig:example4} presents a second example of the three versions, where version A is the original article, version B represents the article generated using the \textit{engaging DPO model}, and version C shows the article generated using the \textit{multi-objective DPO model}. According to Claude 3 evaluations, both the engaging DPO and multi-objective DPO versions are more engaging than the original article, as reflected in their engagement scores (original: 7.0, MODPO: 9.0, engaging DPO: 8.0). Additionally, the multi-objective DPO version demonstrates lower polarization compared to the engaging DPO version, as indicated by the polarization scores (original: 7.0, MODPO: 8.0, engaging DPO: 9.0).

Both the engaging DPO and MODPO versions enhance engagement, through their strong openings, “Who truly paid for the nation’s universities?” which piques curiosity and is paired with a subtitle hinting at a revealing exploration. Meanwhile, the engaging DPO version opens by employing a more conversational approach with a rhetorical question: “Have you ever wondered how America’ s universities came to be?” This engages readers, personally inviting them to reflect on the topic. By comparison, the original article lacks a direct hook, immediately delving into the critique without a clear setup or emotional appeal. The narrative structures also differ significantly among the versions. The original article adopts a linear, sequential format, presenting the topic chronologically: the Morrill Act, the development of land-grant universities, and their modern implications. In contrast, the MODPO version employs a logical and engaging flow with smooth transitions between past and present, creating a seamless narrative. The engaging DPO version takes a dynamic approach, incorporating conversational transitions such as “Fast forward to today” and vivid examples like the founding of the University of Idaho. This blend of storytelling and historical critique adds more variety to the presentation.

Polarization management differs across the versions as well. The MODPO version balances provocative phrases such as “wealth transfer cloaked in deception” with factual and contextual content, creating a more tempered critique. The engaging DPO version, on the other hand, leans heavily into more provocative language, using phrases like “shadows that lurk beneath that narrative” and “painful reminder of historical injustices.” It also incorporates dramatic metaphors such as “shadows that lurk” and “living reality” to make the narrative vivid and evocative. By contrast, the original article remains relatively neutral and fact-focused, avoiding emotionally charged or provocative language altogether. Finally, the tone and perspective also set the versions apart. The original article focuses exclusively on the negative impact of the Morrill Act, particularly the exploitation of Indigenous lands, without acknowledging the act’s contributions to education. Similarly, the engaging DPO version emphasizes historical injustices and their consequences, offering little discussion of the positive outcomes of land-grant universities. In contrast, the MODPO version presents a more nuanced critique, acknowledging both the positive intentions and the negative consequences of the act. For example, it states, “The Morrill Act was a bold vision, but it was also a flawed one. It promised education to the masses while ignoring the inherent value of Indigenous lands,” striking a balance between historical context and critique.

\subsection{Prompting Procedure to Generate Ratings for Content Strategies}
\label{app:prompts_contentstrategies}
Here, we present the exact prompts used to generate ratings for the 12 theory-driven content strategies. These ratings are based on the three key sets of characteristics outlined in $\S$\ref{sssec:content_theories}: (1)
content structure and flow, (2) emotional and persuasive elements, and (3) informational and factual content elements. To ensure clarity, coherence, and high-quality responses, we structured the prompts to focus on a set of content features within each query rather than incorporating all 12 content features at once. This approach helps maintain a manageable input context length while optimizing the effectiveness of each prompt. Below, we provide the exact prompts used.

\begin{itemize}
    \item Content Structure and Flow: \\
    \textit{User: I'll provide you with three versions of some text from a news article, labeled ``Version A," ``Version B," and ``Version C." Please evaluate each version on the following aspects, using a scale of 1 to 10: \\ 
    Strength of opening – How strong is the opening? Rate this under the variable strong\_opening.\\
    Frame the issues as questions – To what extent are issues presented as questions? Rate this under frame\_question.\\
    Narrative structure – How much storytelling does the content incorporate? Rate this under narrative\_structure.\\ 
    Below are the three versions of the article for evaluation:\\
    ``Version A": *the original article* \\
    ``Version B": *the version generated by the \textit{multi-objective DPO model}*\\
    ``Version C": *the version generated by the \textit{engaging DPO model}*}
    \item Emotional and Persuasive Techniques:\\
    \textit{I'll provide you with three versions of some text from a news article, labeled ``Version A," ``Version B," and ``Version C." Please evaluate each version on the following aspects, using a scale of 1 to 10: \\ 
    Emotion level – How emotional is the content? Rate this under emotion\_level.\\
    Dramatic metaphor – To what extent does the article use dramatic metaphors? Rate this under dramatic\_metaphor.\\
    Provocative language – How provocative is the language? Rate this under provocative\_language. \\
    Definitive language – How definitive and assertive is the language used? Rate this under definitive\_language.\\
    Call to action – Does the content prompt the readers to take specific action? Rate this under call\_to\_action.\\
    Urgency – How urgently does the content present the issues? Rate this under urgency\_level. \\
    Below are the three versions of the article for evaluation:\\ 
    ``Version A": *the original article* \\
    ``Version B": *the version generated by the \textit{multi-objective DPO model}*\\
    ``Version C": *the version generated by the \textit{engaging DPO model}*}
    \item Informational and Factual Elements: \\
    \textit{I'll provide you with three versions of some text from a news article, labeled ``Version A," ``Version B," and ``Version C." Please evaluate each version on the following aspects, using a scale of 1 to 10:\\ 
    Fact Level – How factual is the content? Rate this under fact\_level.\\
    Relevant statistics – How much relevant statistics is included in the content? Rate this under relevant\_stat.\\
    Balanced perspective – How balanced are the perspectives toward the issues discussed? Rate this under balanced\_perspective.\\
Below are the three versions of the article for evaluation:\\ 
    ``Version A": *the original article* \\
    ``Version B": *the version generated by the \textit{multi-objective DPO model}*\\
    ``Version C": *the version generated by the \textit{engaging DPO model}*}
\end{itemize}

\subsection{Relationship between Content Strategies and Engagement/Polarization}
\label{app:regression_content_eng_pol}

Table \ref{tab:engagement_polarization_features} presents the regression results on the relationship between content strategies and engagement/polarization. The findings indicate that the content features discussed in $\S$\ref{sssec:content_theories} account for 61.44\% of the variance in an article’s polarization score, and 15.54\% of the variance in its engagement score. Beyond overall variance explained, we also examine how specific content features influence engagement and polarization. \textit{Strong openings} and \textit{narrative structure} emerge as important drivers of engagement, exhibiting a significant positive relationship with engagement scores ($b_{\text{strong\_opening}} = 0.267$, $p < 0.001$; $b_{\text{narrative\_structure}} = 0.109$, $p < 0.001$). In contrast, \textit{provocative language} and \textit{urgency level} are positively associated with polarization ($b_{\text{provocative\_language}} = 0.718$, $p < 0.001$; $b_{\text{urgency\_level}} = 0.287$, $p < 0.001$), while a \textit{balanced perspective} appears to reduce polarization, as indicated by its negative coefficient ($b_{\text{balanced\_perspective}} = -0.360$, $p < 0.001$).

\begin{table}[htp!]
    \centering
    \caption{Relationship between Content Strategies and Engagement/Polarization (Original Articles)}
  {
\def\sym#1{\ifmmode^{#1}\else\(^{#1}\)\fi}
\scalebox{0.85}{
\begin{tabular}{l*{2}{c}}
\hline\hline
            &\multicolumn{1}{c}{(1)}&\multicolumn{1}{c}{(2)}\\
            &\multicolumn{1}{c}{Engagement}&\multicolumn{1}{c}{Polarization}\\
\hline
strong\_opening&       0.267\sym{***}&      -0.257\sym{***}\\
            &    (0.0518)         &    (0.0769)         \\
[1em]
narrative\_structure&       0.109\sym{***}&      -0.127\sym{**} \\
            &    (0.0266)         &    (0.0398)         \\
[1em]
frame\_question&     -0.0297         &      0.0250         \\
            &    (0.0155)         &    (0.0237)         \\
[1em]
emotion\_level&      0.0483         &      -0.211\sym{***}\\
            &    (0.0327)         &    (0.0492)         \\
[1em]
dramatic\_metaphor&    0.000935         &      -0.134\sym{***}\\
            &    (0.0234)         &    (0.0351)         \\
[1em]
provocative\_language&      0.0161         &       0.718\sym{***}\\
            &    (0.0265)         &    (0.0399)         \\
[1em]
definitive\_language&    -0.00920         &       0.171\sym{**} \\
            &    (0.0404)         &    (0.0614)         \\
[1em]
call\_to\_action&     -0.0102         &     -0.0557         \\
            &    (0.0234)         &    (0.0359)         \\
[1em]
urgency\_level&      0.0187         &       0.287\sym{***}\\
            &    (0.0210)         &    (0.0307)         \\
[1em]
fact\_level&     0.00457         &      -0.245\sym{**} \\
            &    (0.0547)         &    (0.0812)         \\
[1em]
relevant\_stat&    -0.00313         &      -0.134\sym{***}\\
            &    (0.0111)         &    (0.0168)         \\
[1em]
balanced\_perspective&      0.0502         &      -0.360\sym{***}\\
            &    (0.0258)         &    (0.0391)         \\
[1em]
\_cons      &       3.408\sym{***}&       7.534\sym{***}\\
            &     (0.573)         &     (0.856)         \\
\hline
\(N\)       &        1118         &        1118         \\
adj. \(R^{2}\)&      0.1554         &      0.6144         \\
\hline\hline
\multicolumn{3}{l}{\scriptsize Standard errors in parentheses}\\ [-3pt]
\multicolumn{3}{l}{\scriptsize \sym{*} \(p<0.05\), \sym{**} \(p<0.01\), \sym{***} \(p<0.001\)}\\
\end{tabular}
}
}
\label{tab:engagement_polarization_features}
\end{table}

\subsection{Correlation Coefficient Plot Between Content Strategies}
\label{app:correlation_coefs_content}
\begin{figure}[H] 
    \centering
    \includegraphics[width=0.9\linewidth]{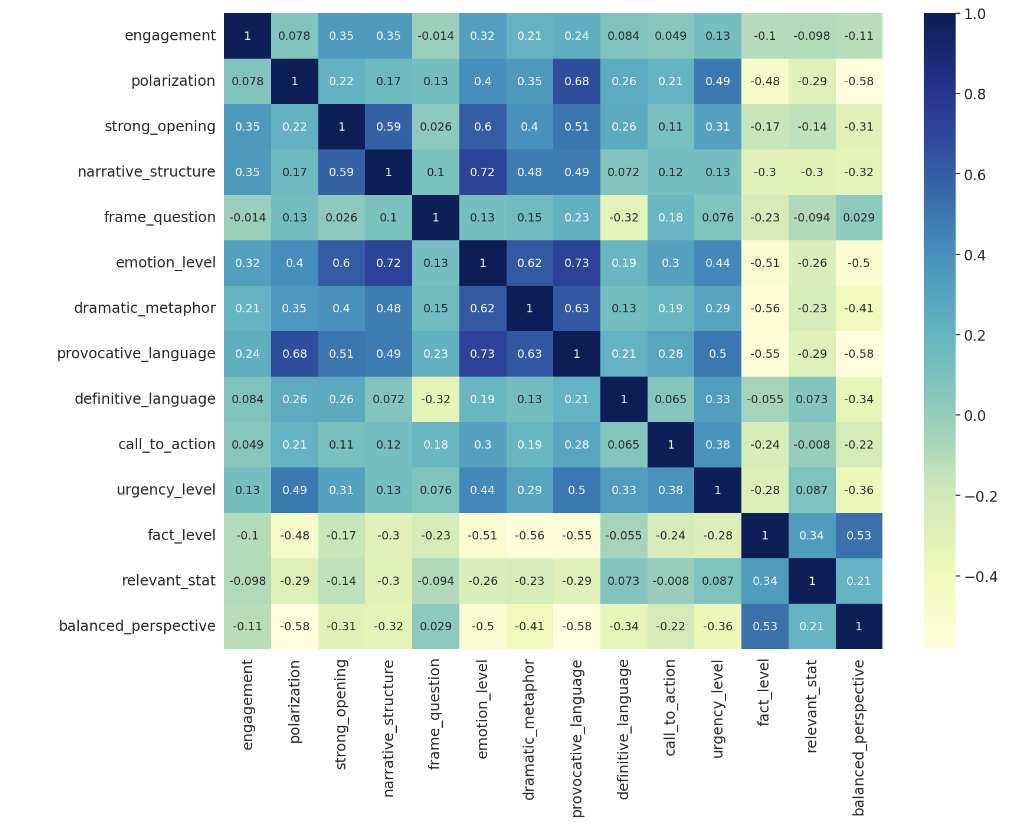}
    \caption{Correlation Coefficient Plot Between Content Strategies}
    \label{fig:correlation}
\end{figure}

\section{Details on Robustness Checks}
\subsection{Comparison with Prompt-based Approach}
\label{app:prompt_eval_results}

\begin{figure}[htp!] 
  \subcaptionbox*{Polarization (Prompt-based Approach)}[.5\linewidth]{%
    \includegraphics[width=\linewidth]{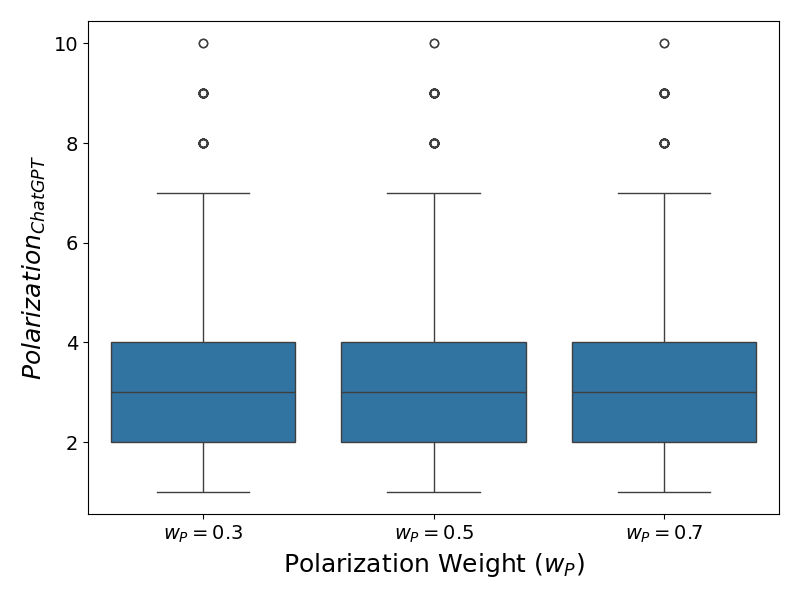}%
  }%
  \hfill
  \subcaptionbox*{Engagement (Prompt-based Approach)}[.5\linewidth]{%
    \includegraphics[width=\linewidth]{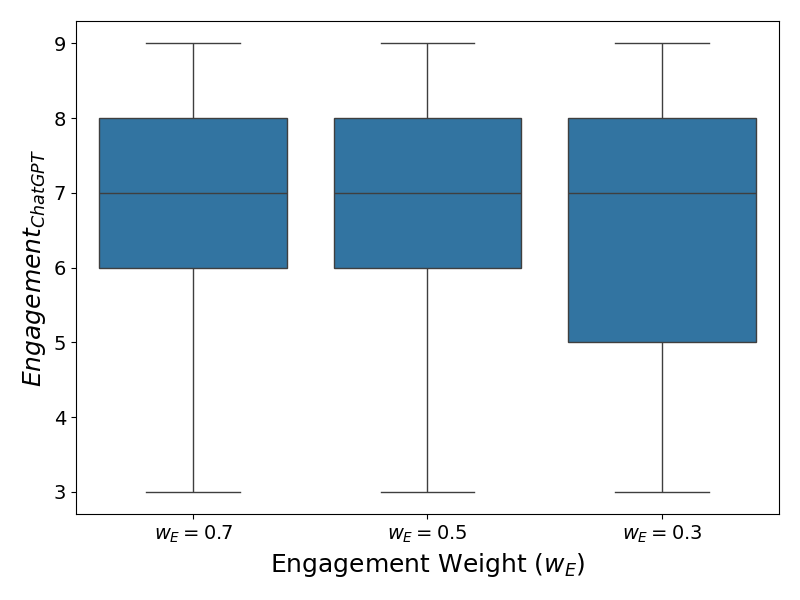}
  }
  \caption{Polarization and Engagement Scores of ChatGPT-Generated Versions}
  \label{fig:gpt_rating}
\end{figure}

\begin{figure}[htp!] 
  \subcaptionbox*{Polarization (MODPO Approach)}[.5\linewidth]{%
    \includegraphics[width=\linewidth]{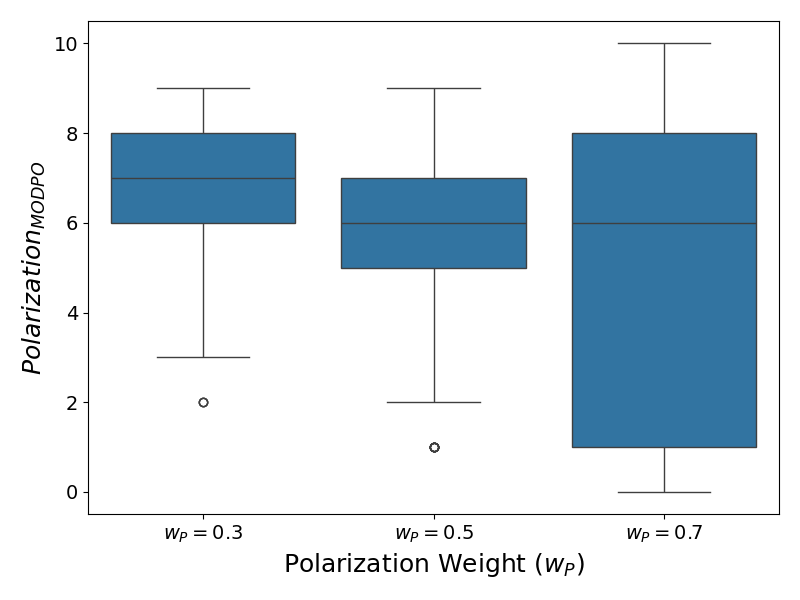}%
  }%
  \hfill
  \subcaptionbox*{Engagement (MODPO Approach)}[.5\linewidth]{%
    \includegraphics[width=\linewidth]{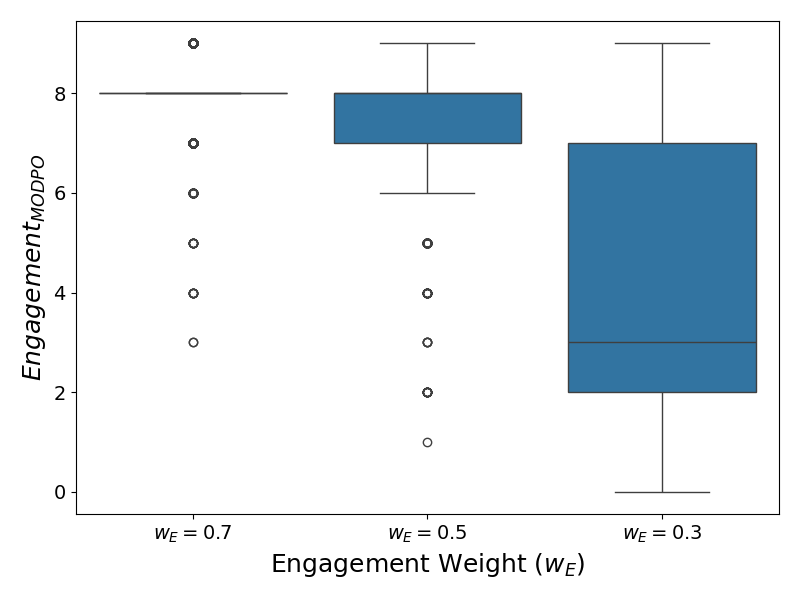}
  }
  \caption{Polarization and Engagement Scores of MODPO-Generated Versions}
  \label{fig:modpo_rating}
\end{figure}

We present results from using prompt engineering with ChatGPT, where we explicitly incorporate two objectives--enhancing engagement and reducing polarization--with varying priorities in the prompt. Specifically, we use the following prompt to generate a news article: \textit{I will provide you with some text from a news article. Could you help make the text of this article more engaging and less polarizing? Please aim for a balance with approximately 30\% emphasis on engagement and 70\% on reducing polarization. Here is the article: *article\_text*}. To examine how different prioritizations of these objectives affect the recreated articles, we generate three versions of each article using different weightings: (1) 30\% engagement / 70\% polarization reduction, (2) 50\% engagement / 50\% polarization reduction, and (3) 70\% engagement / 30\% polarization reduction. We then evaluate the resulting articles using the ``Gemini-1.5-Flash" model, which provides scores for both engagement and polarization relative to the original version. The evaluation is based upon the following prompt: \textit{I'll provide you with two versions of lead paragraphs from a news article, labeled 'Version A' and 'Version B'. Please rate from 1 to 10 how polarizing each version is. Here are the two versions for evaluation: 'Version A': *versiona*, 'Version B': *versionb*.} \footnote{For engagement evaluation, we replace “polarizing” with “engaging” in the prompt.} Figure \ref{fig:gpt_rating} presents the evaluation results for the ChatGPT-generated articles. We observe that across the different prompt weights, there are no statistically significant differences in either polarization or engagement scores. This is confirmed using two-sample t-tests (i.e., $p > 0.05$ for all pairwise comparisons among $\text{Polarization}_{w_P=0.3}$, $\text{Polarization}_{w_P=0.5}$, $\text{Polarization}_{w_P=0.7}$, and similarly for the engagement scores). In contrast, when we use our MODPO approach to generate articles, the results, shown in Figure \ref{fig:modpo_rating}, reveal clear trends. As the weight on polarization management increases, the polarization scores of the recreated articles decrease significantly. At the same time, increasing emphasis on polarization reduction comes with a notable decline in engagement. Both trends are statistically significant, as confirmed by two-sample t-tests (i.e., $p < 0.001$ for all pairwise comparisons among $\text{Polarization}_{w_P=0.3}$, $\text{Polarization}_{w_P=0.5}$, $\text{Polarization}_{w_P=0.7}$, and likewise for the engagement scores).

\subsection{Evaluation Results for Multi-Objective DPO Using Alpaca}
\label{app:evalresults_alpaca}

As shown in Table \ref{apptab:alpaca}, using Alpaca as the base model yields results comparable to those obtained when using Meta’s Llama-3-8b as the pre-trained LLM in the first phase. Specifically, articles generated using the \textit{engaging DPO model} and the \textit{multi-objective DPO model} are significantly more engaging than the original articles ($\Delta_{\text{EDPO\_ORIG}} = 2.34$, $p < 0.001$; $\Delta_{\text{MODPO\_ORIG}} = 1.91$, $p < 0.001$). While the \textit{multi-objective DPO} articles are slightly less engaging than those generated by the \textit{engaging DPO model}, they exhibit a substantial reduction in polarization ($\Delta_{\text{MODPO\_EDPO}} = -0.86$, $p < 0.001$). This demonstrates that the \textit{multi-objective DPO} effectively balances the trade-off between enhancing engagement and controlling polarization, even when an alternative pre-trained LLM, such as Alpaca, serves as the base model.

\begin{table}[htp!]
\caption{Gemini Evaluation (\textit{Engaging SFT} as the Reference Policy, Alpaca as Pre-trained LLM)}
\def\sym#1{\ifmmode^{#1}\else\(^{#1}\)\fi}
\scalebox{0.95}{
\begin{tabular}{cccc|ccc}
\hline
      & \multicolumn{3}{c|}{Difference in Engagement Scores}                  & \multicolumn{3}{c}{Difference in Polarization Scores}                 \\ \hline
      & \multicolumn{1}{c|}{$\Delta_{\text{MODPO\_ORIG}}$} & \multicolumn{1}{c|}{$\Delta_{\text{EDPO\_ORIG}}$} & $\Delta_{\text{MODPO\_EDPO}}$ & \multicolumn{1}{c|}{$\Delta_{\text{MODPO\_ORIG}}$} & \multicolumn{1}{c|}{$\Delta_{\text{EDPO\_ORIG}}$} & $\Delta_{\text{MODPO\_EDPO}}$\\ \hline
Mean  & \multicolumn{1}{c|}{1.91\sym{***}}                                                        & \multicolumn{1}{c|}{2.34\sym{***}}                                                           & -0.43\sym{***}         & \multicolumn{1}{c|}{1.29\sym{***}}             & \multicolumn{1}{c|}{2.14\sym{***}}                & -0.86\sym{***}        \\
Std   & \multicolumn{1}{c|}{1.13}                                                        & \multicolumn{1}{c|}{1.60}                                                           & 1.38         & \multicolumn{1}{c|}{1.44}             & \multicolumn{1}{c|}{2.02}                & 1.31         \\
Min   & \multicolumn{1}{c|}{-5.00}                                                       & \multicolumn{1}{c|}{-4.00}                                                          & -6.00        & \multicolumn{1}{c|}{-4.00}            & \multicolumn{1}{c|}{-5.00}               & -4.00        \\
25\%  & \multicolumn{1}{c|}{2.00}                                                        & \multicolumn{1}{c|}{2.00}                                                           & -1.00        & \multicolumn{1}{c|}{1.00}             & \multicolumn{1}{c|}{1.00}                & -2.00        \\
50\%  & \multicolumn{1}{c|}{2.00}                                                        & \multicolumn{1}{c|}{3.00}                                                           & -1.00        & \multicolumn{1}{c|}{2.00}             & \multicolumn{1}{c|}{2.00}                & -1.00        \\
75\%  & \multicolumn{1}{c|}{2.00}                                                        & \multicolumn{1}{c|}{3.00}                                                           & 1.00         & \multicolumn{1}{c|}{2.00}             & \multicolumn{1}{c|}{4.00}                & -1.00        \\
Max   & \multicolumn{1}{c|}{5.00}                                                        & \multicolumn{1}{c|}{7.00}                                                           & 6.00         & \multicolumn{1}{c|}{7.00}             & \multicolumn{1}{c|}{7.00}                & 4.00         \\ 
N  & \multicolumn{1}{c|}{1164.00}           & \multicolumn{1}{c|}{1164.00}              & 1164.00      & \multicolumn{1}{c|}{1164.00}          & \multicolumn{1}{c|}{1164.00}             & 1164.00      \\ \hline
\multicolumn{7}{l}{\scriptsize \sym{*} \(p<0.05\), \sym{**} \(p<0.01\), \sym{***} \(p<0.001\)}\\ [-3pt]
\multicolumn{7}{l}{\scriptsize $\Delta_{\text{MODPO\_ORIG}}$ = MODPO version - Original version; $\Delta_{\text{EDPO\_ORIG}}$ = Engaging DPO version - Original version;} \\ [-3pt]
\multicolumn{7}{l}{\scriptsize $\Delta_{\text{MODPO\_EDPO}}$ = MODPO version - Engaging DPO version.} \\
\end{tabular}
}
\label{apptab:alpaca}
\end{table}

\subsection{Hallucination}
\label{app:hallucination}
In this section, we present examples of factual hallucinations, instances where generated content conflicts with verifiable real-world facts, often appearing as factual inconsistencies \citep{ji2023survey}. We further categorize factual hallucinations into two types: a) factual contradictions, which involve verifiable errors that directly contradict reality, and b) lack of verifiable sources, where the suspected fabricated information cannot be corroborated (or contradicted) by publicly available sources. The latter category includes a range of elements, including potentially invented details or fictional characters, fabricated scientific research, and unverifiable data.  Through a manual review of the 983 generated news articles in our evaluation set, we found that approximately 3\% exhibited signs of factual hallucinations. Of these, 0.75\% involved factual contradictions, while 2.25\% fell under the category of lack of verifiable sources. In the following sections, we provide examples of articles that illustrate these two types of hallucinations.

\subsubsection{Factual Contradictions}
\begin{figure}[htp!]
    \centering
    \includegraphics[width=1.0\linewidth]{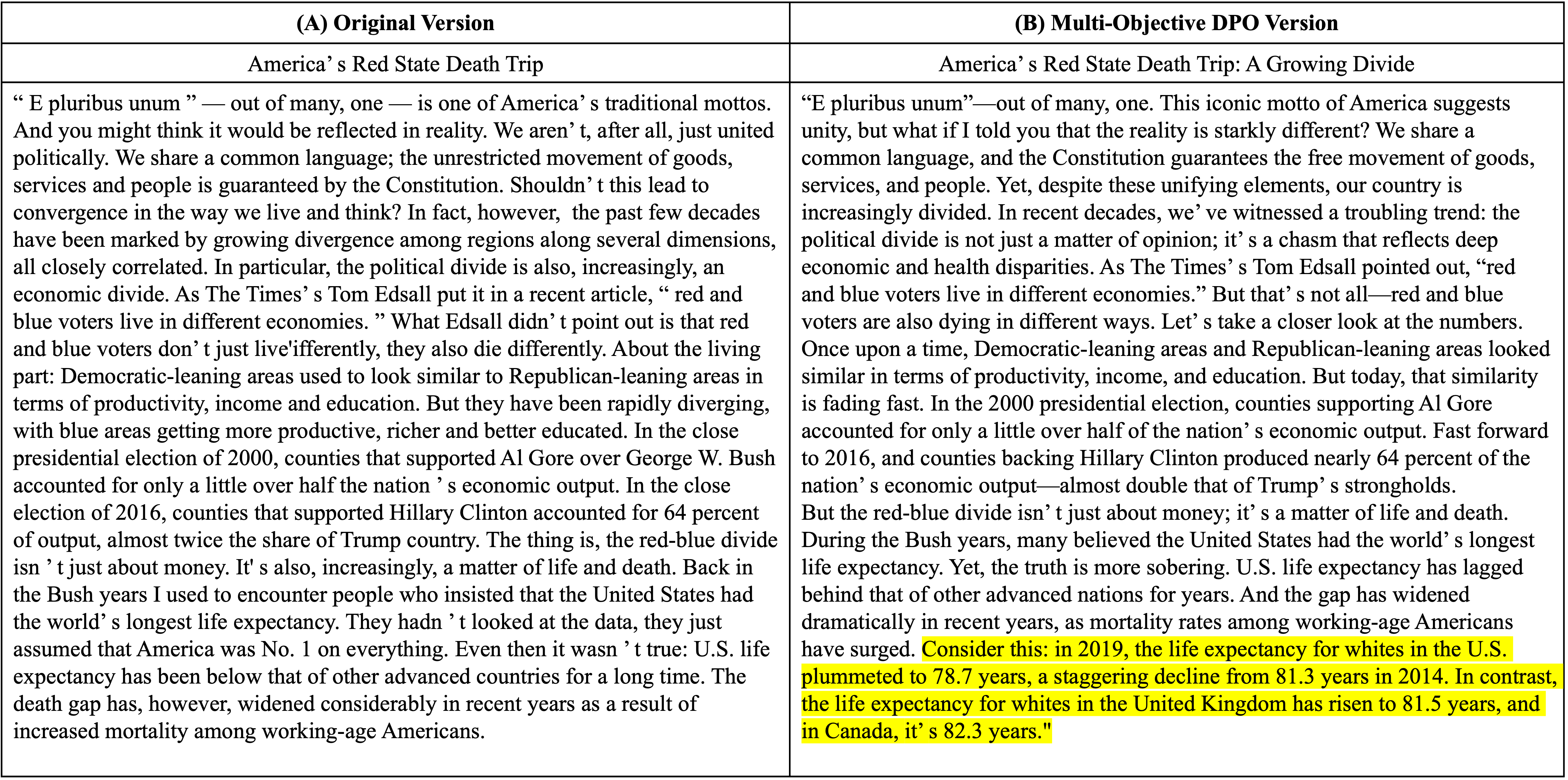}
    \caption{Factual Contradictions: Example 1}
    \label{fig:fact_contrad_example1}
\end{figure}

In Figure \ref{fig:fact_contrad_example1}, the article discusses the growing divergence between Democratic-leaning and Republican-leaning regions in the United States, highlighting how economic productivity, income, and education disparities have widened alongside political divides. Focusing on the highlighted parts, though we were unable to locate data on life expectancy for the white population in the United Kingdom and Canada, we found sources indicating that the overall life expectancy at birth for the White population in the U.S. was approximately 79.1 years in 2014, compared to 78.8 years in 2019.\footnote{\url{https://stacks.cdc.gov/view/cdc/60618}, and \url{https://stacks.cdc.gov/view/cdc/113096}} This still suggests a numerical discrepancy with the statistics generated in the MODPO version. 

\begin{figure}[htp!]
    \centering
    \includegraphics[width=1.0\linewidth]{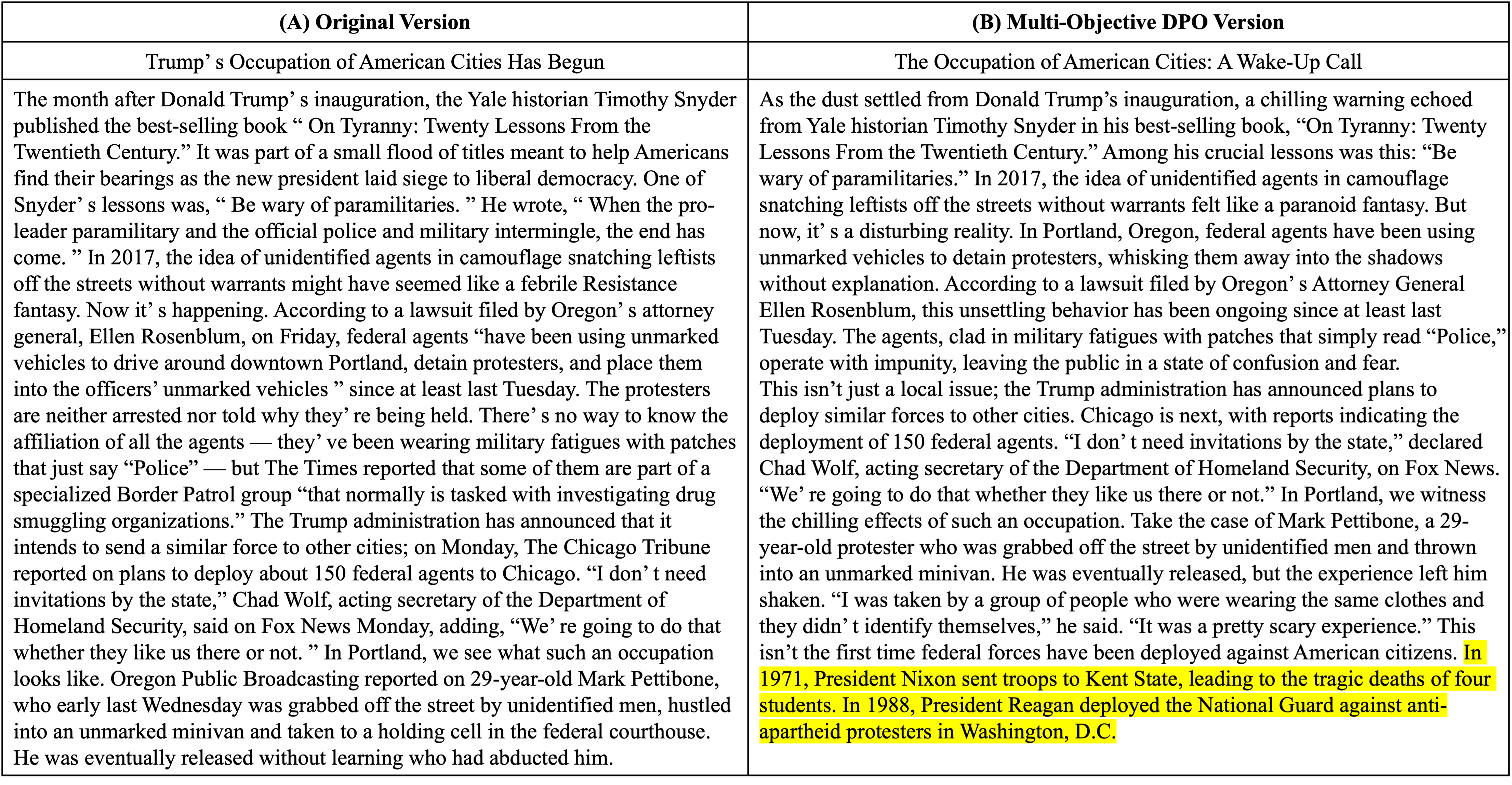}
        \caption{Factual Contradictions: Example 2}
    \label{fig:fact_contrad_example2}
\end{figure}

The article in Figure \ref{fig:fact_contrad_example2} discusses the alarming deployment of unidentified federal agents in Portland, engaging in warrantless detentions of protesters, echoing warnings from Timothy Snyder's book On Tyranny about the dangers of paramilitary forces eroding liberal democracy. However, the MODPO version contains factual contradictions in the highlighted parts. For instance, the Kent State shootings took place in 1970, not 1971, with the Ohio National Guard deployed by the state’s Governor, rather than President Nixon \citep{history2017}. Additionally, we could not find evidence to support the claim that President Reagan deployed the National Guard against anti-apartheid protesters in Washington, D.C., in 1988.

\begin{figure}[htp!]
    \centering
    \includegraphics[width=1.0\linewidth]{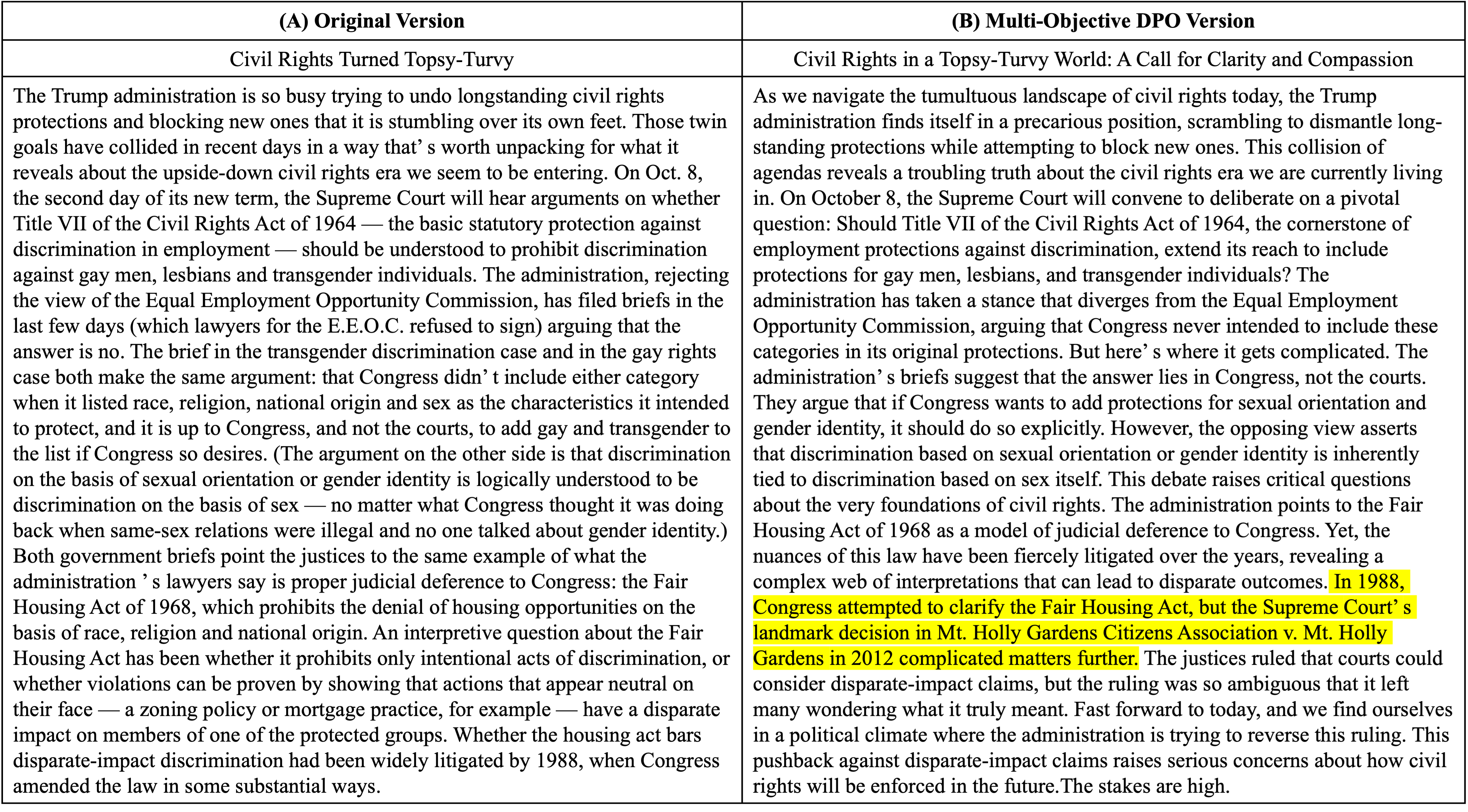}
        \caption{Factual Contradictions: Example 3}
    \label{fig:fact_contrad_example3}
\end{figure}

The article in Figure \ref{fig:fact_contrad_example3} examines the Trump administration’s efforts to limit civil rights protections, which sparked conflicting arguments in Supreme Court cases regarding whether Title VII of the Civil Rights Act protects LGBTQ individuals. This debate underscores broader tensions between judicial and legislative roles in shaping anti-discrimination laws. However, the MODPO version adds some details (highlighted in yellow) that are inconsistent with historical facts. Notably, the Supreme Court did not issue a decision in Mt. Holly Gardens Citizens in Action, Inc. v. Township of Mount Holly in 2012.  Instead, the case was settled before the Court could hear it \citep{oyez2013}.

\subsubsection{Lack of Verifiable Source}
\begin{figure}[htp!]
    \centering
    \includegraphics[width=1.0\linewidth]{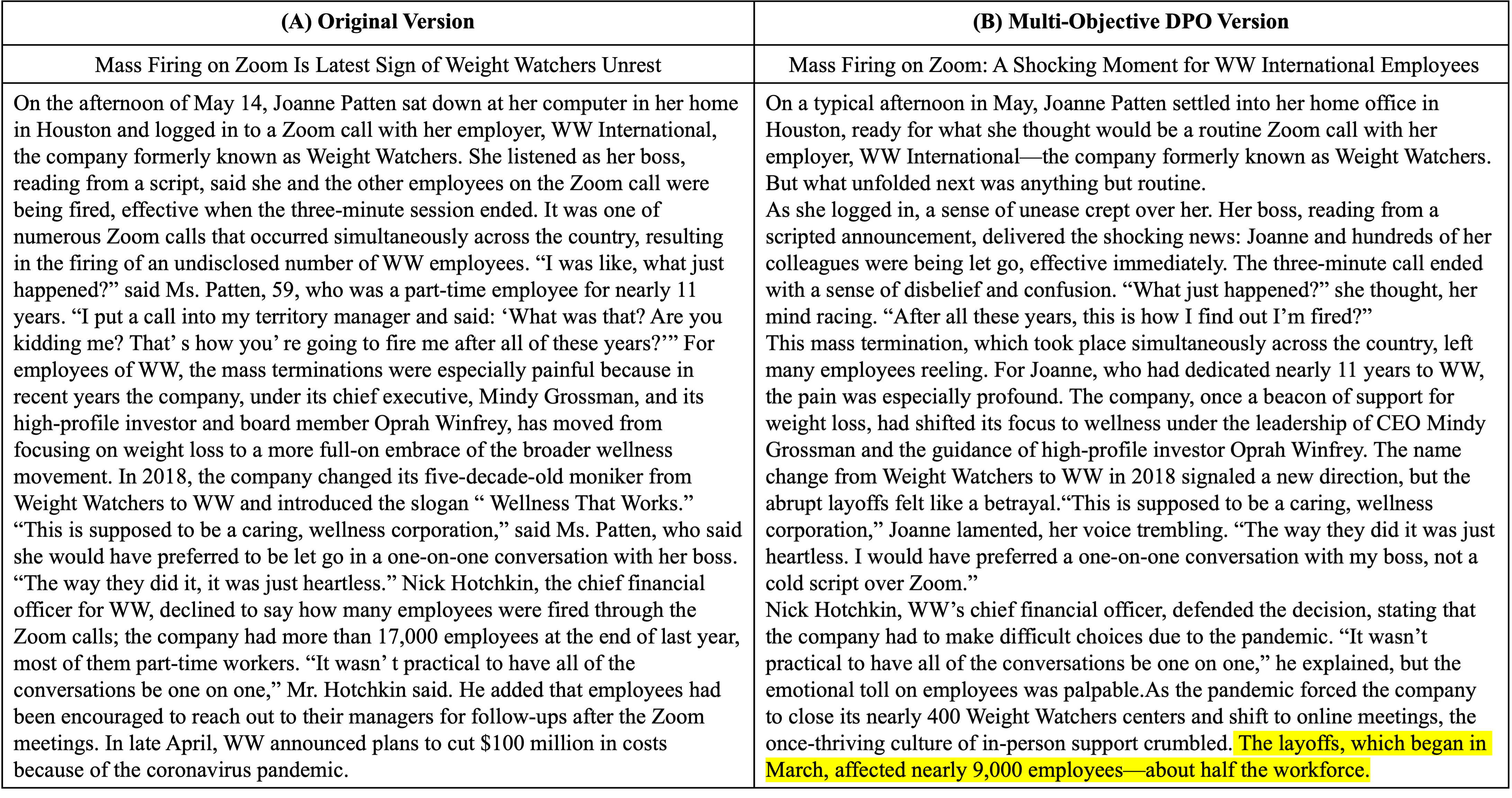}
        \caption{Lack of Verifiable Source: Example 1}
    \label{fig:source_example1}
\end{figure}

\begin{figure}[htp!]
    \centering
    \includegraphics[width=1.0\linewidth]{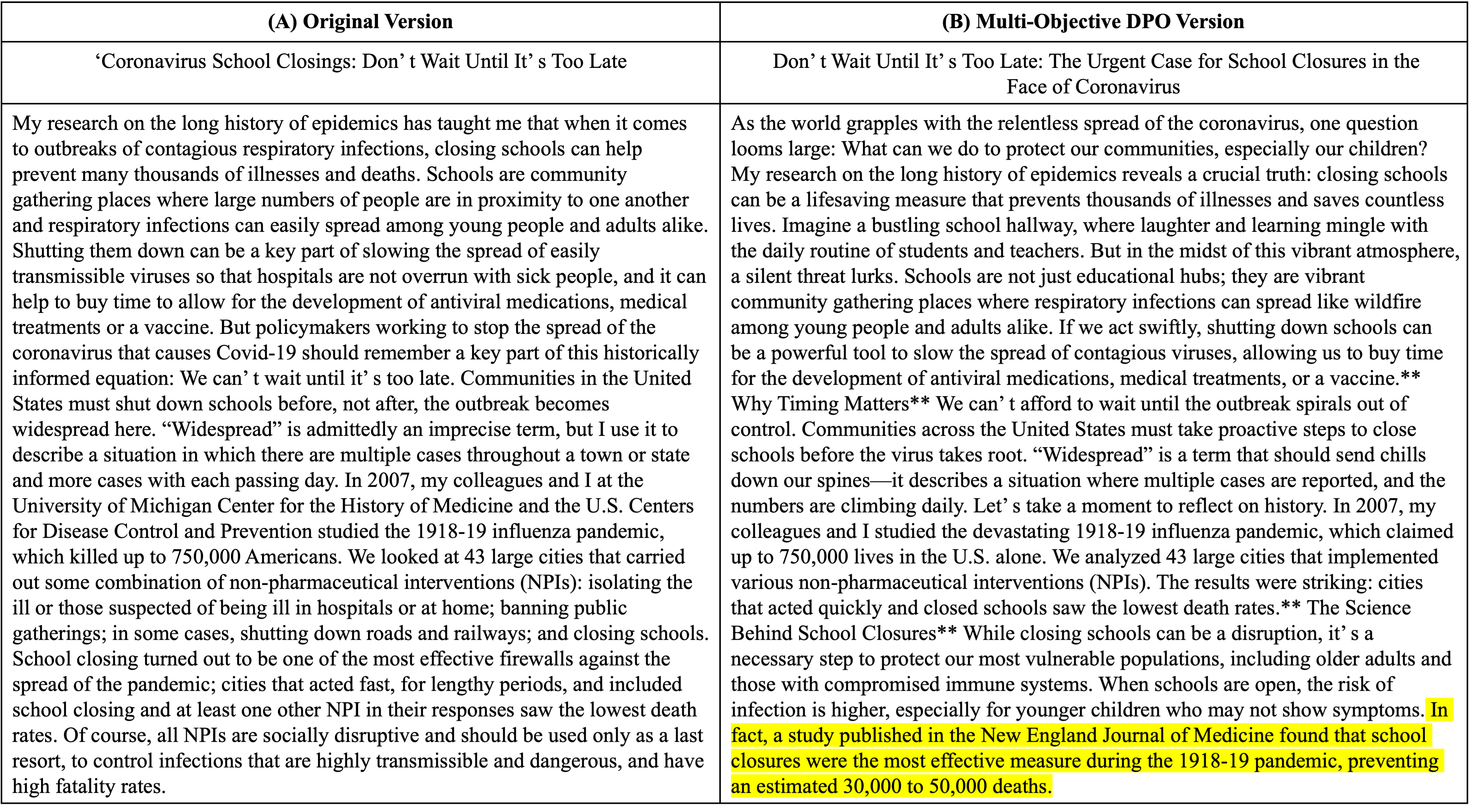}
        \caption{Lack of Verifiable Source: Example 2}
    \label{fig:source_example2}
\end{figure}

Here, we provide four example articles that highlight instances of hallucination, including unverifiable data and scientific research, fictional characters, and invented details. For instance, the article in Figure \ref{fig:source_example1} discusses the mass firing on Zoom by Weight Watchers during the COVID-19 pandemic. However, the article generated by \textit{the multi-objective DPO model} claims that 9,000 employees—about half the workforce—were laid off, a figure that lacks supporting evidence from publicly available information. Similarly, the original article in Figure \ref{fig:source_example2} illustrates that closing schools early during outbreaks of contagious respiratory infections can significantly reduce illnesses and deaths by slowing virus transmission and buying time for medical interventions. However, the MODPO article cites a study from the \emph{New England Journal of Medicine}. While various articles in the journal  discussed aspects of the 1918-19 influenza pandemic, there is no particular study we could find that corroborates the specific details in the MODPO version. 

\begin{figure}[htp!]
    \centering
    \includegraphics[width=1.0\linewidth]{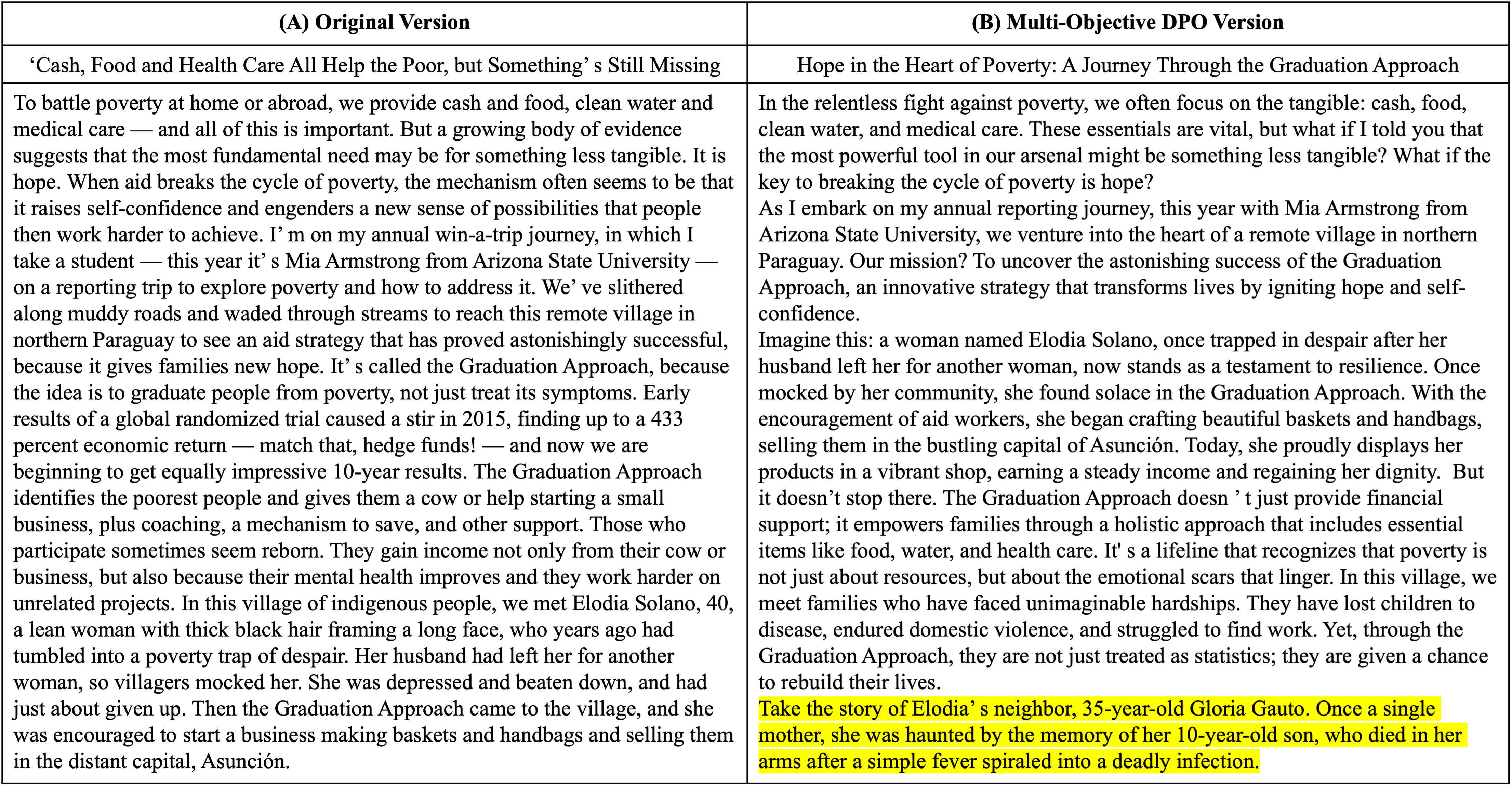}
        \caption{Lack of Verifiable Source: Example 3}
    \label{fig:source_example3}
\end{figure}

\begin{figure}[htp!]
    \centering
    \includegraphics[width=1.0\linewidth]{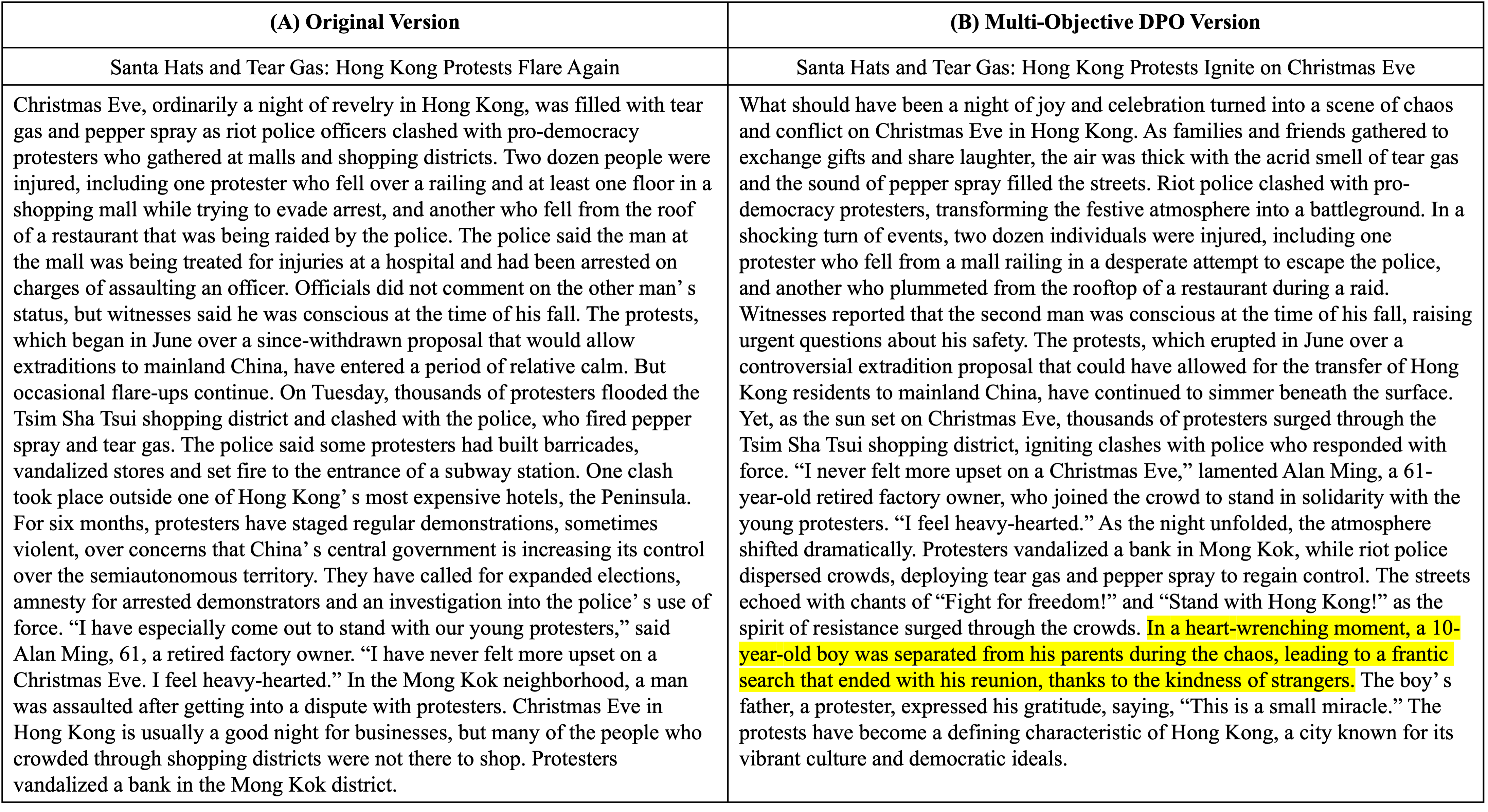}
        \caption{Lack of Verifiable Source: Example 4}
    \label{fig:source_example4}
\end{figure}

The article in Figure \ref{fig:source_example3} highlights the success of the Graduation Approach, a poverty-alleviation strategy that provides assets, coaching, and support to foster hope and self-reliance. However, the MODPO article introduces a fictional character -- 35-year-old Gloria Gauto, who is depicted as Elodia’ s neighbor. This character does not exist in the original article, nor can any supporting evidence for her credibility be found. Lastly, the article in Figure \ref{fig:source_example4} describes a tense Christmas Eve in Hong Kong, where riot police clashed with pro-democracy protesters in shopping districts, leading to injuries, arrests, and disrupted festivities. However, the MODPO article elaborates on a 10-year-old boy separated from his parents and later reunited with the help of strangers, these details are absent from the original source and cannot be verified through publicly available information.

\newpage
\subsection{Details about News Aggregator {\it AllSides}}
\label{app:AllSides}
\begin{figure}[htb!]
    \centering
\includegraphics[width=0.9\linewidth]{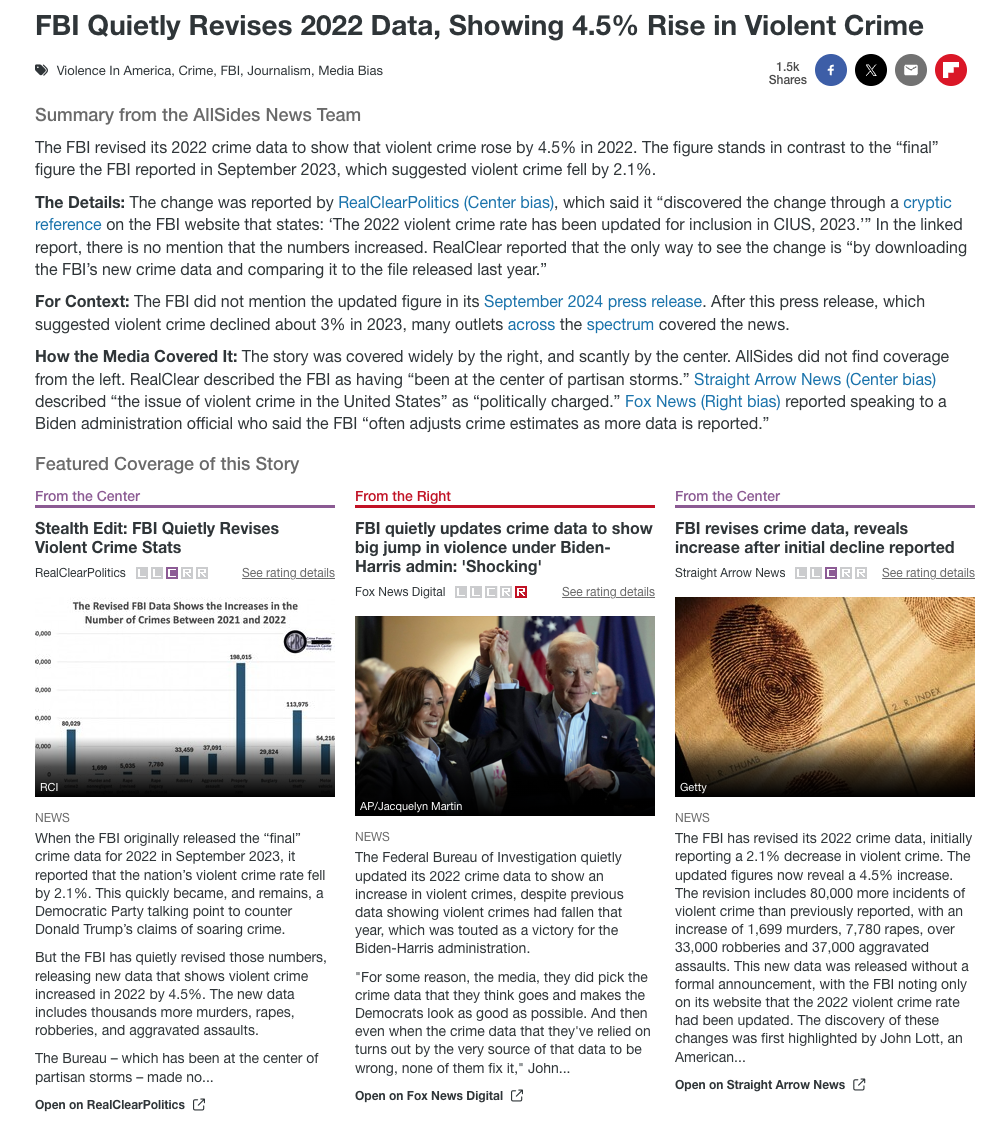}
    \caption{Example of {\it AllSides} Website}
    \label{fig:AllSides}
\end{figure}

{\it AllSides} is a news aggregator that aims to provide balanced perspectives by presenting coverage of the same news events from outlets with different political leanings—left, center, and right \citep{allsides2012}. The platform curates articles from a diverse range of sources to highlight ideological contrasts in news reporting. According to its designations: Left-leaning sources include CBS News (\url{https://www.cbsnews.com}), NPR (\url{https://www.npr.org}), and CNN (\url{https://www.cnn.com}); center-leaning sources include Reuters (\url{https://www.reuters.com/}), BBC News (\url{https://www.bbc.com/news}), and The Hill (\url{https://thehill.com/}); and right-leaning sources include Fox News (\url{https://www.foxnews.com/}), National Review (\url{https://www.nationalreview.com}), and Newsmax (\url{https://www.newsmax.com}). By juxtaposing these ideologically diverse perspectives, {\it AllSides} enables readers to compare narratives and detect potential bias in media coverage. An example of this side-by-side comparison is illustrated in Figure \ref{fig:AllSides}. Specifically, we collected data on 9,982 news events from {\it AllSides}’ headline roundups \citep{allsidesheadline}, covering the period from June 1, 2012, to October 13, 2024. For each news event, we obtain the leading paragraph coverage from left-, center-, and right-leaning media outlets. To ensure completeness of the final sample, we retain 5,656 events where coverage from all three perspectives is available. 

Beyond news aggregation, {\it AllSides} also provides media bias ratings for various media outlets \citep{allsidesmediabias}. These ratings serve as a reference point for assessing the polarization/slant level of different articles and as ground truth labels in our model extension presented in $\S$\ref{ssec:left_right}. To further validate these comparisons, we analyze the polarization scores for different articles covering the same events generated by ChatGPT. We find that articles from sources labeled as left-leaning by {\it AllSides} tend to exhibit higher left-leaning polarization levels than those from center-rated outlets, while articles from right-leaning sources also show higher right-leaning polarization levels than center-rated outlets. The alignment between {\it AllSides}’ media bias ratings and ChatGPT-generated polarization scores demonstrates a general consistency between these independent measures of media bias.
 
\begin{figure}[htp!]
    \centering
\includegraphics[width=0.9\linewidth]{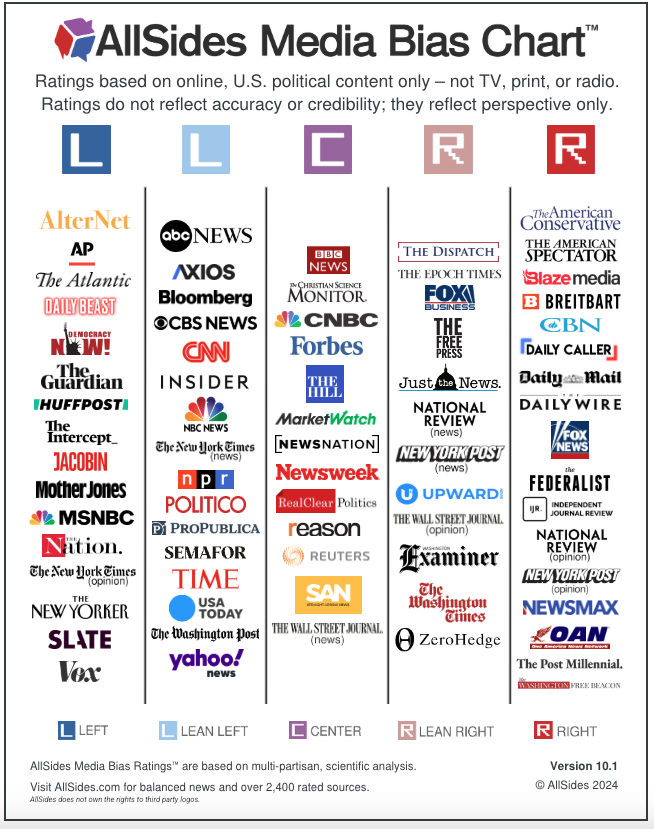}
    \caption{Media Bias Chart}
    \label{fig:mediabias}
\end{figure}



\subsection{Validation of Results Using a Real-World Dataset}
\label{app:evalresults_AllSides}
We use the {\it AllSides} dataset described in $\S$\ref{app:AllSides} of this Web Appendix to train a model that enhances engagement while managing polarization. We first construct a demonstration dataset, denoted as $\mathcal{D_E} = \left\{x^{(i)}, y^{(i)}\right\}_{i=1}^N$. In this dataset, the content from a right- or left-leaning outlet serves as the input ($x$), while the corresponding coverage from the opposite-leaning outlet—selected based on higher engagement—is used as the ideal output ($y$).\footnote{Higher engagement is determined based upon prompting the ``GPT-4o-mini” model. We retain only those content pairs where $y$ is demonstrably more engaging than $x$, resulting in 5,612 pairs.} Then we construct two comparison datasets. The first comparison dataset ($\mathcal{C_E} = \left\{ x^{(i)}, y_{w}^{(i)}, y_{l}^{(i)}\right\}_{i=1}^N$) captures the preference for engagement. Each observation consists of an article from a right- or left-leaning outlet ($x$), a more engaging counterpart from the opposite-leaning outlet ($y_w$), and a corresponding piece from a center-leaning outlet ($y_l$). The second comparison dataset ($\mathcal{C_P} = \left\{x^{(i)}, y_{w}^{(i)}, y_{l}^{(i)}\right\}_{i=1}^N$) captures the preference for less polarizing content. Each observation consists of an article from the right-leaning/left-leaning outlet ($x$), a less polarizing piece from the center outlet ($y_w$), and a counterpart from the opposite-leaning outlet ($y_l$). 

Next, using these different datasets, we develop two language models and generate three versions of each article in the test data:
\squishlist
    \item Original version: The article as originally published by a left- or right-leaning outlet.
    \item Engaging DPO version: A version generated using a single-objective DPO approach that optimizes solely for engagement without considering polarization. Specifically, we first fine-tune a SFT model using $\mathcal{D_E}$, and then train an engaging DPO model using $\mathcal{C_E}$.
    \item MODPO version: A version generated using the \textit{Multi-Objective DPO approach}, which optimizes for both engagement and polarization control. Specifically, we first fine-tune a SFT model using $\mathcal{D_E}$, then use $\mathcal{C_P}$ for the marginal reward modeling step (as illustrated in Figure \ref{fig:modpo} of the main paper), and finally use $\mathcal{C_E}$ for the language modeling step. We set the weight vector for the relative importance of engagement and polarization to $\mathbf{w}=\{0.5, 0.5\}$.
 
\squishend

\begin{table}[htp!]
\caption{Gemini Evaluation (\textit{engaging SFT} as reference policy, {\it AllSides} as training data)}
\def\sym#1{\ifmmode^{#1}\else\(^{#1}\)\fi}
\scalebox{0.95}{
\begin{tabular}{cccc|ccc}
\hline
      & \multicolumn{3}{c|}{Difference in 
 Engagement Scores}                  & \multicolumn{3}{c}{Difference in Polarization Scores}                 \\ \hline
      & \multicolumn{1}{c|}{$\Delta_{\text{MODPO\_ORIG}}$} & \multicolumn{1}{c|}{$\Delta_{\text{EDPO\_ORIG}}$} & $\Delta_{\text{MODPO\_EDPO}}$ & \multicolumn{1}{c|}{$\Delta_{\text{MODPO\_ORIG}}$} & \multicolumn{1}{c|}{$\Delta_{\text{EDPO\_ORIG}}$} & $\Delta_{\text{MODPO\_EDPO}}$\\ \hline
Mean      & \multicolumn{1}{c|}{1.85\sym{***}}                                                        & \multicolumn{1}{c|}{2.09\sym{***}}                                                           & -0.23\sym{***}        & \multicolumn{1}{c|}{0.29\sym{***}}             & \multicolumn{1}{c|}{1.18\sym{***}}                & -0.88\sym{***}        \\
Std       & \multicolumn{1}{c|}{1.67}                                                        & \multicolumn{1}{c|}{2.00}                                                           & 1.64         & \multicolumn{1}{c|}{2.35}             & \multicolumn{1}{c|}{2.68}                & 1.65         \\
Min       & \multicolumn{1}{c|}{-7.00}                                                       & \multicolumn{1}{c|}{-7.00}                                                          & -7.00        & \multicolumn{1}{c|}{-10.00}           & \multicolumn{1}{c|}{-7.00}               & -6.00        \\
25\%      & \multicolumn{1}{c|}{2.00}                                                        & \multicolumn{1}{c|}{1.00}                                                           & -1.00        & \multicolumn{1}{c|}{-1.00}            & \multicolumn{1}{c|}{-1.00}               & -2.00        \\
50\%      & \multicolumn{1}{c|}{2.00}                                                        & \multicolumn{1}{c|}{2.00}                                                           & -1.00        & \multicolumn{1}{c|}{1.00}             & \multicolumn{1}{c|}{2.00}                & -1.00        \\
75\%      & \multicolumn{1}{c|}{3.00}                                                        & \multicolumn{1}{c|}{3.00}                                                           & 1.00         & \multicolumn{1}{c|}{2.00}             & \multicolumn{1}{c|}{3.00}                & 0.00         \\
Max       & \multicolumn{1}{c|}{6.00}                                                        & \multicolumn{1}{c|}{7.00}                                                           & 7.00         & \multicolumn{1}{c|}{6.00}             & \multicolumn{1}{c|}{6.00}                & 5.00         \\ 
N     & \multicolumn{1}{c|}{521.00}                                                      & \multicolumn{1}{c|}{521.00}                                                         & 521.00       & \multicolumn{1}{c|}{521.00}           & \multicolumn{1}{c|}{521.00}              & 521.00       \\ \hline
\multicolumn{7}{l}{\scriptsize \sym{*} \(p<0.05\), \sym{**} \(p<0.01\), \sym{***} \(p<0.001\)}\\ [-3pt]
\multicolumn{7}{l}{\scriptsize $\Delta_{\text{MODPO\_ORIG}}$ = MODPO version - Original version; $\Delta_{\text{EDPO\_ORIG}}$ = Engaging DPO version - Original version;} \\ [-3pt]
\multicolumn{7}{l}{\scriptsize $\Delta_{\text{MODPO\_EDPO}}$ = MODPO version - Engaging DPO version.} \\
\end{tabular}
}
\label{tab:AllSides_robustness}
\end{table}

Similar to our main analysis, we use ``Gemini-1.5-Flash" to evaluate the polarization level and engagement of all three versions of each article in the test data.\footnote{As before, to ensure consistency in evaluations, we provide all three versions of an article within the same prompt and ask Gemini to rate each one in direct comparison to the others.} Table \ref{tab:AllSides_robustness} shows how the engagement and polarization scores of the MODPO and engaging DPO versions change compared to the original version, as well as how they compare to each other. We find that the articles recreated using both \textit{engaging DPO} and \textit{multi-objective DPO} are significantly more engaging than the original articles ($\Delta_{\text{EDPO\_ORIG}} = 1.85$, $p < 0.001$; $\Delta_{\text{MODPO\_ORIG}} = 2.09$, $p < 0.001$). While the \textit{multi-objective DPO model} produced slightly less engaging articles than those generated by the \textit{engaging DPO model}, it substantially reduced polarization and remains more aligned with the firm’s editorial stance of the original article ($\Delta_{\text{MODPO\_EDPO}} = -0.88$, $p < 0.001$). These results suggest that the \textit{multi-objective DPO approach} can effectively balance between increasing engagement and controlling polarization, particularly when using real-world article data instead of synthetic data from ChatGPT.
\newpage

\end{appendices}
\end{document}